\documentclass[onecolumn]{aastex63}

\usepackage[caption=false]{subfig} 
\usepackage{graphicx}
\usepackage{amssymb}
\usepackage{pifont}
\usepackage{hyperref}
\usepackage{color}
\usepackage{xspace}

\newcommand{\eqb}{\begin{eqnarray}}
\newcommand{\eqe}{\end{eqnarray}}
\newcommand{\sth}{\sigma_{\rm T}}
\newcommand{\ath}{ATHE$\nu$A~}
\newcommand{\am}{AM$^3$~}
\newcommand{\paris}{LeHa-Paris~}
\newcommand{\boet}{B13~}
\newcommand{\lehamoc}{LeHaMoC~}

\definecolor{frenchblue}{rgb}{0.0, 0.45, 0.73}
\definecolor{burgundy}{rgb}{0.5, 0.0, 0.13}
\definecolor{darkspringgreen}{rgb}{0.0, 0.65, 0.27}
\definecolor{blue}{rgb}{0.2, 0.2, 0.6}
\definecolor{indigo}{rgb}{0.0, 0.25, 0.42}
\definecolor{bluegray}{rgb}{0.4, 0.6, 0.8} 
\definecolor{lightgray}{rgb}{0.85, 0.85, 0.85}
\definecolor{gray}{rgb}{0.65, 0.65, 0.65}
\definecolor{cadetblue}{rgb}{0.37, 0.62, 0.63}
\definecolor{cambridgeblue}{rgb}{0.64, 0.76, 0.68}
\definecolor{paleaqua}{rgb}{0.74, 0.83, 0.9}
\definecolor{oldlace}{rgb}{0.99, 0.96, 0.9}
\definecolor{platinum}{rgb}{0.9, 0.89, 0.89}
\definecolor{applegreen}{rgb}{0.55, 0.71, 0.0}
\definecolor{cornsilk}{rgb}{1.0, 0.97, 0.86}
\definecolor{lavender}{rgb}{0.9, 0.9, 0.98}
\definecolor{straw}{rgb}{0.89, 0.85, 0.44}

\newcommand{\xmark}{\textcolor{indigo}{\ding{55}}}%
\newcommand{\cmark}{\textcolor{applegreen}{\ding{51}}}%

\received{\today}
\revised{}
\accepted{}
\submitjournal{ApJS}

\graphicspath{{./}{figs/}}

\begin{document}
\title{A Comprehensive Hadronic Code Comparison for Active Galactic Nuclei}

\correspondingauthor{Matteo Cerruti (for the \paris code), Annika Rudolph (for the \am code), Maria Petropoulou (for the \ath code), Markus B\"ottcher (for the \boet code), and Stamatios Stathopoulos (for the \lehamoc code)}
\email{cerruti@apc.in2p3.fr; contact-am3@desy.de;  mpetropo@phys.uoa.gr; markus.boettcher@nwu.ac.za; stamstath@phys.uoa.gr}

\author[0000-0001-7891-699X]{Matteo Cerruti}
\affiliation{Universit\'{e} Paris Cit\'{e}, CNRS, Astroparticule et Cosmologie, F-75013 Paris, France}

\author[0000-0003-2040-788X]{Annika Rudolph}
\affiliation{Deutsches Elektronen-Synchrotron DESY, 
Platanenallee 6, 15738 Zeuthen, Germany}
\affiliation{Niels Bohr International Academy and DARK, 
Niels Bohr Institute, University of Copenhagen, 
Blegdamsvej 17, \\
2100, Copenhagen, Denmark}

\author[0000-0001-6640-0179]{Maria Petropoulou}
\affiliation{Department of Physics, National and Kapodistrian University of Athens, University Campus Zografos, GR 15784, Athens, Greece}
\affiliation{Institute of Accelerating Systems \& Applications, University Campus Zografos, Athens, Greece}
    
\author[0000-0002-8434-5692]{Markus B{\"o}ttcher}
\affiliation{Centre for Space Research, North-West University, Private Bag X6001, Potchefstroom 2520, South Africa}

\author[0000-0002-1344-3754]{Stamatios I. Stathopoulos}
\affiliation{Department of Physics, National and Kapodistrian University of Athens, University Campus Zografos, GR 15784, Athens, Greece}
\affiliation{Institute of Accelerating Systems \& Applications, University Campus Zografos, Athens, Greece}

\author[0000-0002-0525-3758]{Foteini Oikonomou}
\affiliation{European Southern Observatory, Karl-Schwarzschild-Str. 2, Garching bei M{\"u}nchen D-85748, Germany}

\author[0000-0002-3368-3739]{Stavros Dimitrakoudis}
\affiliation{Department of Physics, National and Kapodistrian University of Athens, University Campus Zografos, GR 15784, Athens, Greece}
\affiliation{Institute of Accelerating Systems \& Applications, University Campus Zografos, Athens, Greece}

\author[0000-0003-0102-5579]{Anton Dmytriiev}
\affiliation{Centre for Space Research, North-West University, Private Bag X6001, Potchefstroom 2520, South Africa}
\affiliation{Laboratoire Univers et Th\'{e}ories, Observatoire de Paris, Universit\'{e}  PSL, CNRS, Universit\'{e} Paris Cit\'{e}, 92190 Meudon, France}


\author[0000-0002-5309-2194]{Shan Gao}
\affil{Deutsches Elektronen-Synchrotron DESY, 
Platanenallee 6, 15738 Zeuthen, Germany} 

\author[0000-0003-1096-9424]{Susumu Inoue }
\affiliation{Faculty of Education, Bunkyo University, Koshigaya, Japan}
\affiliation{Astrophysical Big Bang Laboratory, RIKEN, Wako, Japan}
\affiliation{Interdisciplinary Theoretical and Mathematical Sciences Program (iTHEMS), RIKEN, Wako, Japan}


\author[0000-0001-5217-4801]{Apostolos Mastichiadis}
\affiliation{Department of Physics, National and Kapodistrian University of Athens, University Campus Zografos, GR 15784, Athens, Greece}

\author[0000-0002-5358-5642]{Kohta Murase}
\affiliation{Department of Physics; Department of Astronomy \& Astrophysics Penn State University, University Park, USA}
\affiliation{Yukawa Institute for Theoretical Physics, Kyoto, Kyoto 606-8502, Japan}

\author[0000-0001-8604-7077]{Anita Reimer}
\affiliation{Institut f{\"u}r Astro- und Teilchenphysik, Leopold-Franzens-Universit{\"a}t Innsbruck, A-6020 Innsbruck, Austria}

\author[0000-0001-6399-3001]{Joshua Robinson}
\affiliation{Centre for Space Research, North-West University, Private Bag X6001, Potchefstroom 2520, South Africa}

\author[0000-0001-9001-3937]{Xavier Rodrigues}
\affiliation{European Southern Observatory, Karl-Schwarzschild-Straße 2, 85748 Garching bei M\"unchen, Germany}
\affiliation{Excellence Cluster ORIGINS, Boltzmannstr. 2, D-85748 Garching bei M\"unchen, Germany}

 \author[0000-0001-7062-0289]{Walter Winter}
\affil{Deutsches Elektronen-Synchrotron DESY, 
Platanenallee 6, 15738 Zeuthen, Germany}

\author[0000-0002-4388-5625]{Andreas Zech}
\affiliation{Laboratoire Univers et Th\'{e}ories, Observatoire de Paris, Universit\'{e}  PSL, CNRS, Universit\'{e} Paris Cit\'{e}, 92190 Meudon, France}




\begin{abstract}
{We perform the first dedicated comparison of five hadronic codes (AM$^3$, ATHE$\nu$A, B13, LeHa-Paris, and LeHaMoC) that have been extensively used in modeling of the spectral energy distribution (SED) of jetted active galactic nuclei. The purpose of this comparison is to identify the sources of systematic errors (e.g., implementation method of proton-photon interactions) and to quantify the expected dispersion in numerical SED models computed with the five codes. The outputs from the codes are first tested in synchrotron self-Compton scenarios that are the simplest blazar emission models used in the literature. We then compare the injection rates and spectra of secondary particles produced in pure hadronic cases with monoenergetic and power-law protons interacting on black-body and power-law photon fields. We finally compare the photon SEDs and the neutrino spectra for realistic proton-synchrotron and leptohadronic blazar models. We find that the codes are in excellent agreement with respect to the spectral shape of the photons and neutrinos. There is a remaining spread in the overall normalization that we quantify, at its maximum, at the level of $\pm 40\%$. This value should be used as an additional, conservative, systematic uncertainty term when comparing numerical simulations and observations.}

\end{abstract}



\section{Introduction}
\label{sec:intro}
Among the most important open questions in astrophysics is the origin of cosmic rays. They are observed on Earth as a flux of high-energy \citep[up to $10^{20}$ eV,][]{Aab20}, charged particles, mainly protons and light nuclei \citep{AMS21}. Despite extensive searches, the quest for the loci of cosmic-ray acceleration in the Universe remains without a clear answer, especially for the highest energies. Direct searches face the problem that cosmic rays cannot travel on geodesics due to their electric charge, and are deflected in their journey to Earth by magnetic fields. 
Whenever protons (hadrons) are accelerated to relativistic energies they can produce energetic secondary particles (pairs, photons, and neutrinos) through inelastic interactions with low-energy photons and matter that are both commonly found in astrophysical sources. Indirect searches for cosmic-ray accelerators are thus based on the detection of the accompanying high-energy photon fluxes escaping their sources \citep{scienceCTA, AMEGO}.

Indirect cosmic-ray searches face however a major issue: how can we be sure that the photons we are observing have been produced by hadrons? Leptons (mainly electrons and positrons) can also emit high-energy photons and in the large majority of astrophysical sources the two emissions cannot be easily disentangled. The \textit{leptonic} versus \textit{hadronic} discussion is a recurrent topic in high-energy astrophysical studies. This degeneracy can be broken if we also detect neutrinos from the same object. Neutrinos are naturally produced in hadronic interactions together with photons due to the production and decay of pions. On the other hand, they cannot be produced by leptonic radiative processes, and they can thus be seen as the \textit{smoking gun} for the acceleration of hadrons by an astrophysical object, providing key information such as the maximum energy of the accelerated hadrons and their total power. This multi-messenger path has been recently opened with the detection of a diffuse flux of astrophysical neutrinos by IceCube \citep{IceCube2014PRL}. While the gamma-neutrino connection is the most obvious approach to study cosmic-ray accelerators, such multi-messenger observations are still limited: so far the best candidates are the 3$\sigma$ association between a high-energy neutrino detected by IceCube and a gamma-ray flare from the blazar TXS~0506+056 \citep{0506science1}, the 3.5$\sigma$ excess in neutrinos coming from the same source during 2014-2015 \citep{0506science2}, and the 4.2$\sigma$ neutrino excess from the Seyfert galaxy NGC~1068 \citep{1068science}.

Well before the detection of high-energy astrophysical neutrinos, hadronic radiative models have been developed to fit multi-wavelength observations, both in galactic \citep[see][for hadronic emission in supernova remnants]{Drury94, Baring99} and extragalactic sources (see \citet{Mannheim93,Aharonian00, Mucke01} for blazars; see \citet{Bottcher98, Zhang01} for gamma-ray bursts; see \citet{Paglione96, Volk96} for starburst galaxies). 
A considerable effort has been devoted to identify specific spectral and time properties of hadronic models than can be used to discriminate them from leptonic ones \citep{Mucke03, Boettcher05, DMPR12, PM12, Boettcher_2013, MPD13, Diltz15, Zech17}.
Unfortunately, hadronic radiative models are intrinsically more numerically challenging compared to leptonic ones. In the latter, the main radiative processes are the well known synchrotron radiation, inverse Compton scattering, and relativistic bremsstrahlung (the latter being important in the modeling of high-density sources like starburst galaxies and supernova shocks propagating in dense circumburst media). In hadronic processes there is the additional complex task of the computation of the interactions of relativistic protons with low-energy photons and/or matter.  In both cases the final outcome is the production of pions that then decay into photons, leptons, and neutrinos. Photons and leptons can then trigger pair-cascades in the emitting region via photon-photon pair production, thus producing more targets for proton-photon interactions. Hadronic source models are intrinsically non-linear problems, involving physical processes operating typically on vastly different timescales. The numerical complexity of hadronic models in high-energy astrophysics also explains why they have been relatively less investigated compared to leptonic ones. In addition, while a numerical leptonic radiative code can be checked with simple analytical formulae, in hadronic models these first order analytical approximations are not always available or have limited applicability. 

A clear missing step in the scientific literature is a systematic comparison of numerical codes, to estimate the degree of precision reached by current numerical models. This step is particularly urgent at the dawn of the multi-messenger gamma-neutrino astrophysics, now that we can test theoretical models and compare their predicted neutrino rates with the observations performed by current neutrino telescopes. The goal of this work is to perform the first comprehensive comparison of five numerical leptohadronic codes published in the scientific literature, discuss the differences in their numerical implementation of hadronic interactions, and estimate their agreement in various parts of the parameter space. As a first part of this project, we limit ourselves to parameters relevant to blazar modeling. In addition, we include a comparison of the numerical neutrino spectra against those obtained with simple analytical approximations often used in the literature. We release all outputs from the five codes for all the tests we performed, as online material, to be used for benchmarking of new codes. 

The paper is organized as follows: in Section \ref{sec:introonmodels} we describe one-zone blazar emission models and their parameters; in Section \ref{sec:codes} we describe the five codes and the effect of numerical resolution on the outputs of individual codes; in Section \ref{sec:casestudies} we describe the tests we performed; in Section \ref{sec:results} we show the results of the code comparison; in Section \ref{sec:otherscenarios} we discuss non-linear scenarios; in Section \ref{sec:discussion} we discuss the results of the code comparison, and we summarize the conclusions of this work. In appendices we provide details on the semi-analytical neutrino calculation (Appendix \ref{app:analytical}); additional tests we performed (Appendix \ref{app:other}); and also, details on how Bethe-Heitler pair production is implemented in \ath and \lehamoc (Appendix \ref{app:BH}). \\

\section{One-zone blazar emission models}
\label{sec:introonmodels}

Blazars are a subclass of jetted (radio-loud) active galactic nuclei (AGNs) characterized by non-thermal emission from radio to gamma rays, a high degree of polarization, and rapid variability, which may be as short as a few minutes \citep{HESS2155}. Blazars are observed at all wavelengths and are remarkably the most common extra-galactic source type in the gamma-ray sky. The name \textit{blazar} is a portmanteau word, from BL Lacertae (a source initially incorrectly classified as peculiar star, hence its variable star name) and quasars, explicitly reminding us of their blazing behavior. In the framework of the AGN unified model, blazars are understood as the observational effect of a radio-loud AGN whose relativistic jet of plasma launched by the super-massive black hole points towards the observer \citep{Urry95}. Due to the relativistic motion of the emitting plasma, the emission is Doppler boosted in the observer's frame, making these objects much brighter than their parent radio galaxies. Given that only the emission from the jet is boosted, this interpretation explains why blazar emission is dominated by a non-thermal component that outshines all other known AGN radiative components such as the thermal emission from the accretion disk, the broad and narrow emission lines, and the X-ray corona \citep{2017A&ARv..25....2P}. 

Spectral energy distributions (SEDs) of blazars are also peculiar among astrophysical objects: the non-thermal emission, while spanning almost 20 orders of magnitude in energy, is pretty simple with two well separated components, the first one peaking in infrared-to-X-rays, the second one peaking in gamma rays (MeV-to-TeV). The position of the peaks varies not only within the blazar population (hence prompting further classifications in X-ray and radio selected blazars \citep{PadovaniGiommi}, or in low-intermediate-high frequency peaked blazars \citep{Fermi10}), but also in a single blazar as a function of the flux state \citep[see][for a study of Mrk 421]{Balokovic16}. The origin of the low-energy SED component is well understood and interpreted as synchrotron radiation by a population of electrons (and positrons) in the relativistic jet. This interpretation is observationally supported by the detection of a polarized component from radio up to X-rays \citep[see][for Mrk 421]{digesu22}, by the broad-band measurements of the spectral shape of the emission, and by the detection of the core shift effect due to synchrotron self-absorption. The origin of the second, high-energy, SED component is less clear, and falls into the \textit{leptonic} versus \textit{hadronic} dispute discussed in the Introduction. In leptonic blazar models the high-energy emission is due to inverse Compton scattering, while in hadronic blazar models it is due to proton synchrotron radiation, and/or synchrotron radiation by secondary particles produced in proton-photon or proton-proton interactions \citep[see][for a recent review]{Cerruti20}. It is important to underline that even in hadronic models the first SED component is still explained as synchrotron emission by leptons: the \textit{hadronic} term refers here to the type of accelerated particles responsible (directly or indirectly) for the high-energy part of the blazar SED. 

The simplest model for blazar emission considers an emitting region in the jet moving towards the observer with Doppler factor $\delta$. To further simplify the modeling, a spherical geometry (in the jet rest frame) is commonly assumed, with the radius $R$ as its only parameter. To explain blazar emission two additional ingredients are needed: the magnetic field strength $B$ in the emission region, and a population of relativistic particles. As discussed above, in the radio band the synchrotron emission is absorbed due to synchrotron self-absorption, and emission from this single region in the jet cannot reproduce radio observations. Therefore, \textit{one-zone} blazar models consider a single zone in the jet for producing the observed SED, with the notable exception of the radio that is treated as an upper flux limit in the fit. It is important to underline that the theoretical framework behind one-zone models is not that there is a single, immutable, radiating sphere of plasma in the jet, but rather that, at any given moment, the blazar SED is \textit{dominated} by the emission from one zone. In their simplicity, one-zone models have been successful in explaining the emission from blazars, both for the broad-band SEDs, and for the time variability and multi-wavelength correlations. Exceptions exist, and in particular the detection of objects with unusual SEDs, orphan flares \citep[eruptions seen only in a limited energy band, see e.g.][]{Krawczynski04}, or the observations of peculiar multi-wavelength correlations~\citep[e.g.][]{2020ApJS..248...29A}, often result in extensions towards multi-zone emission scenarios \citep[e.g.][]{2022PhRvD.105b3005W}. \\

\section{The codes} \label{sec:codes}
In this section we give a brief alphabetically ordered overview of the codes participating in this study.  We also provide notes on the development history and astrophysical applications. 

\subsection{\am}

\am \citep{Klinger:2023zzv} is a time-dependent code that solves the coupled kinematic equations of electrons, positrons, photons, protons, neutrons, neutrinos, muons and pions homogeneously distributed in a comoving volume $V$ of radius $R$. The code solves the time- and energy-dependent density distribution of each particle species $i$ following a differential equation of the form

\eqb
    \partial_t n_i(\gamma, t) = - \partial_\gamma\dot{\gamma_i}(\gamma,t)n_i(\gamma,t)  
    -\alpha_i(\gamma, t) n_i(\gamma, t) + Q_i(\gamma, t),  
\label{equ_kinematic_AM$^3$}
\eqe
where the $\dot{\gamma}$ terms encode energy loss processes, as listed below, the $\alpha_i$ are sink terms that encode processes leading to particle disappearance, including physical escape from the region and any interactions that destroy particles of species $i$, and the $Q_i$ are source terms that encode particle injection, including the injection of primary species by the user and  interactions that create particles of species $i$. In this notation, $\alpha_i$, $Q_i$ and $\dot{\gamma_i}$ represent the sum over all processes $j$ for species $i$ ($\alpha_i = \sum_j \alpha_i^j$, and so on).
For electrons, positrons, and photons, the dimensionless energy is given by $\gamma = E / (m_e c^2)$; and for protons, neutrons, neutrinos, pions, and muons, by $\gamma = E / (m_{p\ [GeV]})$.

 The $\dot\gamma$-, $\alpha$-, and $Q$-terms encode the following processes:
\begin{enumerate}
    \item injection of primary particles with an arbitrary distribution ($Q^\mathrm{inj}$) 
    \item particle escape ($\alpha^\mathrm{esc}$)
    \item synchrotron emission by charged particles, including pions and muons ($\dot{\gamma}_\mathrm{e^\pm}^\mathrm{syn}$, $\dot{\gamma}_\mathrm{p}^\mathrm{ syn}$, $\dot{\gamma}_\mathrm{\pi^\pm}^\mathrm{syn}$, $\dot{\gamma}_\mathrm{\mu^\pm}^\mathrm{syn}$, $Q_\mathrm{phot}^\mathrm{syn}$ from $e^{\pm}$, $\pi^{\pm}$, $\mu^{\pm}$ and $p$)
    \item synchrotron-self absorption ($\alpha_\mathrm{phot}^\mathrm{ssa}$)
    \item inverse Compton radiation from $e^{\pm}$ and $p$ ($\dot{\gamma}_\mathrm{e^\pm}^\mathrm{ic}$,s $\alpha_\mathrm{phot}^\mathrm{ic}$, $Q_\mathrm{phot}^\mathrm{ic}$)
    \item photon-photon annihilation ($\alpha_\mathrm{phot}^\mathrm{pp}$, $Q_\mathrm{e^\pm}^\mathrm{pp}$)
    \item proton-photon pion production ($\alpha_\mathrm{phot}^\mathrm{p\gamma\pi}$, $Q_\mathrm{phot}^\mathrm{p\gamma\pi}$, $\alpha_\mathrm{p}^\mathrm{p\gamma\pi}$, $Q_\mathrm{p}^\mathrm{p\gamma\pi}$, $Q_\mathrm{\pi}^\mathrm{p\gamma\pi}$)
    \item neutron-photon pion production ($\alpha_\mathrm{phot}^\mathrm{n\gamma\pi}$, $Q_\mathrm{phot}^\mathrm{n\gamma\pi}$, $\alpha_\mathrm{n}^\mathrm{n\gamma\pi}$, $Q_\mathrm{n}^\mathrm{n\gamma\pi}$, $Q_\mathrm{\pi}^\mathrm{n\gamma\pi}$)
    \item proton-photon (or Bethe-Heitler, BH) pair production ($\alpha_\mathrm{phot}^\mathrm{BH}$, $\dot{\gamma}_\mathrm{p}^\mathrm{BH}$, $Q_\mathrm{e^\pm}^\mathrm{BH}$)
    \item pion and muon decay kinematics and resulting neutrino emission ($\alpha_{\mu^\pm}^\mathrm{dec}$, 
    $\alpha_{\pi^\pm}^\mathrm{dec}$,
    $Q_{e^\pm}^\mathrm{dec}$, 
    $Q_{\mu^\pm}^\mathrm{dec}$, 
    $Q_{\nu}^\mathrm{dec}$)
    \item adiabatic cooling of charged particles ($\alpha^\mathrm{ad}$, $\dot{\gamma}^\mathrm{ad}$)
\end{enumerate}
For the numerical computation, Equation~\ref{equ_kinematic_AM$^3$} is re-written in terms of logarithmic energy $x = \ln\gamma$:
\eqb
    \partial_t n(x, t) = - \partial_x A(x,t)n(x,t)
    -\alpha(x, t) n(x, t) + \epsilon(x, t) \, , 
\label{equ_kinematic_AM$^3$_log}
\eqe
with the differential number density $n(x) = \gamma n(\gamma)$. The physical processes outlined above are implemented as $A$, $\alpha$, and $\epsilon$ terms an Equation~\ref{equ_kinematic_AM$^3$_log}, which can be derived from the initial physical quantities in  Equation~\ref{equ_kinematic_AM$^3$} as

\eqb
A(x) = \frac{\dot{\gamma}}{\gamma} , ~ 
\epsilon(x) = \gamma Q(\gamma), \, ~ \alpha(x) = \alpha(\gamma).
\eqe

\am solves Equations (\ref{equ_kinematic_AM$^3$_log}) for discrete, equally spaced values of $x$, resulting in logarithmically spaced values of $\gamma$. The spacing is fixed to $x_{i+1}-x_i\equiv\ln (\gamma_{i+1}/\gamma_{i}) = 0.1$, corresponding to 23 bins per energy decade. The code uses three separate energy grids for three different particle groups: a leptonic one for electrons and positrons, a hadronic one for protons, neutrons, pions, muons, and neutrinos, and one for photons. The leptonic and hadronic grids start from cold particles, i.e. with $\gamma=1$ ($x_0=\ln(1)$). The maximum value of each of the three grids, as well as the minimum value of the photon grid, can be defined by the user upon initialization. The particle densities are computed in units of $\mathrm{cm}^{3}$ for protons, neutrons, pions, muons, and neutrinos, and in units of $\sigma_T^{3/2}$ for electrons, positrons and photons, where $\sigma_T$ is the Thomson scattering cross-section.

The system is evolved linearly in time $t$. The size of the time step $\Delta t$, is defined by the user upon initialization, and can be subsequently redefined at any moment during the simulation. At every time $t$, \am automatically selects the algorithm to be deployed when calculating the subsequent step, as detailed further in \citet{Klinger:2023zzv}. In summary, if the energy loss terms in \ref{equ_kinematic_AM$^3$_log} dominate over the sink terms or vice-versa ($A(x,t)\gg\alpha(x,t)$ or $A(x,t)\ll\alpha(x,t)$), the non-dominant terms are neglected and an analytical approximation is deployed when calculating $n(x,t+\Delta t)$, increasing the efficiency of the code. If both terms are of comparable magnitude, the full numerical solver is deployed using a tri-diagonal matrix method based on a semi-implicit scheme in $x$ \citep{chang1970a} and a Crank-Nicolson scheme in $t$. The solver updates the coefficients for each species at time $t$, including the integration over energy distributions of target particles when applicable, as in the case of Compton scattering or hadronic interactions. The particle spectrum is then evolved forward by $\Delta t$. The new density value $n(x,t+\Delta t)$ is calculated, as well as the new coefficients $A(x,t+\Delta t)$ and $\alpha(x,t+\Delta t)$. The user is responsible for choosing a sufficiently small value of  $\Delta t$ to maintain accuracy, especially in optically thick regimes, which requires testing the stability of the solution.

The treatment of processes 1-6 for leptons, positrons and protons is described in detail in \citet{Vurm:2008ue, Gao_2017}. The synchrotron photon emission and respective cooling timescale also include the quantum synchrotron regime \citep[following][]{Brainerd_1987}, which is relevant in the presence of strong magnetic fields.
Bethe-Heitler pair production is calculated following the treatment by \citet{Kelner_2008}, with the cross-section given by Eq. (10) of \citet{BG70}, valid in the Born limit \citep[see][]{Klinger:2023zzv} and the expression for the energy loss term given by Eq. (9.35) of \citet{2009herb.book.....D}.  
For photon-proton and photon-neutron pion production, the code follows the procedure outlined by \citet{Hummer:2010vx}, based on {\sc sophia} \citep{Mucke_2000}. In Sec.~\ref{sec:hadronictest} we provide further details on the approach, including the methods used for performance optimization. Neutral pions are not included as an individual species but are assumed to decay instantaneously, while charged pions are treated as individual species that undergo escape and cooling, emit synchrotron radiation, and decay. The same is true for the unstable, intermediate muons emerging from pion decay. The neutrino emission rates are calculated at each time step based on the current pion and muon densities.

The code has been applied to time-dependent multi-wavelength studies of AGN \citep[e.g.][]{Gao_2017, Gao:2018mnu,2023ApJ...958L...2F,Rodrigues:2020fbu,Rodrigues:2024fhu,Rodrigues:2023vbv}, and more recently to the prompt and afterglow phases of gamma-ray bursts (GRBs) \citep{Rudolph:2021cvn,2023ApJ...950...28R,2023ApJ...944L..34R,2024arXiv240313902K} and tidal disruption events (TDEs) \citep{2023ApJ...956...30Y,2024ApJ...969..136Y,2024ApJ...974..162Y}. \\

\subsection{\texorpdfstring{\ath}{ATHENA}}

\ath is a time-dependent leptohadronic radiative transfer code, which was first presented in \cite{MK95}. Since then, it has been updated in various ways and evolved to its current form \citep[for details, see][]{MPK05, DMPR12, PGD14}. The numerical code solves a system of coupled integro-differential equations that describe the evolution of five particle distributions within a fixed spherical volume (of radius $R$), namely relativistic protons, relativistic neutrons, relativistic electrons and positrons,  photons, and neutrinos. The equation of each one of the above particle distributions can be written in the following compact form
\eqb 
\frac{\partial n_{i}(x_i,t)}{\partial t} & + &  \frac{n_{i}(x_i,t)}{t_{i,esc}(x_i)} + \sum_j\mathcal{L}^{j}_i(x_i,t)  =   \sum_{j}\mathcal{Q}^j_i(x_i,t) + \mathcal{Q}^{inj}_i(x_i,t),
\label{equ_kinematic_athena}
\eqe 
where $t$ is time (in units of $R/c$), $n_{i}$ is the differential number density (normalized to $\sth R$) of particle species $i$, $x_i$ is the particle dimensionless energy (in units of $m_i c^2$), $t_{i, esc}$ is the particle escape timescale (also in units of $R/c$), $\mathcal{L}^{j}_i$ is the operator for particle losses (sink term) due to process $j$, $\mathcal{Q}^j_i$ is the operator of  particle injection (source term) due to process $j$, and $\mathcal{Q}^{inj}_i$ is the operator of a generic external injection. The coupling of the equations happens through the energy loss and injection terms for each particle species. When the differential energy loss rate for a particle species $i$ due to process $j$ is integrated over the particle energy grid it equals the bolometric energy injection rate to particle species $i'$ due to the same process,
\eqb
\int {\rm d}x \ x \mathcal{L}^{j}_i(x,t) = \int {\rm d}x \ x \mathcal{Q}^{j}_{i'}(x,t).
\label{eq:kinetic}
\eqe
This guarantees, at least formally, that the total energy lost by one particle species (e.g. protons) equals the energy transferred to other particles (e.g., pairs, neutrinos, and photons)\footnote{Because the integration is performed on discretized functions, Eq.~\ref{eq:kinetic} cannot ensure exact energy conservation. In practice, the two integrals agree to within approximately 1\%-10\% (depending on the physical process and input parameters) for the default code settings. If stricter energy conservation is required, the user can choose to renormalize the energy injection rates to match the corresponding energy loss rates for all processes.}.

The physical processes that are included in the code are:
\begin{enumerate}
    \item electron synchrotron radiation ($\mathcal{L}^{syn}_e, \mathcal{Q}^{syn}_{\gamma}$)
    \item synchrotron self-absorption  ($\mathcal{L}^{ssa}_\gamma$)
    \item electron inverse Compton scattering ($\mathcal{L}^{ic}_e, \mathcal{Q}^{ic}_{\gamma}$)
    \item electron-positron annihilation ($\mathcal{L}^{ann}_{e}, \mathcal{Q}^{ann}_{\gamma}$)
    \item photon-photon ($\gamma \gamma$) pair production ($\mathcal{Q}^{\gamma \gamma}_e, \mathcal{L}^{\gamma \gamma}_{\gamma}$)
    \item photon Compton down-scattering on cold pairs
    \item triplet pair production ($\mathcal{L}^{tpp}_e, \mathcal{Q}^{tpp}_{e}$)
    \item proton synchrotron radiation ($\mathcal{L}^{syn}_p, \mathcal{Q}^{psyn}_{\gamma}$)
    \item proton-photon (Bethe-Heitler) pair production ($p \gamma e$; $\mathcal{L}^{p\gamma e}_p, \mathcal{Q}^{p\gamma e}_{e}$)
    \item proton-photon pion production ($p \gamma \pi$; $\mathcal{L}^{p\gamma\pi}_p, \mathcal{Q}^{p\gamma\pi}_{\gamma}$, $\mathcal{Q}^{p\gamma\pi}_{e}, \mathcal{Q}^{p\gamma\pi}_{\nu}$, $\mathcal{Q}^{p\gamma\pi}_{n}$)
    \item neutron-photon pion production ($n \gamma \pi$; $\mathcal{L}^{n\gamma\pi}_n, \mathcal{Q}^{n \gamma \pi}_p$)
    \item neutron decay
\end{enumerate}
Leptonic processes 1-6 are modeled as described in \cite{MK95}, except for the photon production and electron energy loss operators for inverse Compton scattering. The inverse Compton photon production operator is calculated according to Eq.~(2.48) of \cite{BG70}, which applies to the Thomson and Klein-Nishina scattering regimes. The electron loss operator due to scatterings in the Klein-Nishina regime is given by the third and fourth terms in the left hand-side of Eq.~(5.7) in \cite{BG70}. The probability $P(E,E')$ of an electron with energy $E$ to emit a photon of energy $E-E'$ that enters in these terms is computed according to Eq.~(5.17) of \cite{BG70}. Triplet pair production was modeled according to \cite{Mastichiadis91} \citep[see also,][]{Petroetal2019}. Proton-photon (photohadronic) interactions are modeled using the results of Monte Carlo simulations. In particular, for Bethe-Heitler pair production the Monte Carlo results by \cite{PJ96} were used (for more details on the implementation see \cite{MPK05} and Appendix~\ref{app:BH}).  Photopion interactions were incorporated in the time-dependent code by using the results of the Monte Carlo event generator {\sc sophia} \citep{Mucke_2000}, which takes into account channels of multi-pion production for interactions much above the threshold \citep[for more details see][]{DMPR12}.

We also include the effects of kaon, pion and muon synchrotron losses, albeit in a way that does not require the use of additional kinetic equations. Pion ($\pi^\pm, \pi^0$), charged and neutral ($K^0_S$ and $K^0_L$) kaon production rates from photomeson interactions have been computed by the {\sc sophia} event generator \citep{Mucke_2000}. For each particle energy, we calculate the energy lost to synchrotron radiation before it decays. The remainder of that energy is then instantaneously transferred to the particle's decay products, whose yields have also been computed by the {\sc sophia} event generator. Since the secondary particles from kaon decay include pions, we first calculate charged kaon decay and then charged pion decay. Finally, the same process is applied to the resulting muons. The photons, electrons and neutrinos resulting from kaon, pion and muon decay are added as production rates to their respective kinetic equations, as are the photons from kaon, pion and muon synchrotron radiation. Neutral kaons ($K^0_S$ and $K^0_L$) and pions ($\pi^0$) are, as in \cite{DMPR12}, assumed to decay instantaneously, therefore directly contributing their decay products to the kinetic equations.

The remaining particle distributions are modeled by coupled kinetic equations. The code uses equally spaced logarithmic grids for $x_i$ of particle species $i$ (for neutrons (neutrinos) we use the same energy grid as protons (electrons); see Table~\ref{tab:grids} where the energy grid parameters of each code are specified. An array $y(x, t) = \{n_{\rm p}(x_{\rm p},t), n_{\rm e}(x_{\rm e},t), n_{\rm \gamma}(x_{\gamma},t), n_{\rm n}(x_{\rm n},t), n_{\rm \nu}(x_{\rm \nu},t) \}$ is defined, where $x = \{x_{\rm p}, x_{\rm e}, x_{\rm \gamma}, x_{\rm n}, x_{\rm \nu} \}$ denotes the combined energy grid for all species. 
The code uses the routine \href{https://support.nag.com/numeric/nl/nagdoc_26/nagdoc_fl26/html/d02/d02ejf.html}{d02ejf} from the NAG Fortran library, which integrates a stiff system of first-order ordinary differential equations (ODEs) over an interval with suitable initial conditions. The routine uses a variable-order, variable-step method implementing the Backward Differentiation Formulae (BDF). More specifically, the routine solves a set of equations of the form $y'_i = f_i(x, y_1, y_2, ..., y_n)$ for $i=1,2,..., n$. Here, the prime denotes the time derivative and $n$ is the number of ODEs, which equals the length of $x$. The function $f_i$ contains all the injection and loss terms in Eq.~(\ref{eq:kinetic}) and is evaluated in time $t$ for  $i=1,2,..., n$.

The adopted numerical scheme is ideal for studying electromagnetic cascades in both  linear  and  non-linear regimes\footnote{If the energy  density  of  the  secondary  photons  is lower than that of the synchrotron photons from primary electrons (and/or external radiation fields), the cascade is considered to be linear, i.e., the interactions between secondary pairs and photons can be neglected. Otherwise, the cascade is characterized as non-linear or self-supported.}, and for simulating time-dependent problems, such as flares. The code has been extensively used in the steady-state or time-dependent modeling of non-thermal radiation from AGN \citep[e.g.,][]{MK97, DPM14, Petroetal14, Petroetal15, Petroetal16, Petroetal20, karavola, 2025JCAP...04..075K}, applied to multi-messenger emissions from gamma-ray bursts \citep{PDMG14, PGD14, Florou21}, used for the study of pair cascades in electrostatic gaps \citep{Petroetal2019}, and of radiative instabilities in leptohadronic relativistic plasmas \citep{MPK05, PM12, PDMG14, PM18,  Mastichiadis20}.\\

\subsection{\boet}

The leptohadronic code of \cite{Boettcher_2013} (\boet)  describes the multi-wavelength behavior of blazars in steady state.
This homogeneous one-zone model starts with the injection of a power-law distribution of ultra-relativistic electron-positron pairs and protons into a spherical emission region of size $R$, which moves with a constant relativistic speed $\beta_{\Gamma}$ corresponding to bulk Lorentz factor $\Gamma$.
Cooling is performed via synchrotron and Compton emission on various photon fields:
The co-moving electron-synchrotron (SSC) and direct accretion-disk radiation as well as an arbitrary external radiation field, assumed to be isotropic in the AGN rest frame. For the latter, a separate routine is used to write the desired isotropic external photon spectrum into an input file which is then read by the main code. The full Klein-Nishina cross section for inverse Compton scattering is used, adopting the analytical solution of \cite{Boettcher_1997}.
In a pure leptonic scenario, the self-consistent radiative output is calculated based on a temporary equilibrium between particle injection $Q(\gamma)$, radiative cooling at a rate $\dot\gamma$ and electron escape on a timescale of $t_{esc} = \eta_{esc} R/c$. Given a critical Lorentz factor $\gamma_c$, where $\dot\gamma (\gamma_c) = \gamma_c / t_{\rm esc}$, an analytical approximation to the equilibrium solution ($\partial_t n (\gamma, t) = 0$) to Eq. \ref{equ_kinematic_AM$^3$}, i.e., 
\begin{equation}
0 = Q(\gamma) - {\partial \over \partial\gamma} \left( \dot\gamma \, n [\gamma] \right) - {n(\gamma) \over t_{\rm esc}} 
\label{B13_kinetic}
\end{equation}
is evaluated as
\begin{equation}
n(\gamma) = \cases{
f_c \, Q(\gamma) \, t_{\rm esc}  & for $\gamma \le \gamma_c$ \cr 
{1 \over \dot\gamma} \int\limits_{\gamma}^{\infty} Q(\gamma) \, d\gamma & for $\gamma > \gamma_c$ \cr}
\label{B13_equilibrium}
\end{equation}
where $f_c$ is a normalization constant to ensure continuity of the paricle distribution at $\gamma = \gamma_c$. In the fast-cooling regime, where $Q(\gamma)$ has a low-energy cut-off at $\gamma_{\rm min} > \gamma_c$, the $\gamma \le \gamma_c$ branch of Eq. \ref{B13_equilibrium} is empty and the integral in the $\gamma >  \gamma_c$ branch becomes constant for $\gamma \le \gamma_{\rm min}$, i.e., the solution reduces to $n (\gamma) = {\rm const.} / \dot\gamma$ for $\gamma_c \le \gamma \le \gamma_{\rm min}$. The resulting particle distribution is normalized to a kinetic power of $L_{\rm kin} = \Gamma^2 \, \beta_{\Gamma} \, m c^3 \, \pi R^2 \, \int_1^{\infty} \gamma \, n(\gamma) \, d\gamma$ which is an input parameter. As SSC cooling depends on the self-consistently produced synchrotron radiation field, which, in turn,  depends on the equilibrium particle distribution, the code employs an iterative scheme: Starting with an inital guess of $\dot\gamma_{\rm SSC} = \dot\gamma_{\rm sy}$, it evaluates a first-guess equilibrium distribution, which is then used to calculate the corresponding synchrotron spectrum, based on which a new $\dot\gamma_{\rm SSC}$ is evaluated, which is used to calculate a new equilibrium particle distribution. This procedure is repeated until the solution converges to a self-consistent equilibrium. 
In a leptohadronic scenario, the radiative output in a temporary equilibrium is evaluated for both primary electrons and protons, taking internal $\gamma\gamma$ pair production and cascades into account.

For leptohadronic processes (in a stationary state), the code employs the semi-analytical expressions of \cite{Kelner_2008} for the final decay products (electrons, positrons, photons, and neutrinos) from  photopion interactions. Bethe-Heitler pair production is not included. A semi-analytical approach, described in detail in \cite{Boettcher_2013}, 
is used to evaluate the output due to pair cascades initiated by pairs from pion and muon decay and from photon-photon attenuation of primary $\gamma$-rays. The code has been designed to describe stationary SEDs, thus short-term variability is averaged out over the integration time of the SED. 
All particle energies and photon frequencies are discretized on regular logarithmic grids, as specified in Table \ref{tab:grids}, and numerical integrations over particle and frequency distributions are carried out using standard finite-difference schemes. This code has shown robust results on the modeling of \textit{Fermi}-LAT detected blazars \citep[e.g.,][]{Boettcher_2013}. \\

\subsection{\paris}

The  \paris code \citep{Cerruti15} is a steady-state numerical code which computes photon and neutrino emission by a population of electrons and protons at equilibrium in a spherical emitting region in a relativistic jet. The code is built upon the leptonic (SSC) code by \cite{Kata01}. A population of primary electrons at equilibrium in the emitting region (assumed spherical, with parameters $\delta$, $B$, and $R$) is parameterized by a broken-power law distribution (with parameters $\gamma_{\min}, \gamma_{\rm break},\gamma_{\max}$, $\alpha_{1,2}$ and normalization $K$). The leptonic part of the code computes synchrotron emission and SSC emission \citep[using the kernel by][]{Jones68}. Photon and particle energy vectors are discretized with a logarithmic step, as specified in Table \ref{tab:grids}. Numerical integrals are computed with a Gauss-Legendre algorithm.
A first development of the leptonic part of the code has been the computation of pair injection from $\gamma \gamma$ pair production, which is implemented following \citet{Aha83}. Their steady state distribution is computed taking into account synchrotron cooling using the integral solution of \citet{Inoue_1996} (that solves the differential equation \ref{equ_kinematic_AM$^3$} for $\partial_t n_i(\gamma, t) = 0$).\\

The hadronic part of the code computes synchrotron radiation by protons (which is implemented as for the primary electrons) and proton-photon interactions. The latter are computed assuming that the target photon field is synchrotron emission by primary electrons and protons, and SSC. The energy distributions of secondary particles injected via photo-meson interactions are computed running the Monte-Carlo code {\sc sophia} \citep{Mucke_2000}. With respect to the public version of  {\sc sophia}, the code is modified to take as input arbitrary target photon fields, to extract as well the muon spectra, and to include the synchrotron cooling of kaons, pions, and muons.  {\sc sophia} outputs are considered as injection terms to compute the steady state distributions at equilibrium. Photon and neutrino spectra are obtained by multiplying their injection rates times their escape time-scale; for leptons the steady state distribution are computed as for gamma-gamma pairs. For muons, the steady state solution needs to include their decay time, and the explicit integral solution is provided in \citet{Cerruti15}. Pair injection from Bethe-Heitler pair-production are computed following \citet{Kelner_2008} and \citet{BG70}, and their steady state is computed as for other leptons. Pair cascades are computed iteratively: the code computes the steady state distribution of a leptonic population, its synchrotron and inverse-Compton radiation, then computes the pair injection from $\gamma\gamma$ pair production, its steady state, its radiative emission, and so on, until the contribution of the n$^{th}$ generation is negligible. For all applications in this work the cascades are computed up to the 5$^{th}$ generation included. The explicit assumption is that the cascade is never self-supported and always driven by the synchrotron and SSC photons from the primary electrons, or by the synchrotron photons from the primary protons. In the same way, the emission from the cascade does not affect the distribution of primary electrons and protons.\\

In the original version of the code, the simulation is limited to a scenario in which external photon fields are negligible, meaning that the dominant soft photon field for $\gamma\gamma$ and p$\gamma$ interactions is the synchrotron radiation by primary electrons, and that synchrotron cooling is always dominating over inverse-Compton cooling. It is thus suited to model emission from high-frequency-peaked BL Lac objects. Besides the original application to extreme blazar, the code has been used to study the potential detection of hadronic spectral features in the TeV band with CTA  \citep{Zech17}, to model the emission from the peculiar high-luminosity blazar PKS~1424+240 \citep{Cerruti17}, and to model the blazar neutrino candidate TXS~0506+056 \citep{Cerruti19}. As a major upgrade for this project, the code has been extended to be able to deal with external photon fields: an arbitrary photon field (expressed as photon energy density as a function of frequency, in the reference frame of the emitting region) provided by the user is read into the code, and used for the calculation of all radiative processes (inverse Compton emission, Bethe-Heitler pair production and photo-meson interactions). 
Additionally, the computation of particle distributions at equilibrium takes into account inverse-Compton cooling. The augmented version of the code, that now also includes p-p interactions, has been already applied to study blazar neutrino candidates \citep{Acciari22, Acharyya23} and the multi-messenger emission from NGC~1068 \citep{Inoue1068}.\\

\begin{deluxetable}{l  c  c  c  c c}

\tablewidth{0pt}
\tablecaption{Physical processes included in the numerical codes. \label{tab:processes}} 

\tablehead{Physical Processes & \multicolumn{5}{c}{Codes}}

\startdata
  & \am  & \ath & \boet & \paris & \lehamoc \\ 
electron synchrotron radiation &   \cmark & \cmark  &  \cmark &  \cmark & \cmark  \\
synchrotron self-absorption & \cmark  &  \cmark &   \cmark &  \cmark & \cmark\\ 
electron inverse Compton scattering &  \cmark &   \cmark &   \cmark &  \cmark & \cmark\\ 
electron-positron annihilation &   \xmark  &  \cmark &  \cmark &  \xmark &  \xmark\\ 
photon-photon pair production & \cmark  & \cmark &  \cmark &  \cmark &  \cmark \\ 
triplet pair production & \xmark &  \cmark &  \xmark   &  \xmark  &  \xmark\\ 
proton synchrotron radiation & \cmark  &  \cmark & \cmark   &  \cmark &  \cmark \\ 
proton inverse Compton scattering &  \cmark & \xmark &  \xmark & \xmark & \xmark \\
proton-photon pair production & \cmark   & \cmark & \xmark    &  \cmark &  \cmark \\ 
proton-photon pion production & \cmark   & \cmark & \cmark    &  \cmark &  \cmark \\ 
proton-proton pion production & \xmark   & \xmark & \xmark    &  \cmark &  \cmark \\ 
neutron-photon pion production & \cmark &   \cmark & \xmark &   \xmark  &   \xmark \\
neutron decay &  \xmark &  \cmark & \xmark & \xmark & \xmark \\
kaon synchrotron radiation &\xmark &  \cmark &  \xmark &   \xmark  &   \xmark\\ 
pion synchrotron radiation & \cmark &  \cmark & \xmark &  \xmark &  \xmark \\ 
muon synchrotron radiation & \cmark &   \cmark &  \xmark &   \cmark &  \xmark \\ 
\enddata

\end{deluxetable}

\subsection{\lehamoc}
\lehamoc\footnote{\url{https://github.com/mariapetro/LeHaMoC/}} is an open-source, time-dependent leptohadronic code described in \citet{lehamoc}. It is designed to solve complex partial differential equations (PDEs) that model the interactions and emissions of particles (kinetic equations) in high-energy astrophysical environments such as blazars. The PDEs account for particle injection, escape, and losses due to physical processes such as synchrotron radiation, inverse Compton scattering, and photohadronic interactions. The code accounts for various particle species, including relativistic electrons, positrons, protons, photons and neutrinos, within an assumed spherical and potentially expanding region of radius $R$. 

\lehamoc tackles the coupled integrodifferential equations governing these particle populations (see Eq.~\ref{equ_kinematic_athena}) by using a combination of finite difference methods and an implicit time-stepping scheme \citep{chang1970a}, which ensures stability and accuracy over long simulation periods. Stiff differential equations
commonly arise in the context of AGN modeling, since the various timescales associated with physical processes can differ by many orders of magnitude across the energy domain. The Chang \& Cooper scheme is well suited for the numerical solution of the said problems. To discretize the kinetic equations in time, the user selects a time step that is related to the characteristic time scale of the system. The default choice is 1 $R/c$, but shorter time steps may be needed. The use of an implicit scheme ensures that the solution at any time step is stable, which is useful since the time scales of different physical processes can vary widely. The discretization in energy is achieved by using a logarithmic energy grid that allows us to accurately capture the behavior of the distribution over a wide range of energies. The energy grid size is a parameter that the user can define. Each discretized equation forms a tridiagonal matrix that we solve using the Thomas algorithm \citep{thomas1949}.

In \lehamoc relativistic pairs and protons are injected into power-law distributions, featuring either sharp or exponential cut-offs. The time evolution of these particles is governed by the solution of the coupled kinetic equations. The leptonic processes modeled include synchrotron radiation, synchrotron self-absorption \citep{dermer2009high}, electron inverse Compton scattering \citep{BG70,MK95}, and photon-photon annihilation \citep{MK95}. The hadronic processes encompass proton synchrotron radiation, Bethe-Heitler pair production \cite{BG70,Kelner_2008}, proton-photon pion production \citep{Kelner_2008}, and the inelastic collisions of relativistic protons with ambient cold protons \citep{Kelner_2006}.

The leptonic part of the code has been used in the modeling of the afterglow emission of GRB~221009A \citep{2024arXiv240515855B} and for the calculation of the pair signatures and distribution in the magnetospheric current sheets in M87* \citep{2024arXiv240601211S}. The full leptohadronic code, with the inclusion of external photon fields, was employed by \citep{2025arXiv250708680S} to investigate the multi-messenger emission of blazar jets powered by magnetic reconnection at various distances from the central supermassive black hole. The time-dependent features of the code are presented in \cite{2024Univ...10..392C}. This work explores the time-variable behavior of blazar Mrk 501, focusing on identifying hadronic signatures on its photon emission when varying some of the model parameters (i.e. electron/proton luminosity, magnetic field, and the power-law indexes of the distributions) according to simulated light curves. \\

\subsection{Resolution in codes}\label{app:resolution} 

In this section we discuss the effects of energy grid resolution and time step on the results obtained by the different codes, whenever applicable.

In \ath the routine used for the integration of kinetic equations {\tt d02ejf}\footnote{\url{https://support.nag.com/numeric/fl/nagdoc_fl24/pdf/d02/d02ejf.pdf}} uses an adaptive time step which cannot be controlled by the user. However, the resolution in the energy grid of electrons (and protons) is set by the user through the parameter {\tt npdec} (i.e. number of grid points per decade).
The photon energy grid is designed to have twice this number of points per decade. While the minimum energy in the proton ($\gamma_{p,\min}$) and electron ($\gamma_{e,\min}$) grids is typically some factor of 10, the minimum energy in the photon grid is set to $x_{\min} = b \gamma_{e,\min}^2$, where $b=B/B_{c}$ and $B_c = {{m_e^2c^3} / {e\hbar}}$. That creates a seamless correspondence of bins when photons are produced by synchrotron radiation but is less ideal for other processes. In neutral pion decay, in particular, the mismatch between the proton and photon grids introduces a small error in the photon spectrum around its peak, since it usually involves only a few bins. Nevertheless, when applying the proton-photon pion production process, {\tt npdec} is always set to 10, so as to align with the {\sc sophia} tabulated particle distributions with a small degree of smoothing. The latter come in a binning of 20 grid points per decade, so by using 10 points in \ath for most particles and 5 points for photons we are able to add together the values of two particle bins and four photon bins from {\sc sophia}. Moreover, the Bethe-Heitler implementation (see Appendix~\ref{app:BH}) is designed only for {\tt npdec} = 10. Therefore, the energy resolution in \ath can only be changed in purely leptonic scenarios.  We use the example of the inverse Compton catastrophe (Sec.~\ref{subsec:iccatastrophe}) to illustrate the effects of {\tt npdec}.

\begin{figure}
    \centering
    \includegraphics[width=0.9\textwidth]{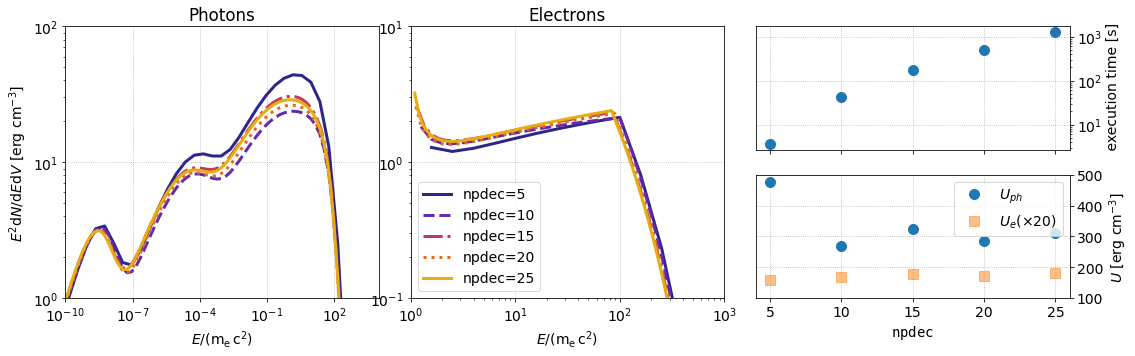}
    \caption{Dependence of results on the resolution of particle energy arrays ({\tt npdec}) used in the \ath code for the Compton Catastrophe scenario discussed in Sec.~\ref{subsec:iccatastrophe}. \textit{Left panel:} Photon energy density distribution. \textit{Middle panel:} Electron energy density distribution. \textit{Top right:} CPU time as a function of {\tt npdec}. \textit{Bottom right:} Integrated photon and electron energy densities plotted against {\tt npdec}. The electron energy density is multiplied by a factor of 20 for display purposes.}
    \label{fig:cc-resolution}
\end{figure}

\begin{figure}
    \centering
    \includegraphics[width=\textwidth]{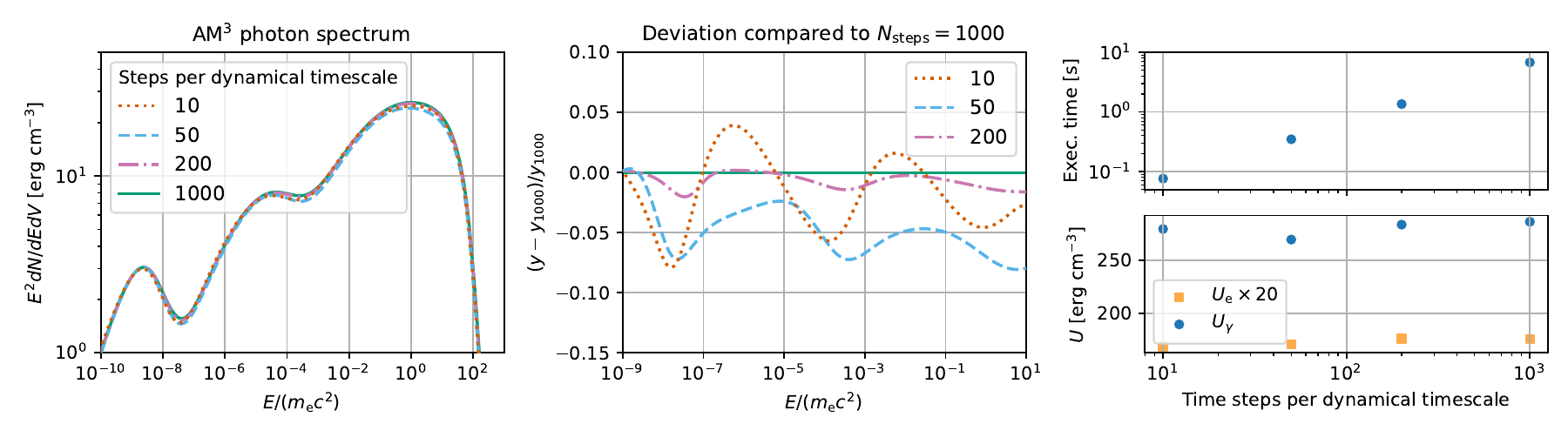}
    \caption{Dependence of AM$^3$ results on the size of the solver time step selected by the user. Like in Fig. \ref{fig:cc-resolution}, we simulate the Compton Catastrophe scenario of Sec.~\ref{subsec:iccatastrophe}. \textit{Left:} Steady-state photon energy density, obtained using 10, 50, 200, and 1000 steps per dynamical timescale, which in this case corresponds to the light-crossing time of the system, $R/c$. \textit{Middle:} Relative ratio between the fluxes shown in the left panel, using as reference the result obtained with 1000 time steps. \textit{Top right:} CPU time as a function of the number of time steps. \textit{Bottom right:} Integrated photon and electron energy densities.}
    \label{fig:am3-timestep}
\end{figure}

In \am the resolution of the energy grid is fixed to $\Delta\log(E)=0.1$, but the time step of the solver can be selected by the user. The effect of the time step on the accuracy of the results is summarized in Fig. \ref{fig:am3-timestep}. We have simulated the Compton Catastrophe scenario, using the same parameters as in Sec. \ref{subsec:iccatastrophe}, and evolving the system to 10 times the dynamical timescale, that is up to $10\,t_\mathrm{esc}=10\,R/c$, to guarantee a steady-state result. For this particular purpose, we ignore the effect of synchrotron self-absorption, which would otherwise lead to significant attenuation of the emitted photon flux, due to the extreme electron density. We tested four different time step sizes, from $\Delta t=0.1\,t_\mathrm{esc}$, which is the least computationally demanding down to $10^{-3}\,t_\mathrm{esc}$, which is the most accurate. The resulting photon spectra and their relative deviations are shown in the left and middle panels, respectively. As we can see, the simulation performed using only 10 time steps per dynamical timescale provides the least accurate results overall, with relative error fluctuating from -8\% to +4\%. At 200 time steps per dynamical timescale, the maximum absolute error has been reduced to $2\%$. For any number of steps beyond 1000, the result remains practically unchanged. As we can see on the upper right panel, the number of time steps performed by the solver is directly proportional to the computation time; in that sense, the choice of time step in \am reflects a trade-off between accuracy and computational efficiency.

\begin{figure}
    \centering
        \includegraphics[width=0.45\textwidth]{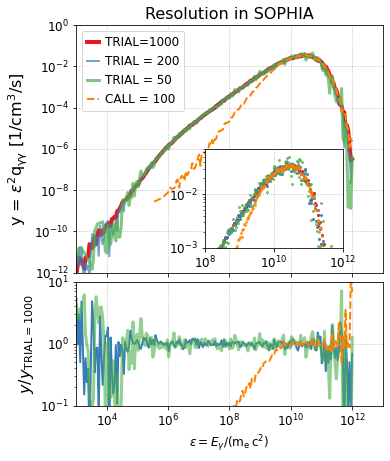}
    \caption{Dependence of results from \paris on the resolution of {\sc sophia}.}
    \label{fig:sophia-resolution}
\end{figure}

In \paris there are two resolution effects, one related to the grid used for particles and photons, one related to the {\sc sophia} simulations that are run on the fly. The first effect is similar to the other codes, and we do not show it here (for reference, all results presented here have been performed using 200 points for particle and photon distributions). We discuss here the effect of changing the resolution of the {\sc sophia} calls. \paris uses five parameters to call {\sc sophia}: the proton distribution is sampled {\sc SOPHIA\textunderscore CALL} times, from the maximum proton energy downwards with a {\sc SOPHIA\textunderscore STEP} step; for each call, {\sc sophia} execute {\sc SOPHIA\textunderscore TRIALS} independent Monte-Carlo realizations of the proton-photon interaction; the outputs for each proton energy are characterized by a {\sc SOPHIA\textunderscore NBINS} number of bins with a {\sc SOPHIA\textunderscore DELX} step. To correctly sum all outputs, the ratio {\sc SOPHIA\textunderscore STEP} over {\sc SOPHIA\textunderscore DELX} has to be an integer. In the following we show the effect of degrading the resolution for the benchmark test p$\gamma$-PLPL, looking at the photon injection only. The result shown in Fig. \ref{fig:hadronic_rates_pl} have been obtained using 
{\sc SOPHIA\textunderscore CALL}= 200; {\sc SOPHIA\textunderscore STEP}=0.05; {\sc SOPHIA\textunderscore TRIAL}=1000; {\sc SOPHIA\textunderscore NBINS}=200; {\sc SOPHIA\textunderscore DELX}=0.025. In Fig. \ref{fig:sophia-resolution} we show the effect of changing {\sc SOPHIA\textunderscore TRIAL} to 200 and 50, as well as reducing {\sc SOPHIA\textunderscore CALL} to 100.

\begin{deluxetable}{l c c c c c}

\tablewidth{0pt}

\tablecaption{Main features of numerical codes and implementation of hadronic processes. \label{tab:summary}}

\tablehead{ & \multicolumn{5}{c}{ Codes}}
\startdata
\hline
Features& \am & \ath & \boet  & \paris & \lehamoc \\ 
\hline 
steady state & \cmark &\cmark &  \cmark &   \cmark &   \cmark\\
time dependent &\cmark & \cmark & \xmark  &  \xmark &   \cmark\\
linear EM cascades &  \cmark  & \cmark &  \cmark &  \cmark &   \cmark\\
non-linear EM cascades &  \cmark &  \cmark &   \xmark &  \xmark &   \cmark\\
\hline
Implementation & \multicolumn{5}{c}{}\\ 
\hline
Photo-pion process &  following Ref.$^a$&  tabulated {\sc sophia}$^b$ & following Ref.$^c$ &  running {\sc sophia}$^b$ & following Ref.$^c$\\
Photo-pair process &  following Ref.$^c$   &  tabulated from Ref.$^d$ &   n/a  &   following  Ref.$^c$  & following Ref.$^c$\\
\enddata
\tablerefs{ $^a$\cite{Hummer:2010vx}, $^b$\cite{Mucke_2000}, $^c$\cite{Kelner_2008}, $^d$\cite{PJ96}}
\end{deluxetable}

\section{Case studies}\label{sec:casestudies} 

Before discussing the details of the test cases we investigated, we need to comment on several differences in the code definitions and implementations that have to be considered to have a fair comparison of the output results.  

\am and \ath are time-dependent codes that take as input the injection rate of particle species $i$ per unit volume, $Q_i^{\rm inj}(\gamma, t)$ (see equations~\ref{equ_kinematic_AM$^3$} and \ref{equ_kinematic_athena}). The particle injection rate translates to a bolometric kinetic injection luminosity, $L^{\rm inj}_i$, as
\begin{equation} 
L_i^{\rm inj} = \frac{4}{3}\pi R^3 m_i c^2 \int_1^\infty d\gamma \, (\gamma-1) Q_i^{\rm inj}(\gamma)
\label{eq:Linj-Q}
\end{equation}
where $Q_i^{\rm inj}$ may depend on time in the general case. The particle injection compactness, a dimensionless measure of the particle luminosity, is then defined as
\begin{equation}
\ell_i^{\rm inj} \equiv \frac{\sth L_i^{\rm inj}}{4 \pi R m_i c^3} =  \frac{\sth R^2}{3 c}  \int_1^\infty d\gamma \, (\gamma-1) Q_i^{\rm inj}(\gamma)
\label{eq:comp-Q}
\end{equation}
\boet and \paris are steady-state codes, hence they take as input the number density of particle species at steady state,  $n_i^{\rm ss}$. In the absence of cooling, the latter quantity is related to the injection compactness as follows
\begin{equation}
\ell_i^{\rm inj} = \frac{\sth R^2}{3c t_{ i, \rm esc}} \int_{\gamma_{i,\min}}^{\gamma_{i,\max}} d\gamma \, (\gamma-1) n_i^{\rm ss}(\gamma)
\label{eq:comp-nss}
\end{equation}
They take as input the number density at $\gamma=1$, $n_i^{\rm ss}|_{\gamma=1}$. So, for the commonly used power-law distribution with slope $s_{i}$ between $\gamma_{i, \min}$ and $\gamma_{i, \max}$, Eq.~(\ref{eq:comp-nss}) reads
\begin{eqnarray}
 n_i^{\rm ss}|_{\gamma=1} =\frac{3c t_{i, \rm  esc} \ell_i^{\rm inj}}{\sth R^2} \left\{ 
 \begin{tabular}{cc}
  $\frac{2-s_i}{\gamma_{i,\max}^{2-s_i}-\gamma_{i,\min}^{2-s}}$    &  $s_i \neq 2$ \\ \\ 
$\ln^{-1}\left(\frac{\gamma_{i, \max}}{\gamma_{i, \min}}\right)$      &  $s_i =2$
 \end{tabular} 
 \right.
 \label{eq:nss-uncool}
\end{eqnarray}
The normalization parameters in \am and \ath are originally implemented assuming that the integrands in Eq. \ref{eq:Linj-Q} and \ref{eq:comp-Q}  are injected power-law functions with a sharp cutoff at $\gamma_{\max}$. For the purpose of this project, all tests have been performed assuming that all primary particle distributions are defined with injected exponential cut-offs at $\gamma_{\max}$, with the notable exception of the tests with monoenergetic protons. Still, the normalization of the particle distributions is computed by integrating up to $\gamma_{\max}$, even though the exponential cut-off is present in the particle distributions.

Therefore, a major difference between the two approaches lies in the way the primary particles are treated: if the injected primary distributions are cooled by either synchrotron or inverse Compton radiation, this effect will be automatically included in \am, \ath, and \lehamoc, while \boet and \paris need to parametrize the cooled primary distributions at equilibrium. In the case of cooling the expression relating $n_i^{\rm ss}|_{\gamma=1}$ and the injection rate of primaries is given by \citep{Inoue_1996}
\begin{equation}
\label{eq:inoue}
    n_i^{\rm ss}|_{\gamma=1} = \gamma_c t_{ i, \rm esc} e^{-\gamma_c} \int_1^\infty d\gamma Q_i^{\rm inj}(\gamma)e^{\gamma_c/\gamma}, 
\end{equation}
where $\gamma_c$ is the cooling Lorentz factor that is defined by balancing the radiative loss and escape timescales ($t_{i, \rm cool}(\gamma)=t_{i, \rm esc}$). 

Another difference among the codes is the fact that \paris and \boet implements the geometric correction described in \citet{Gould79, Kataoka99}: the photon densities that are used for computing inverse-Compton scattering and proton-photon interactions are multiplied by a factor 3/4 to account for the fact that the photon densities in one-zone codes are spatial averages; even if the emissivity of a process is isotropic in a spherical source, the photon densities are radially dependent \citep{Gould79}. 
This correction term is not implemented in the other codes, and in this work the outputs from \paris that are affected by this correction are multiplied by 4/3 at the plotting stage.

\begin{deluxetable}{l c c c c c c c c}

\tablewidth{0pt}
\tablecaption{Input parameter values used for each scenario for the code comparison. \label{tab:parameters}} 
    \tablehead{& &  \footnotesize{SYN-cool} & \footnotesize{SSC-TH} & \footnotesize{SSC-KN}& \footnotesize{p$\gamma$-MONOGB}  & \footnotesize{p$\gamma$-PLPL} & \footnotesize{PS} & \footnotesize{LeHa} \\ 
    \multicolumn{2}{c}{\phantom{a}} & \multicolumn{3}{c}{\textit{Leptonic scenarios}} & \multicolumn{2}{c}{\textit{Hadronic kernel tests}} & \multicolumn{2}{c}{\textit{Hadronic blazar-like scenarios}} }
\startdata
    Input parameters &  Symbol [Units] &  \multicolumn{7}{c}{Values} \\
\hline   
    \footnotesize{Emission Region Radius}     & $R$ \footnotesize{[cm]} & $10^{15}$ & $10^{15}$ & $10^{15}$ & $10^{15}$& $10^{15}$ & $10^{15}$ & $10^{16}$\\ 
    \footnotesize{Magnetic field strength} & $B$ \footnotesize{[G]} & 50 & 0.1 &  0.01 &10 & & 10 & 0.1 \\ \hline 
    \footnotesize{Min. e$^-$ Lorentz factor} & $\gamma_{e, \min}$ &  1  & 1 & 1 & $-$& $-$& 1 & 1 \\
    \footnotesize{Max. e$^-$ Lorentz factor} & $\gamma_{e, \max}$ & $10^4$ & $10^4$ & $10^6$ &$-$ &$-$ & $10^3$ & $3 \times 10^5$\\
    \footnotesize{e$^-$ power-law index} & $s_e$ & 1.9 & 1.9 & 1.9 & $-$& $-$&1.9 & 2.0 \\ 
    \footnotesize{e$^-$ injection luminosity$^a$} & $L^{\rm inj}_{e}$ \footnotesize{[erg s$^{-1}$]} & $-$ & $-$  & $-$ & $-$ & $-$&$1.6 \times 10^{38}$ &  $3.7 \times 10^{40}$\\ 
    \footnotesize{e$^-$ injection compactness$^b$} & log($\ell^{\rm inj}_{e}$) & $-4.5$ & $-4.47$ & $-4.18$ & $-$& $-$& $7.47$& $-5.1$ \\
    \footnotesize{Steady-state $e$ density$^c$} & $n_{e}^{\rm ss}|_{\gamma=1}$ \footnotesize{[cm$^{-3}$]} & $1.65\times10^4$ & $10^4$ & $10^4$ & $-$& $-$& $12.5$ &  282 \\
    \footnotesize{e$^-$ escape timescale} & $t_{e, \rm esc}$ \footnotesize{[$R/c$]} & 1 &1 &1 &$-$ &$-$ & 1 & 1 \\ \hline
    \footnotesize{Min. p Lorentz factor} & $\gamma_{p, \min}$ & $-$ &$-$ & $-$ & $10^{6 (7)}$ & 1 & 1& 1 \\
    \footnotesize{Max. p Lorentz factor} & $\gamma_{p, \max}$ & $-$ & $-$ &$-$ & $10^{6.2 (7.2)}$ & $10^8$ & $10^8$& $10^7$\\
    \footnotesize{$p$ power-law index} & $s_p$ & $-$ & $-$ & $-$ & 1.9 & 1.9 & 1.9 & 2.0 \\ 
    \footnotesize{$p$ injection luminosity$^a$} & $L^{\rm inj}_{p}$ \footnotesize{[erg s$^{-1}$]} & $-$ & $-$ &$-$ &$8.5 \times 10^{43}$ & $8.5 \times 10^{43}$& $10^{44}$ &  $2.8 \times 10^{46}$ \\ 
    \footnotesize{$p$ injection compactness$^b$} & log($\ell^{\rm inj}_{p}$) & $-$ & $-$ & $-$& $-4.0$ & $-4.0$ & $-4.93$ & $-2.5$ \\
    \footnotesize{Steady-state $p$ density$^c$} & $n_{p}^{\rm ss}$ \footnotesize{[cm$^{-3}$]} & $-$ & $-$ &$-$ & $2.4 (1.9) \times 10^5$ & $8490$ & $1000$ & 108300\\
    \footnotesize{$p$ escape timescale} & $t_{p, \rm esc}$ \footnotesize{[$R/c$]} & $-$ & $-$ & $-$& 1 & 1 & 1 &1 \\ 
\enddata
\tablecomments{$^a$\am and \lehamoc,  $^b$\ath, $^c$\paris and \boet. Particle cooling neglected in SSC-TH, SSC-KN, p$\gamma$-MONOGB, p$\gamma$-PLPL, and $\gamma -\gamma$-annihilation was omitted in PS. p$\gamma$-MONOGB: grey-body external photon field of compactness $\ell_{\gamma} = 8.1 \times 10^{-6}$ and temperature $T_{\gamma} = 10^6$~K. p$\gamma$-PLPL: Power-law external field of compactness $\ell_{\gamma} = 10^{-5}$ between $E_{\gamma, \mathrm{min}} =  10^{-6} m_e c^2$ and $E_{\gamma, \mathrm{max}} =  10^{-1} m_e c^2$, with power-law index $p_\gamma = 2.0$. A Doppler factor of 30 was used to transform the luminosities from the jet comoving frame to the observer's frame.}
\end{deluxetable}

\begin{table}[]
    \centering
    \caption{Energy grid specifications of each code, used in the case studies presented in this work.}
    \begin{tabular}{l l l l l l}
    \hline
    & \multicolumn{1}{c}{\am} & \multicolumn{1}{c}{\ath} & \multicolumn{1}{c}{\boet}  & \multicolumn{1}{c}{\paris} & \multicolumn{1}{c}{\lehamoc} \\
    \hline
\textbf{Electron energy grid} &&&&&\\
Number of points or points per decade & 23 / decade & 10 / decade & 150 pts. & 200 pts. & 30 / decade\\
Minimum grid energy / ($m_e\,c^2$)  & 1.0 &  $10^{0.1}$ & 1.0 & 1.0 & 1.0\\
Maximum grid energy / ($m_e\,c^2$)  & $5.3\times10^{14}$ & $10^{13}$ & $1.0 \times 10^{12}$ & $1.0 \times 10^{8}$ & $10^{11}$ \\

\textbf{Proton energy grid} &&&&&\\ 
Number of points or points per decade & 23 / decade & 10 / decade & 300 pts. & 200 pts. & 10 / decade\\
Minimum grid energy / ($m_p\,c^2$)  & 1.0  & $10^{0.1}$ & 1.0 & 1.0 & 1.0\\ 
Maximum grid energy / ($m_p\,c^2$)  & $4.3 \times 10^{13}$ & $10^{11}$ & $1.0 \times 10^{10}$ & $1.0 \times 10^{10}$ & $10^{9}$ \\ 

\textbf{Photon energy grid} &&&&&\\ 
Number of points or points per decade & 23 / decade & 20 / decade & 250 pts. & 200 pts. & 10 / decade \\ 
Minimum grid frequency [Hz] & $8.53 \times 10^{7}$ & $4.5\times 10^7$ & $1.2 \times 10^7 \times B/{\rm G}$ & $1.0 \times 10^{9}$ & $10^{7.5}$\\ 
Maximum grid frequency [Hz] & $6.55 \times 10^{34}$ & $1.1\times 10^{33}$ & $1.2 \times 10^{32}$ & $1.0 \times 10^{45}$ & $10^{32}$\\ 

\hline
    \end{tabular}
    \label{tab:grids}
\end{table}

\section{Results}\label{sec:results}
We start by comparing the leptonic parts of our codes to gauge the level of agreement  concerning the computation of synchrotron radiation and inverse Compton scattering (Sec.~\ref{sec:leptonic}). We then investigated the agreement for the hadronic interactions in terms of secondary production spectra and their resulting emission (Sec.~\ref{sec:leptohadro}). The parameter values for all tests presented in this section are listed in Table \ref{tab:parameters}.

\subsection{Leptonic scenarios}\label{sec:leptonic}
\begin{figure}
\centering
\includegraphics[width = .46 \textwidth]{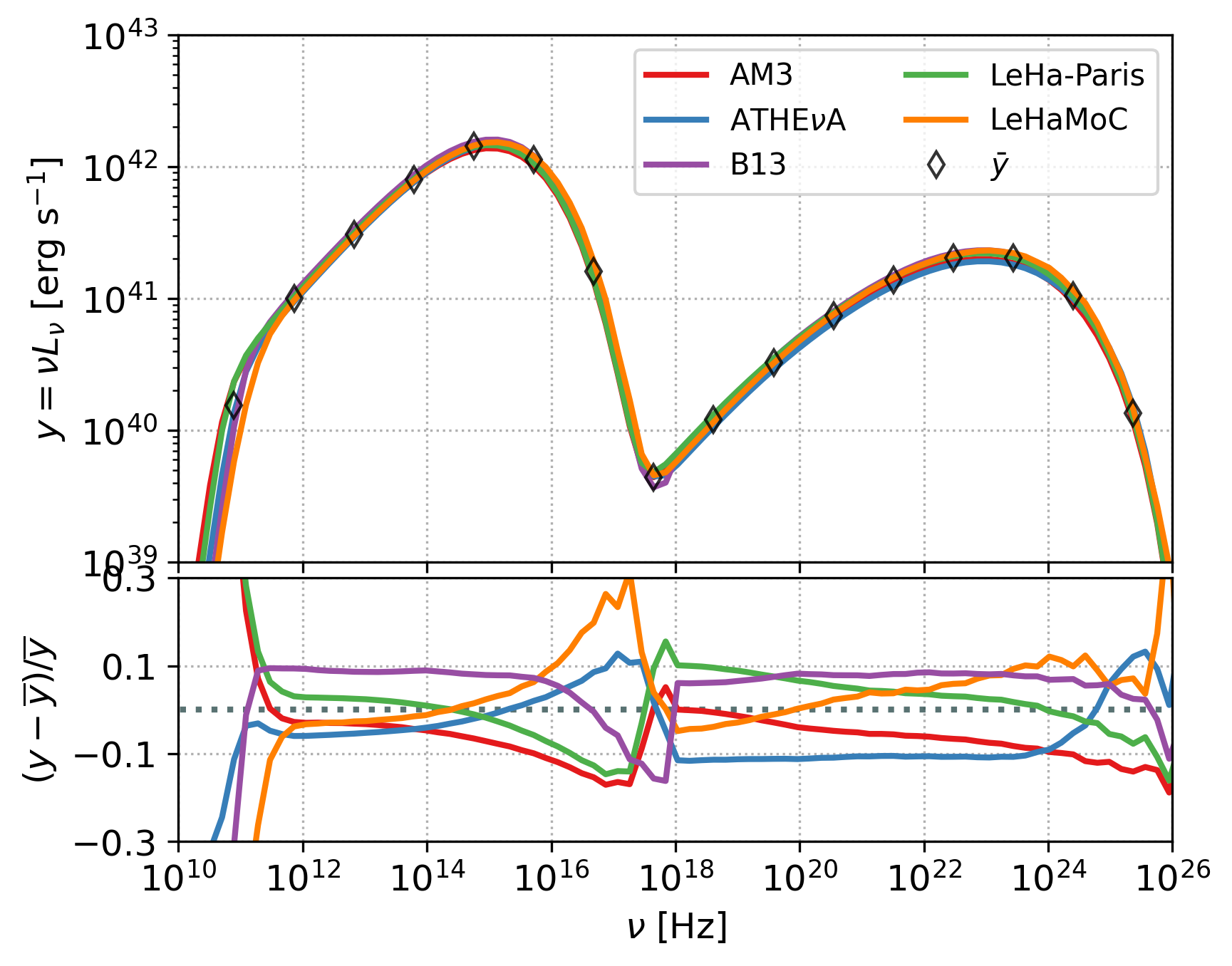}
\includegraphics[width = .45 \textwidth]{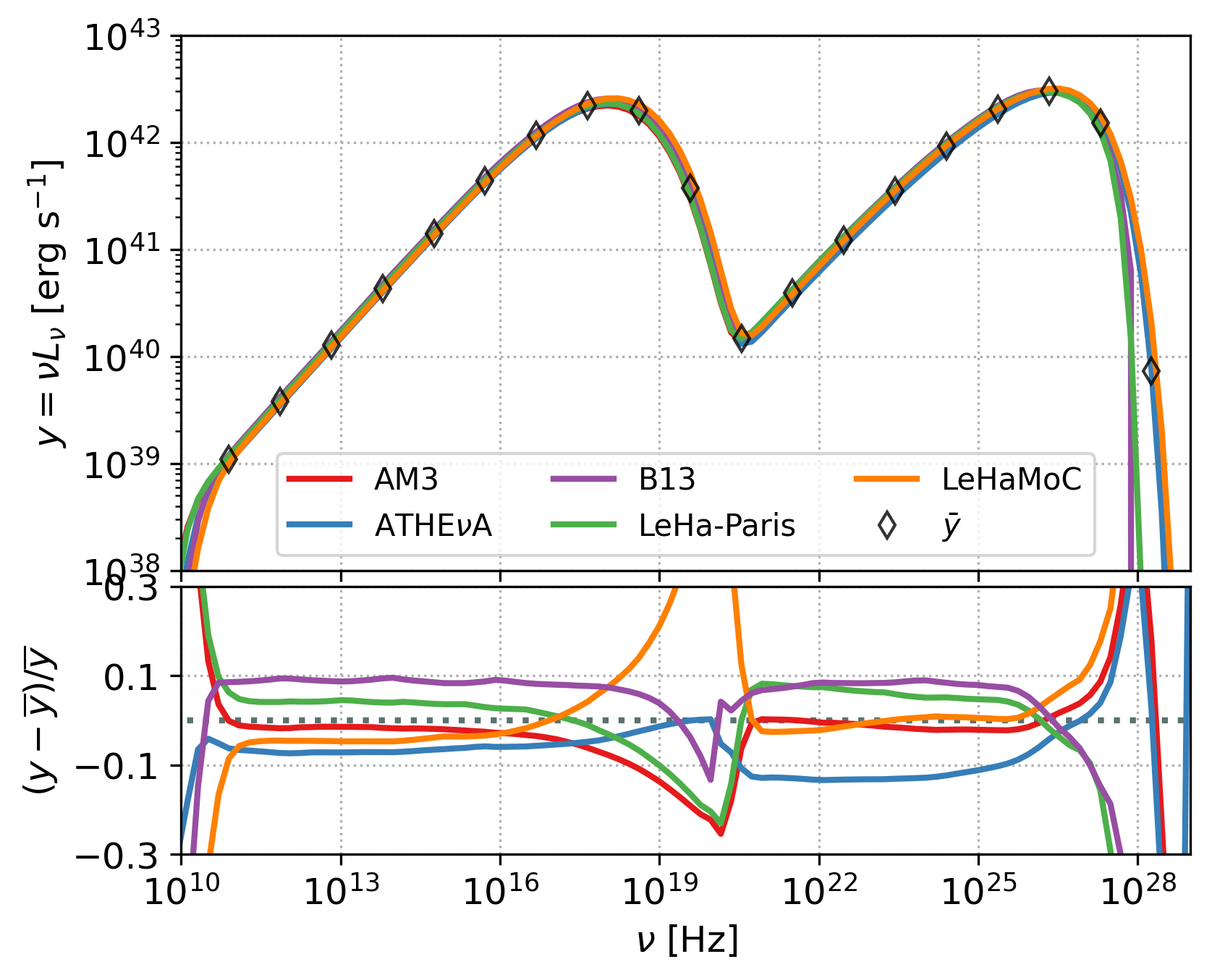}
\caption{Spectral energy distributions (in the observer's frame) computed for the SSC-TH (left) and SSC-KN (right) scenarios using the five codes described in Sec.~\ref{sec:codes} (see inset legend for details). The mean model is shown with open markers. The residuals of each code with respect to the mean value of the results are plotted in the bottom panels. }
\label{fig:leptonic_sed}
\end{figure}

The first tests are done for a purely leptonic synchrotron-self-Compton (SSC) scenario. We simulate emission for parameters typical of high-frequency-peaked BL Lacertae objects (HBLs) (see Table \ref{tab:parameters} for details). To investigate if there are any differences in the emissivity of the Compton scattering specific to the Thomson or the Klein-Nishina regime, we test for two different values of $\gamma_{e,max} = 10^4$ (SSC-TH) and $10^6$ (SSC-KN). For the adopted parameters, the steady state electron distribution is uncooled, and Eq.~(\ref{eq:nss-uncool}) is used for setting the initial conditions. The results of the modeling are provided in Fig.~\ref{fig:leptonic_sed}, in which we show, for both tests, the SED output from the five codes, together with their average\footnote{The average of $\nu F_{\nu}$  or $\gamma^2 dN/d\gamma dV$ is computed on a fixed grid on which the code outputs are interpolated.}. The subplots show the relative error of each model with respect to the average. 

The leptonic simulations indicate that the shape of the synchrotron and SSC components are well reproduced, and that the spread in normalization is within $\pm10\%$ over a broad part of the SEDs. Larger deviations are seen at the cutoffs of each spectral component. For instance, differences in the low-energy cutoff of the synchrotron component can be traced back to different ways of computing the synchrotron self-absorption coefficient. By comparing the SSC-TH and SSC-KN simulations we observe no significant differences in either spectral component.

\begin{figure}
\centering
\includegraphics[width = .47 \textwidth]{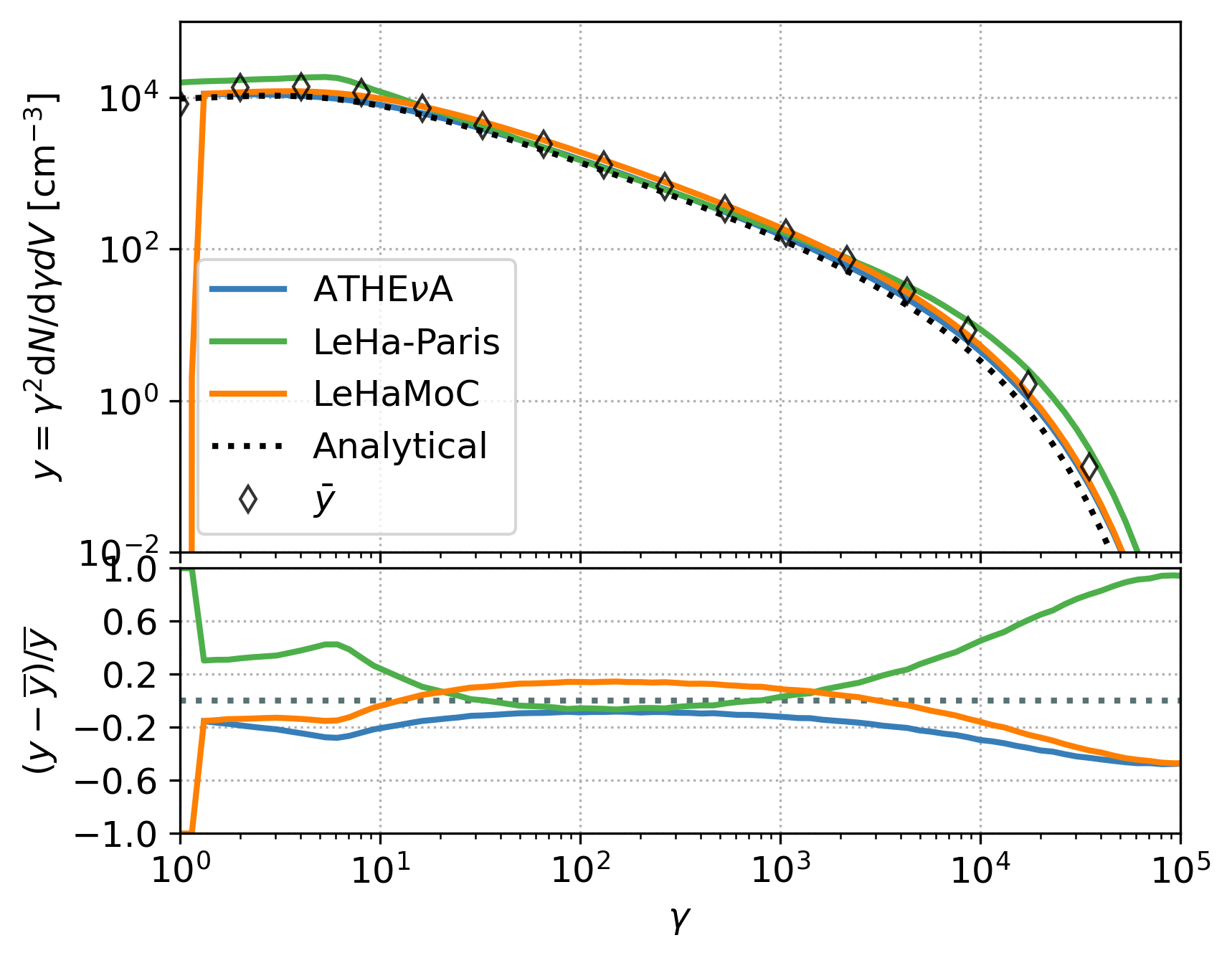}
\includegraphics[width = .47 \textwidth]{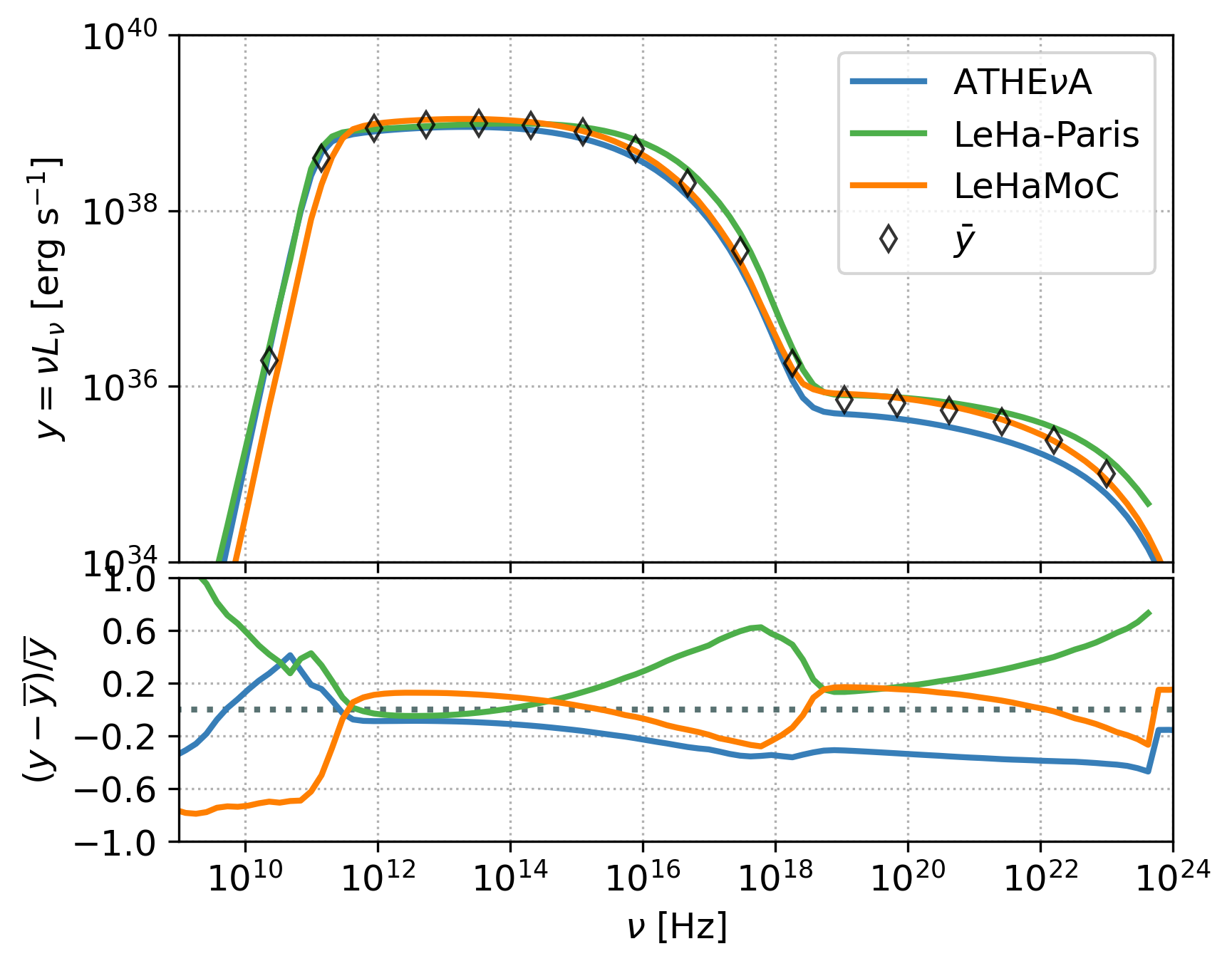}
\caption{Steady state electron distribution (left) and SSC spectra in the jet comoving frame (right) computed for the SYN-cool case using three of the codes described in Sec.~\ref{sec:codes}. Dotted line on the left panel shows the analytical steady-state electron distribution \citep{Inoue_1996}.  
}
\label{fig:leptonic_cooled_electrons}
\end{figure}

We then compare the performance of time-dependent and steady-state codes for a case where radiative cooling is important. For power-law injections with slope $p \gtrsim 2$ the effect of synchrotron cooling, or inverse Compton cooling in the Thomson limit, is to create a break in the energy distribution; if the power law distribution at injection is much harder ($p\ll 2)$, then the particle distribution at equilibrium shows a pile-up instead. As an indicative example we show a case where electrons cool due to synchrotron radiation (SYN-cool), and a cooling break is formed in their distribution. In Fig. \ref{fig:leptonic_cooled_electrons} we show, on the left panel, the electron distributions at equilibrium from \ath and \lehamoc, the corresponding analytical solution following Eq.~\ref{eq:inoue}, and a simple broken power-law parametrization used in \paris. Both solutions match in the power-law segment of the cooled distribution, but differ around the cooling break at $\gamma\sim 10$ and the high-energy cutoff; the parametrization used in \paris produces a sharp cooling break that overestimates the steady-state electron distribution in that energy range. The difference in the high-energy cutoff of the electron distribution is mapped to the cutoff of the synchrotron spectrum. Because electrons with $\gamma \sim 10$ emit synchrotron photons at $\sim 10^{10}$ Hz (i.e., below the synchrotron self-absorption frequency), we do not directly observe in the synchrotron spectra the difference of the respective particle distributions at the cooling break. 
We also observe that the synchrotron self-absorption computed with \lehamoc is higher by a factor of 4 compared to the other codes. As pointed out in \cite{lehamoc} such differences can be attributed to the method of computing the self-absorption coefficient. For example, if we use in \lehamoc the $\delta-$function approximation for the particle emissivity in the calculation of the absorption coefficient as done in \ath, the ratio between \lehamoc and \ath is less than $\sim 1.5$ in the self-absorbed part of the spectrum, as shown in subsection 3.2.1 in \cite{lehamoc}.\\

\subsection{Hadronic scenarios: kernel test cases}
\label{sec:hadronictest}

\begin{figure}[t!]
\centering
\includegraphics[width = .45 \textwidth]{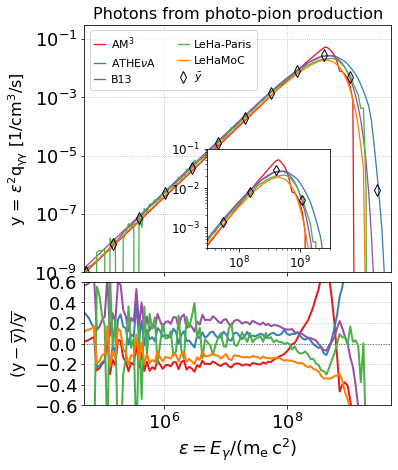}
\includegraphics[width = .45\textwidth]{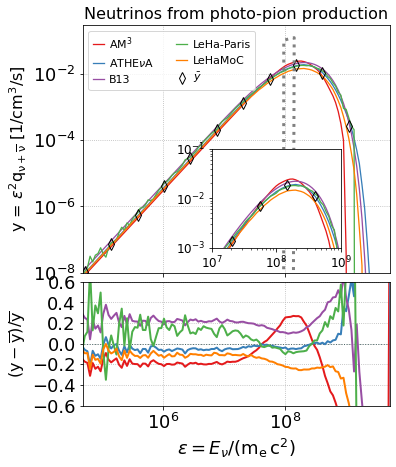}
\includegraphics[width = .45 \textwidth]{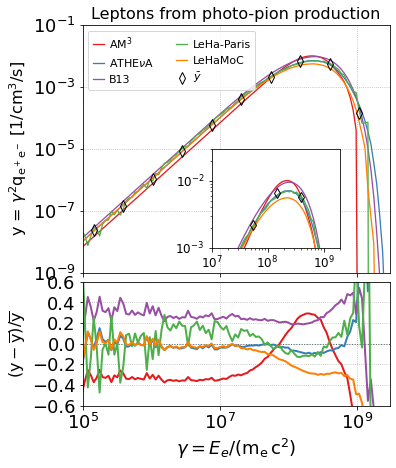}
\includegraphics[width = .47 \textwidth]{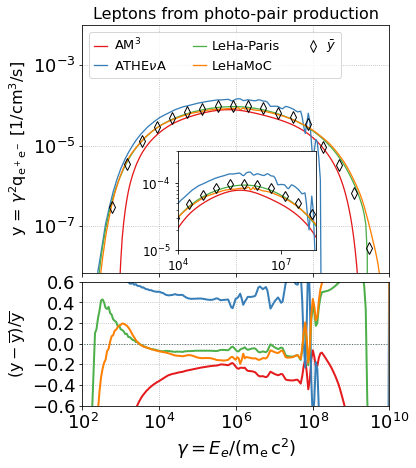}
\caption{Production rates per unit volume of secondaries (photons, leptons, and neutrinos of all flavors) from photopion production and photopair production for the case of a quasi mono-energetic proton distribution with $\gamma_p =10^{6}$ interacting with a grey-body radiation field (see p$\gamma$-MONOGB in Table~\ref{tab:parameters}).  Inset panels show a zoom into the energy range around the peak of the curves. The residuals of each code with respect to the mean value of the results are plotted in the bottom panels.   
The dotted grey line in the top right panel shows the analytical neutrino spectrum computed as explained in Appendix~\ref{app:analytical}.}
\label{fig:hadronic_rates_mono_6}
\end{figure}

\begin{figure}[ht!]
\centering
\includegraphics[width = .45 \textwidth]{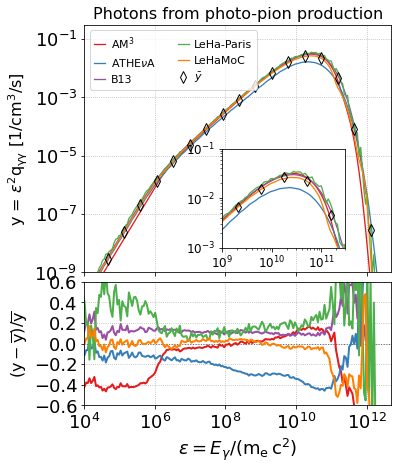}
\includegraphics[width = .45 \textwidth]{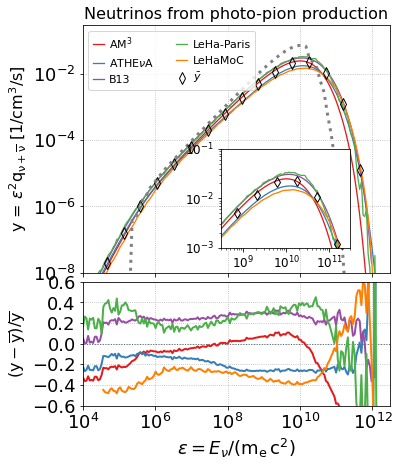}\\
\includegraphics[width = .45 \textwidth]{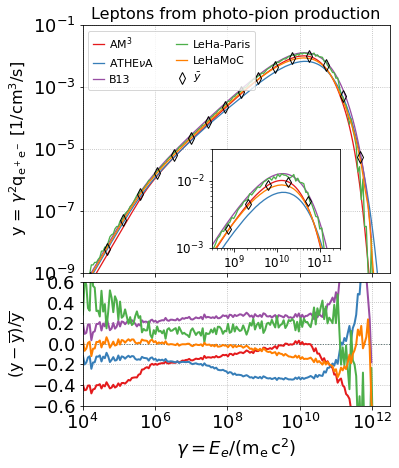}
\includegraphics[width = .45 \textwidth]{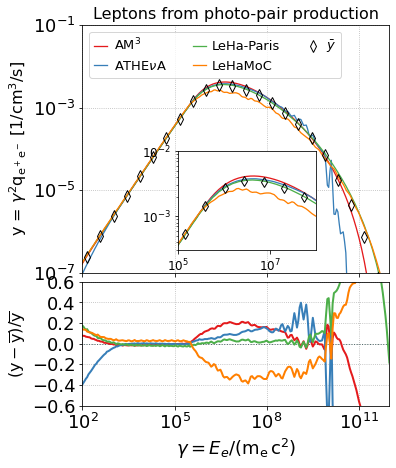}
\caption{Same as in Fig.~\ref{fig:hadronic_rates_mono_6} but for the case of a power-law proton distribution interacting with a power-law radiation field  (see p$\gamma$-PLPL in Table~\ref{tab:parameters}). The dotted grey line in the top right panel shows the analytical neutrino spectrum computed as explained in Appendix~\ref{app:analytical}. } 
\label{fig:hadronic_rates_pl}
\end{figure}

After investigating the differences in the leptonic modules of the codes, we can compare the implementation of hadronic interactions. To better identify potential discrepancies, we study first the injection spectra of secondary particles produced in interactions of protons with non-evolving (fixed) photon distributions.  The purpose of this comparison is to test the implementation of the conventional kernel functions for the production spectra of secondaries.

We start by studying proton-photon interactions between mono-energetic\footnote{A truly mono-energetic distribution cannot be implemented in codes that are expecting as input parameters the minimum/maximum energy of the particle distribution and the spectral index. A mono-energetic distribution is therefore approximated by the most narrow box-like energy distribution allowed by the grid resolution.} protons and photons following a grey-body distribution\footnote{A radiation field with a spectrum following the Planck function of temperature $T$, but with a total energy density smaller than $u_{\rm BB} = a T^4$.}. We choose protons with $\gamma_p = 10^6$ and a grey-body photon distribution of temperature $T_\gamma = 10^6$ K and compactness $\ell_{\gamma} = 8.1 \times 10^{-5}$; this is a dimensionless measure of the radiation energy density $u_{\gamma}$ and is defined as $\ell_{\gamma} = u_{\gamma} \sigma_T R /(3 m_e c^2)$. For the adopted parameters, photomeson interactions of protons with photons from the peak of the grey-body distribution, $\epsilon_{pk} \sim 3 k_B T$, happen close to the $\Delta^+$ resonance. Due to much lower energy threshold for Bethe-Heitler interactions ($\sim m_e/m_{\pi}$), the selected parameters lead to far-from threshold Bethe-Heitler interactions. These are characterized by broad (almost flat) energy injection spectra \citep{karavola}.

The results of this test (p$\gamma$-MONOGB) are shown in Fig.~\ref{fig:hadronic_rates_mono_6}, in which we provide the energy injection spectra for photons from $\pi^0$ decay, leptons from $\pi^\pm$ decay, and leptons from Bethe-Heitler pair-production\footnote{Results for a higher proton energy are provided in Appendix~\ref{appendix:monoenergetic7}.}. When comparing \ath, \paris, and \boet we find that the pionic $\gamma$-ray spectra and all-flavor neutrino spectra agree within $\pm20\%$ over a large part of the spectrum, with divergence only at cutoffs. The difference grows to $\pm30\%$ when considering the leptonic production spectra for a wide range of particle energies.
We observe here that especially \am exhibits a very different behavior from the other cases. This is not unexpected: its photohadronic interaction module is based on \citet{Hummer:2010vx}, which was optimized for power-law spectra balancing performance and precision. More concretely, the results are based on efficient single integrations by discretizing the integral over the secondary re-distribution functions (based on different physics processes) -- instead of using the usual Greens function/kernel approach. This yields unwanted spikes and features if quasi-monochromatic protons or target photons (or sharp cutoffs) are used, see also Fig. 9 in \citet{Hummer:2010vx}. An updated approach has been published in \citet{Biehl:2017zlw} (appendix A.3), where the discretization is automatized and the center-of-mass energy dependency of the re-distribution functions (leading to even stronger discrepancies at higher energies, see Appendix~\ref{appendix:monoenergetic7}) is improved -- at the expense of losing the relationship to the underlying physics processes. It is useful to inspect Fig. 23 (upper right panel) of that article: the total spectrum is composed of different spectra added together with different values of the secondary to primary energy ratio (instead of integrating over it). Omitting certain contributions of that parameter leads to an under-prediction of the secondary spectrum in certain ranges, which can be also seen at low and high energies in Fig.~\ref{fig:hadronic_rates_mono_6}. The original approach in \citet{Hummer:2010vx} compensates this partially by choosing larger multiplicities, which leads to the sharp features.

The agreement for the Bethe-Heitler injection is considerably better than the photomeson part at the flat part of the spectrum well within $\pm 15\%$, for three of the codes. \ath overestimates the Bethe-Heitler injection at all energies: this behavior is understood as an effect of the fixed energy grid resolution in the code that is not optimized for very narrow proton distributions; in Appendix \ref{app:BH} we provide more details about this aspect and show that the agreement is recovered when the proton distribution is widened.\\

\begin{figure}[t!]
    \centering
    \includegraphics[width=0.9\textwidth]{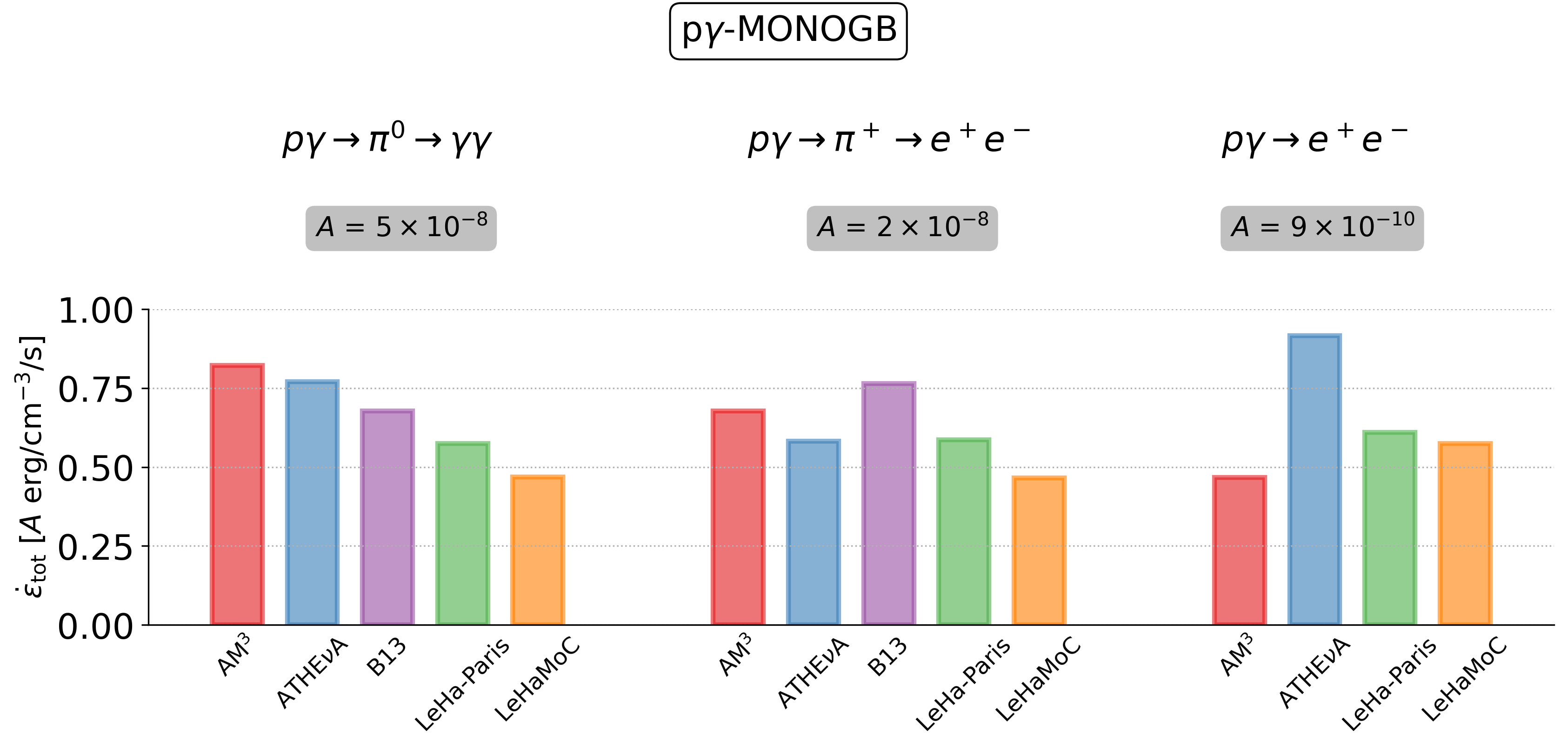} 
    \vspace{1cm}
    \includegraphics[width=0.9\textwidth]{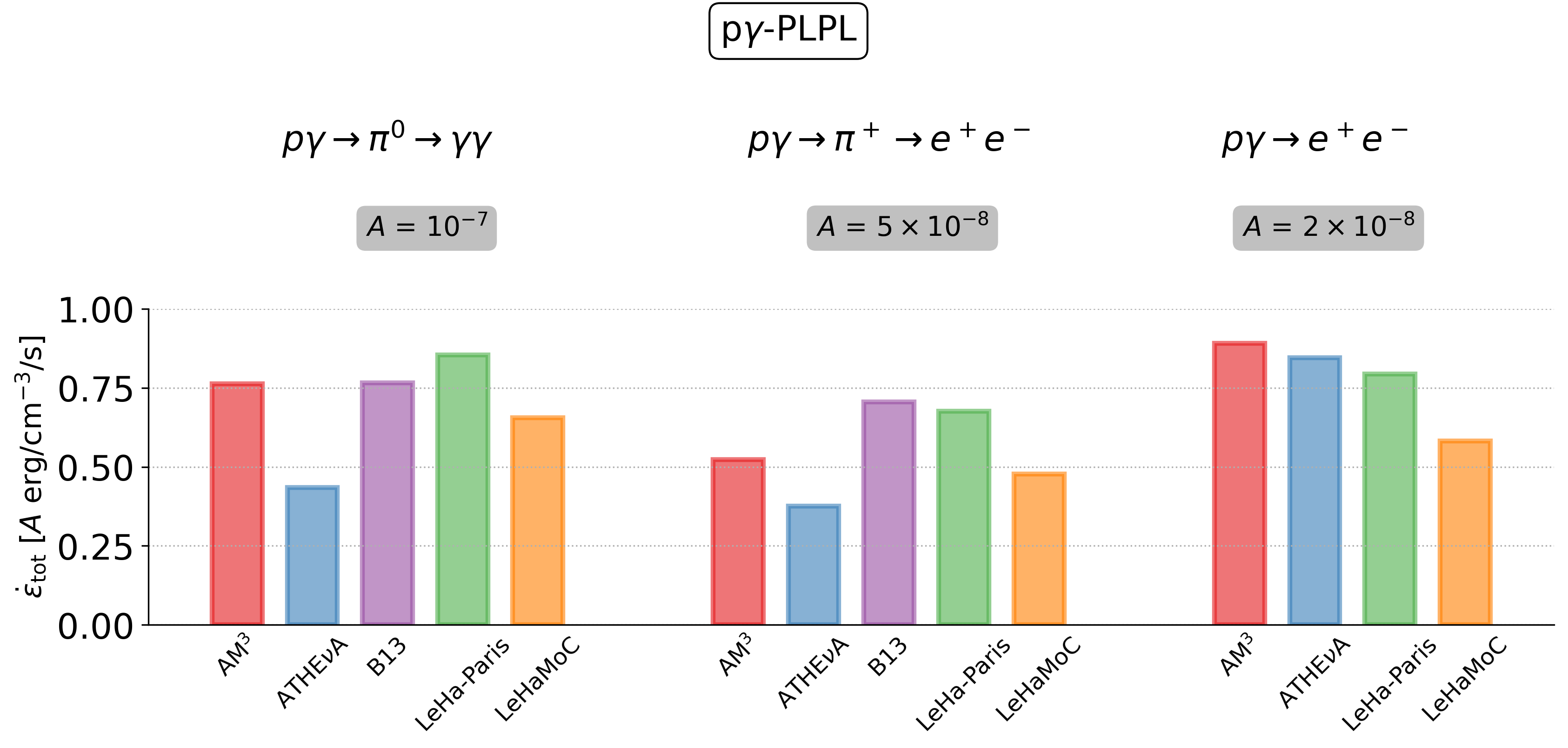}
    \caption{Integrated energy density injection rates of secondaries from neutral pion decay, charged pion decay, and photopair production computed for proton interactions with a grey-body radiation field for  $\mathrm{log_{10}}(\gamma_p) = 6$ (top) and a  power-law proton distribution interacting with a  power-law radiation field  (bottom). To infer the actual energy injection rate, the height of each bar should be multiplied by the appropriate normalization factor $A$. }
    \label{fig:integrated_rates}
\end{figure}

We then consider the case of interacting power-law proton with power-law photon distributions, which can be of more general applicability to astrophysical systems. For the adopted parameters (see p$\gamma$-PLPL in Table~\ref{tab:parameters}), the energy of the proton distribution is carried by the most energetic particles of the distribution with $\gamma_{p, \max}=10^8$. Therefore, the secondary spectra will be determined by the interactions of the most energetic protons of the power law. These protons will interact just above the pion-production energy threshold $\bar{\epsilon}_{\rm th, p\gamma}\simeq 0.145$~GeV with the lowest energy photons from the power law, which are the most numerous since $E_{\gamma} \mathrm{d}N/\mathrm{d}E_{\gamma} \propto E^{-1}_{\gamma}$. Due to the lower energy threshold of the Bethe-Heitler pair production process, $\bar{\epsilon}_{\rm th, BH} \simeq 1$~MeV, the most energetic protons will interact far from the threshold with the lowest energy photons of the power-law distribution. 

The results for the  p$\gamma$-PLPL are presented in Fig.~\ref{fig:hadronic_rates_pl}. Inspection of the $\gamma$-ray, lepton, and neutrino spectra from pion decays shows that the differences between codes are between $\pm 30\%$ to $\pm40\%$ for the different species. The spectral shapes are similar around and below the peaks, with differences becoming larger at the cutoffs. The ordering of codes in terms of their relative difference with respect to the mean is the same for all particle species. At first thought the larger differences found here compared to the p$\gamma$-MONOGB cases may seem unexpected. However, in Appendix \ref{app:other} we show that even in the p$\gamma$-MONOGB cases the relative differences increase up to $\pm40\%$ as the typical interaction energy increases (compare Figs.~\ref{fig:hadronic_rates_mono_6} and \ref{fig:hadronic_rates_mono_7}). Therefore, by summing up the contributions from near and far from threshold interactions, such differences tend to accumulate. The maximum relative difference will depend on the power-law slopes and the energy limits of both distributions, as these variables determine essentially the relative contribution of far and near threshold interactions to the total spectrum. Interestingly, the differences in the pair production spectra are much smaller, less than $20\%$ at the peak energy. Contrary to the broad and almost flat pair-production energy spectra computed for the p$\gamma$-MONOGB cases, the spectra exhibit a clear peak at  $\gamma \sim 10^6$, which is relatively narrow. These spectral properties suggest that maximum energy injection rate is determined by near-threshold interactions of protons with $\gamma_p = 10^6$ with the lowest energy photons of the power law \citep[for more details, see][]{karavola}. It is interesting to note that for AM$^3$, the earlier found seemingly large discrepancies in the photo-meson injection with respect to the mean model disappear close to the peaks and the spectra almost match perfectly with the other codes, which comes from the before-mentioned optimization for power-law spectra.

We additionally compare the volumetric energy injection rates of the five codes by integrating the differential production spectra (per unit volume) over the photon energies and lepton Lorentz factors. The results are displayed in Fig.~\ref{fig:integrated_rates} in the form of bar charts. Because the energy injection spectra of all particle species produced in photomeson interactions have a well defined peak, the differences displayed around the maximal values in the differential energy spectra are also reflected to the differences shown in the bar charts.\\

\subsection{Hadronic blazar-like scenarios}\label{sec:leptohadro}

We now investigate the agreement among codes for more \textit{blazar-like} cases, and we study in particular two emission scenarios that have been put forward to explain blazar SEDs: the first one is a proton-synchrotron (PS) solution, in which the high-energy SED peak is ascribed to synchrotron emission by primary protons in the emitting region, and in which the emission by secondary particles produced in proton-photon interactions emerges only at higher energies, and is subdominant with respect to the proton synchrotron one; the second one is a hybrid, leptohadronic (LeHa) solution, in which the high-energy SED component is due to both primary electrons (via SSC) and radiation by secondary leptons produced in proton-photon interactions. With respect to the previous tests, there are no external photon fields, and the proton-photon interactions happen (mainly) between primary protons and synchrotron photons by primary electrons. For both tests, we are now interested in comparing the multi-messenger photon/neutrino SEDs (in luminosity, $\nu L (\nu)$), and the injection rates are not shown. We only show here the results from the four codes that include the Bethe-Heitler process.  

\begin{figure}[t!]
    \centering
    \includegraphics[width=0.9\textwidth]{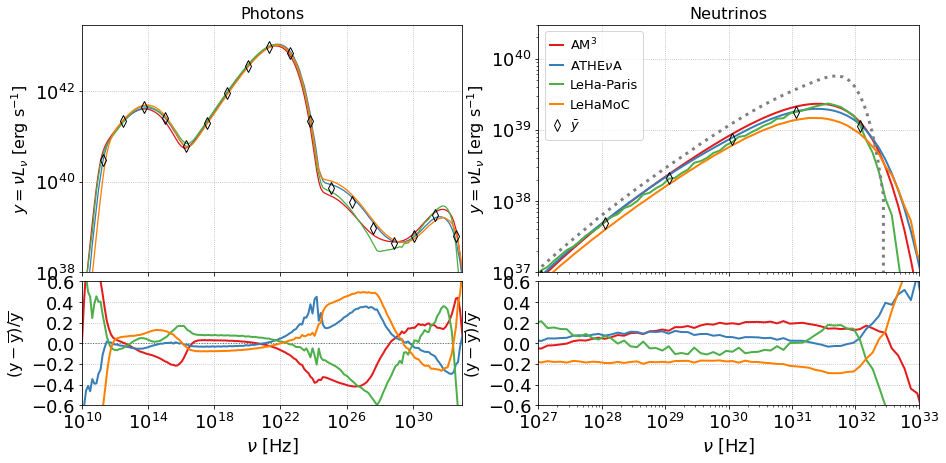}
    \caption{Spectral energy distributions (in the observer's frame) of photons and neutrinos (of all flavors) computed for the PS scenario (see Table~\ref{tab:parameters}). The mean model is shown with open markers. The residuals of each code with respect to the mean value of the results are plotted in the bottom panels. The dotted grey line in the top right panel shows the analytical neutrino spectrum computed as explained in Appendix~\ref{app:analytical}. }
    \label{fig:psyn}
\end{figure}

\begin{figure}[h!]
    \centering
        \includegraphics[width=0.9\textwidth]{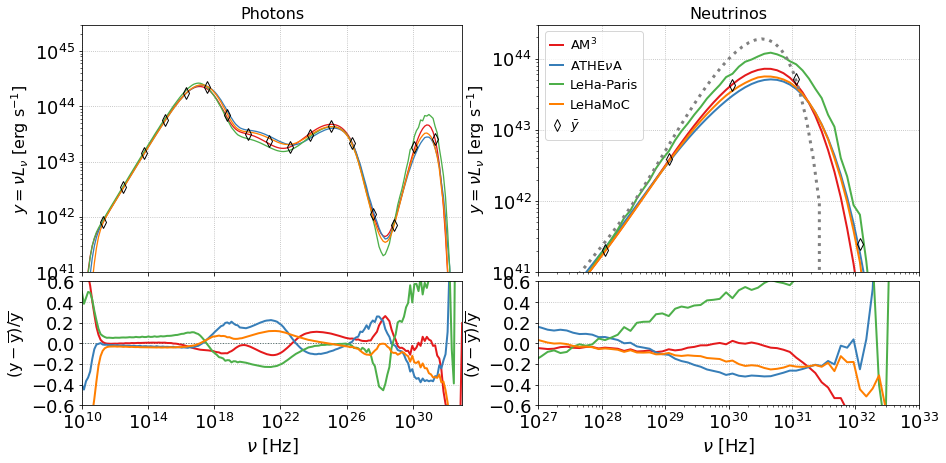}
    \caption{Same as in Fig.~\ref{fig:psyn} but for the LeHa scenario (see Table~\ref{tab:parameters}).
}
    \label{fig:lepto-hadronic}
\end{figure}

The proton synchrotron test has the same proton distribution as the previous test ($p\gamma$-PLPL), i.e. a power-law proton distribution between $\gamma_{\rm p, min}=1$ and $\gamma_{\rm p, max}=10^8$ and index of 1.9; in the leptohadronic test, in order to get an SED more typical of a blazar, we adopt a softer index ($2.0$) and lower maximum proton Lorentz factor ($10^7$). The details of the model parameters are provided in Table \ref{tab:parameters}, and the multi-messenger SEDs are shown in Fig.~\ref{fig:psyn} (for the PS test) and \ref{fig:lepto-hadronic} (for the LeHa test). 

The proton synchrotron SED shows four distinct components in order of increasing energy: synchrotron by primary electrons, synchrotron by primary protons, synchrotron by Bethe-Heitler pairs and photons from $\pi^0$ decay. To ease the comparison and the identification of the various processes, $\gamma-\gamma$ absorption is turned off for this test only. As can be seen from Fig.~\ref{fig:psyn} the photon SEDs agree at the level of $\pm15\%$ in the energy ranges where electron and proton synchrotron emission dominates. The differences among the codes increase to about $\sim 20\%$ at the highest energies, where the $\pi^0$ bump emerges. The largest difference ($\pm 40\%$) is found at energies where the synchrotron emission from Bethe-Heitler pairs dominates.
For the neutrino spectrum we also show the result from a semi-analytical calculation described in Appendix \ref{app:analytical}. While the semi-analytical result captures the peak of the neutrino SED well, it produces a narrower spectrum overall, because it assumes a one-to-one mapping between the energies of the parent proton and the produced neutrino (see Eq.~\ref{eq:Lnu-analytic}). In other words, the semi-analytical approach does not account for the spread in neutrino energies produced by a single proton.

The leptohadronic SEDs also show four components in order of increasing energy: synchrotron by primary electrons, synchrotron by Bethe-Heitler pairs, a superposition of synchrotron-self-Compton and synchrotron by photo-meson pairs, and photons from $\pi^0$ decay. As can be seen from Fig. \ref{fig:lepto-hadronic}  the photon SEDs show a remarkable agreement within $\pm20\%$, with the exception of the $\pi^0$ component that opens up to $-40\%$ to $60\%$. The same larger spread can be seen in the neutrino spectra that show $-30\%$ to $60\%$. The dispersion is driven by the \paris result that overestimates the others. It is not completely clear why this particular test shows a larger spread in the photo-meson injection: part of it could be related to the fact that with these model parameters the primary electrons are cooled, and \paris parametrizes primary particles with a broken-power law function (see Section \ref{sec:leptonic} and Fig.~\ref{fig:leptonic_cooled_electrons}); the over-estimation by about $15\%$ of the peak of the electron-synchrotron implies a larger target density for photo-meson interactions and thus a brighter neutrino emission by the same amount. Finally, the neutrino spectrum from the semi-analytical approach matches the peak of the neutrino SEDs computed with \am, \ath, and \lehamoc, but produces a narrower spectrum.

\section{Other scenarios}
\label{sec:otherscenarios}

In this section we test the performance of the time-dependent codes \am and \ath in computing the spectral and temporal behavior of high-density emitting regions, using a leptonic (Sec.~\ref{subsec:iccatastrophe}) and purely hadronic scenario (Sec.~\ref{subsec:pps}). While these test cases have not been directly applied to AGN modeling, they are ideal for highlighting the non-linear coupling among different particle species.

\subsection{Non-linear electron cooling: the case of inverse Compton catastrophe}
\label{subsec:iccatastrophe}
 The term ``inverse Compton catastrophe'' refers to the dramatic rise in the luminosity of inverse Compton scattered photons that would occur due to the rapid cooling of electrons via inverse Compton scattering~\citep{Longair_2011}.  

In a synchrotron-emitting source Compton catastrophe can be realized when the energy density of the magnetic field, which determines the rate of electron cooling through synchrotron radiation, is much lower than the energy density of synchrotron photons. A runaway process is therefore possible to develop: low-energy (e.g. radio) photons produced by synchrotron radiation are scattered to higher energies (e.g. in X-rays) by the same relativistic electron population. If the energy density of these high-energy photons is larger than that of synchrotron photons, the electrons would suffer even greater energy losses by up-scattering them to even higher energies, e.g. in $\gamma$-rays. These photons would have in turn greater energy density than the X-ray photons, and so on. This process would eventually cease when the highest order inverse Compton scatterings would take place in the Klein-Nishina regime~\citep[e.g.][]{PPM15}.


Here, we compare two of the codes (\am and \ath) that can treat non-linear cooling of the electron population. For this purpose, we use parameters that lead the source into the Compton catastrophe regime, namely $B=10$~G, $R=10^{16}$~cm,  $\ell_{e}^{\rm inj}=1$ or $L_{e}^{\rm inj}=4.6\times10^{45}$~erg s$^{-1}$ (see Eqs.~\ref{eq:Linj-Q} and \ref{eq:comp-Q}), $s_e=2$, $\gamma_{e, \min}=10^{1.9}$, and $\gamma_{e, \max}=10^{2.1}$; an exponential cutoff at $\gamma_{e, \max}$ was also used.

\begin{deluxetable}{l c | l c}
\tablewidth{0pt}
    \tablecaption{Parameters for the other scenarios.    \label{tab:special_scenarios}}
    \tablehead{\multicolumn{2}{c|}{Inverse Compton Catastrophe} &  \multicolumn{2}{c}{PPS loop}}
    \startdata
         Parameter & Value & Parameter & Value \\ \hline
         $B$ [G] & 10 & $B$ [G]& 31.6 \\
         $R$ [cm] & $10^{16}$  & $R$ [cm] & $10^{15}$ \\
         $\gamma_{e, \mathrm{min}}$& $10^{1.9}$ & $\gamma_{p, \mathrm{min}}$ & $10^{0.1}$\\
         $\gamma_{e, \mathrm{max}}$& $10^{2.1}$ & $\gamma_{p, \mathrm{max}}$ & $10^{5}$\\
         $p_e$& 2.0& $p_p$ & 2.0 \\
         log($\ell_e$)& $0.0$ & log($\ell_p$) & $\{-3.2, -2.9, -2.6, -2.3 \}$\\
         $t_{e, \mathrm{esc}}$ [$R/c$] & 1 & $t_{p, \mathrm{esc}}$ [$R/c$]& $10^{3}$ \\
    \enddata
    \tablecomments{ \textit{Inverse Compton catastrophe:} Synchrotron self-absorption and $\gamma\gamma$ -annihilation were not taken into account. \textit{PPS loop:} Synchrotron self-absorption, leptonic and hadronic inverse Compton, photo-pion production and $\gamma\gamma$-annihilation were not taken into account.}
\end{deluxetable}

\begin{figure}
    \centering
    \includegraphics[width=0.47\textwidth]{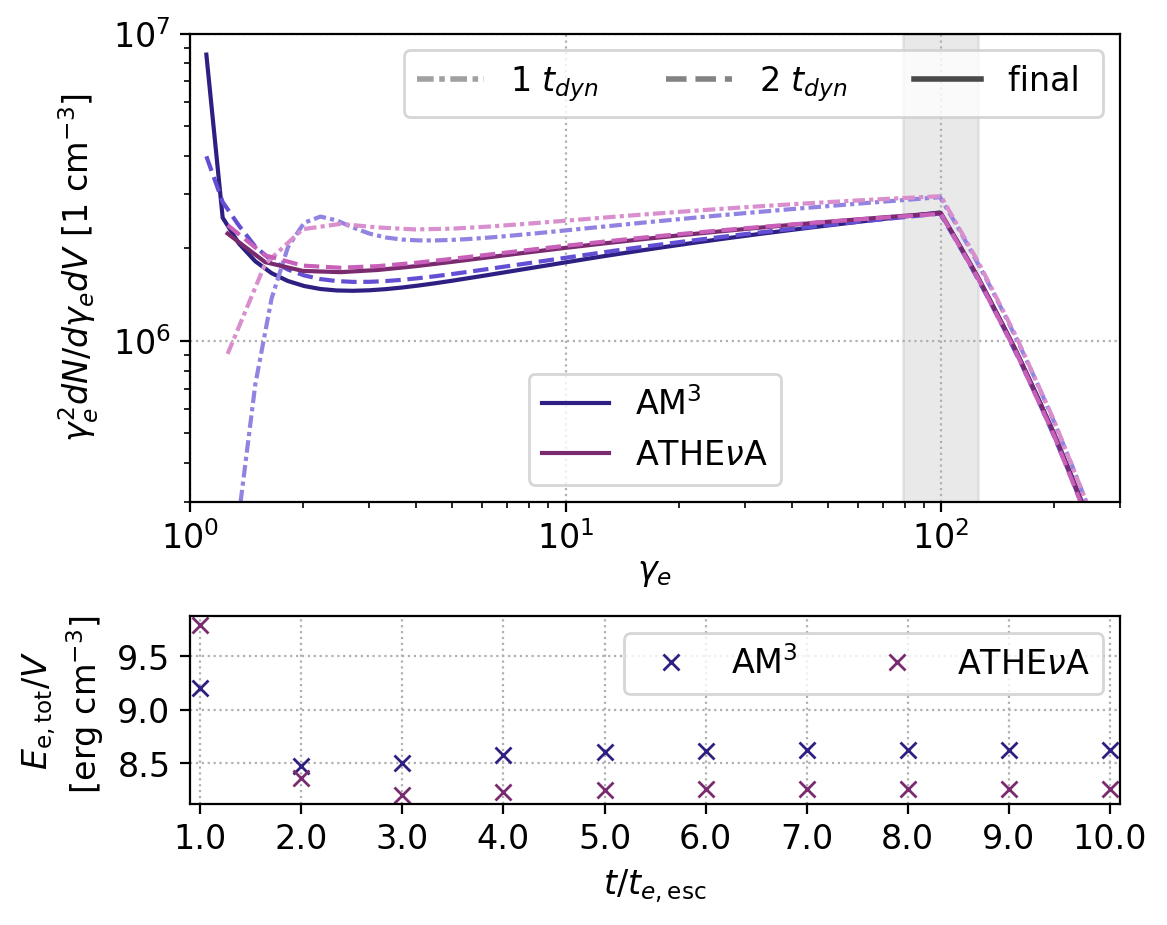}
    \includegraphics[width=0.48\textwidth]{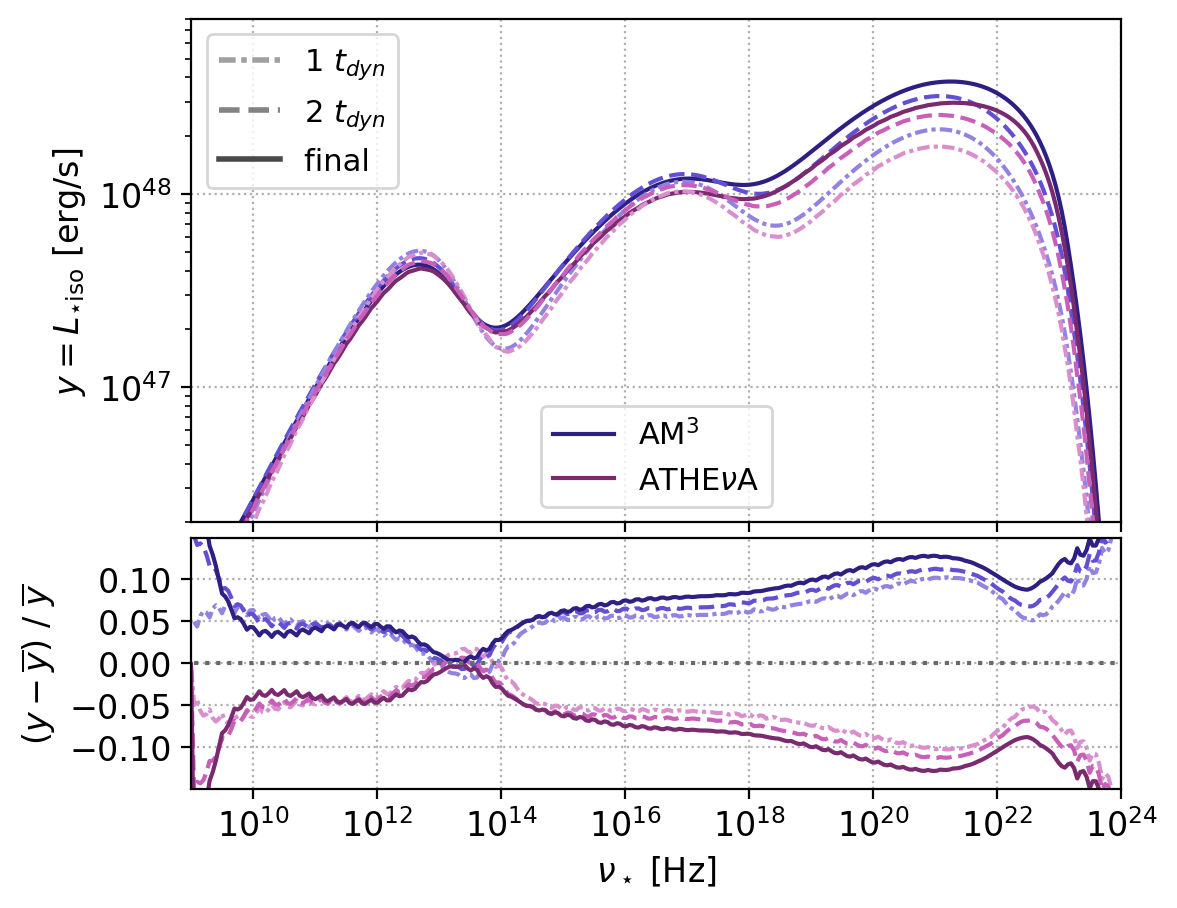}
    \caption{Results for an inverse Compton catastrophe scenario computed using AM$^3$ and \ath. \textit{Left panel:} Electron energy distributions. Thick solid lines show the steady-state distributions, while thin dash-dotted and dashed lines show the distributions after one and two dynamical times, respectively. The grey-shaded region indicates the energy range of injected electrons. The bottom panel shows the temporal evolution of the electron energy density. \textit{Right panel:} Photon spectral energy distributions for the same time stamps shown on the left panel. Residuals with respect to the mean are shown in the bottom panel. }
    \label{fig:CC}
\end{figure}

For the selected parameters electrons are fast cooling, as indicated by the extension of the electron distribution to $\gamma_e \ll \gamma_{e, \min}$ within one dynamical timescale (see left top panel in Fig.~\ref{fig:CC}). 
Differences in the shape of the electron distribution at  $\gamma_e < 2$ are mainly caused by differences in the energy grid resolution of the two codes, the default choice for \ath being 10 points per decade in energy compared to 20 points per decade used in \am. We refer the reader to Appendix~\ref{app:resolution} for details on the impact of the energy grid and temporal resolution used in the two codes.  Moreover, there is a good agreement between the codes in terms of the total energy carried by the electron distribution over the course of 10 dynamical times as shown in the bottom left panel of Fig.~\ref{fig:CC}. Finally, the broadband photon spectra are displayed in the right panel of the figure (top) with the relative difference also shown in the bottom panel. Despite the non-linearity of the physics problem at hand the two codes predict fluxes that differ at most by 10\%, similarly to the  SSC model presented in Fig.~\ref{fig:leptonic_sed}. 

\subsection{Non-linear proton cooling: the Pair-Production-Synchrotron (PPS) loop}\label{subsec:pps}
Relativistic protons can under certain conditions participate in various types of radiative instabilities~\citep[see][for a comprehensive study]{Mastichiadis20}. When the proton energy density in the source exceeds some critical value, a runaway process is initiated resulting in the explosive transfer of the proton energy into electron-positron pairs, radiation, and neutrinos. The runaway also leads to an increase of the radiative efficiency (defined as the ratio of the photon luminosity to the injected proton luminosity). 

One of these hadronic radiative instabilities, known as the Pair-Production-Synchrotron (PPS) loop, was first studied  \cite{KM92}. We consider a source containing relativistic protons and magnetic fields (hence, this is a purely hadronic scenario). If the conditions (i.e. magnetic field strength and proton Lorentz factor) are such so that the pairs produced by photo-pair production radiate synchrotron photons that are also targets for photo-pair production, a closed loop of physical processes is formed. This loop can be self-sustained if at least one of the synchrotron photons produces a pair before escaping the source. This condition translates to a critical proton density \citep[see equation 2 in][]{KM92}, which whenever surpassed leads to an explosive transfer of proton energy to photons. 

While the growth rate of the instability can be computed semi-analytically \citep{KM92}, a study of the instability in the non-linear regime requires a full numerical treatment \citep{MPK05, Mastichiadis20}. The PPS loop is therefore an excellent case study for comparing the performance of the two time-dependent codes \am and \ath.

\begin{figure}
    \centering
\includegraphics[width=0.9\textwidth]{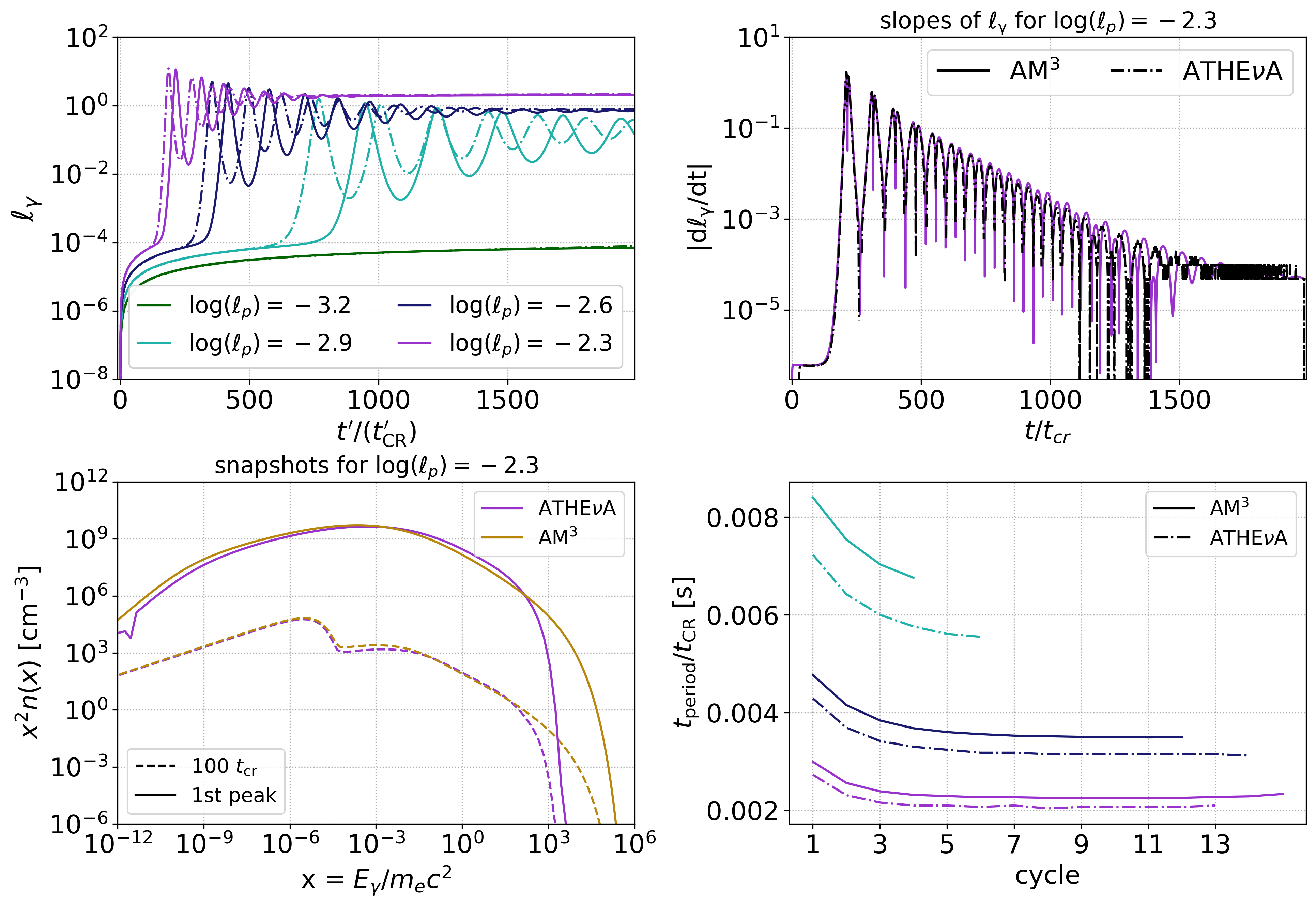}
    \caption{Results for the Pair-Production-Synchrotron (PPS) scenario computed using AM$^3$ and \ath. From top left and in clockwise order we show (i) time series of the photon compactness $\ell_\gamma$ (i.e. dimensionless light curves) computed for four values of the proton injection compactness $\ell_{p}$ (see inset legend), (ii) time derivative of the light curve computed for the highest value of $\ell_p$ (after applying a shift to the light curves to make them overlap), (iii) two snapshots of the photon spectra for the highest value of $\ell_p$,  and (iv) evolution of the period with the number of cycles for the three light curves in the first panel exhibiting damped oscillatory behavior (same color coding used).}
    \label{fig:pps}
\end{figure}

An illustrative example of the PPS loop is presented in Fig.~\ref{fig:pps} for $B=10^{1.5}$~G, $R=10^{15}$~cm, and $t_{p, esc}=10^3~R/c$. The top left panel shows the bolometric photon compactness as a function of time computed for different proton compactnesses, ranging from $10^{-3.2}$ (green curves) to $10^{-2.3}$ (purple curves) with logarithmic increments of 0.3. For proton densities below a critical value the system reaches a steady state that corresponds to a constant photon compactness (green curves). Above the critical proton density, however, non-linear feedback between protons and synchrotron photons from Bethe-Heitler pairs becomes relevant, and the photon light curve exhibits outbursts that correspond to times of efficient proton cooling. Both numerical codes can capture the transition from the linear to the non-linear regime (green and cyan lines) for the same $\ell_p$ values, as well as the damped oscillatory  behaviour of the proton-photon system that was first presented in \citep{MPK05}. This is the first time that the oscillations can be reproduced with a code other than \ath.

There is a constant offset between the peak times of the outbursts, whose origin could not be pinned down. Similar offsets have been observed in numerical runs performed with \ath, after changing, for example, the difference scheme used to replace the partial derivatives. Regardless, we find an excellent agreement between the codes in terms of period and period evolution (see lower right panel). In the top right panel we also show the time derivative of the light curve as a function of time, after correcting for the offset. There is very good agreement between the codes in describing the shape of the light curve. Finally, the broadband photon spectra computed at two indicative times (at the onset of the instability, and at the peak time of the first outburst) are presented in the bottom left panel of the same figure. There is good agreement in terms of shape and normalization. The difference in the high-energy cutoff of the synchrotron spectra is to be expected, since the pair injection spectra in \ath cut off abruptly (see also Fig.~\ref{fig:hadronic_rates_mono_6}). 

\section{Summary and Conclusions}
\label{sec:discussion}

We have presented the results from the first comparison among five hadronic codes developed to model photon and neutrino emission from blazars. We started by comparing the leptonic part of the codes (synchrotron and SSC) showing that we have a general agreement at the level of $\pm10\%$. The hadronic part is first checked by looking at tests for monoenergetic/power-law proton distributions interacting with simple photon fields (grey-body or power-law). In these cases we are interested in comparing the injection rates of secondary particles produced in p-$\gamma$ interactions. The level of agreement is at the level of $\pm20\%$ to $\pm40\%$. It is more difficult here to provide a simple value, as the dispersion depends on the specific test, the process we are evaluating, and the energies (with large deviations observed at cut-offs). A $\pm40\%$ systematic spread represents a conservative value for the typical largest envelop we observe (excluding the cut-offs where anyhow the flux is fast dropping).  The final comparisons are performed on two realistic, blazar-like, tests in which we show the agreement in terms of photon and neutrino SED. In these two cases as well the agreement ranges from  $\pm20\%$-$\pm40\%$. It is again difficult to gauge what will be the systematic uncertainty for another test performed elsewhere in the parameter space to model a future gamma-neutrino source, but we consider that $\pm40\%$ can be considered a conservative, energy-independent, value to be adopted when comparing a single realisation from a single code to observations.\\
Some important caveats should be highlighted here. The code comparison presented here has been done by putting all codes in the same conditions, to isolate only the differences coming from the implementation of hadronic processes. There are other assumptions and hypotheses in the numerical codes that have an impact on the overall normalization, and thus adds up to the spread quantified here. The most relevant for this study is the correction for the radial dependency of the photon distribution in a sphere, that contributes for an extra $25\%$. In a similar way, the hypothesis on the equilibrium distribution of primary electrons also impacts the photon output, and not just in a direct way by modifying their synchrotron emission, but also indirectly by changing the target photon field for p-$\gamma$ interactions.
In addition, in some part of the parameter space, some processes that are included in some codes but not in others become relevant if not dominant (i.e. Bethe-Heitler pair production, or synchrotron by muons) and this aspect should be carefully checked when exploring a large parameter space.
In an effort to facilitate reproducibility and to benchmark any future code development in the community (that already happened since the TXS 0506+056 event and while this comparison project was in preparation, see e.g. \citet{katu, soprano, zacharias}), we release all output files from all codes used to produce the comparison plots in this paper as online material at \url{https://doi.org/10.5281/zenodo.17177642}.\\


\acknowledgments
 M.P. and S.D. acknowledge support from the Hellenic Foundation for Research and Innovation (H.F.R.I.) under the ``2nd call for H.F.R.I. Research Projects to support Faculty members and Researchers'' through the project UNTRAPHOB (Project ID 3013). X.R. was supported by the Deutsche Forschungsgemeinschaft (DFG, German Research Foundation) through grant SFB 1258 ``Neutrinos and Dark Matter in Astro- and Particle Physics'' and by the Excellence Cluster ORIGINS which is funded by the DFG under Germany's Excellence Strategy - EXC 2094 - 390783311.

\bibliography{ref}

\begin{thebibliography}{}
\expandafter\ifx\csname natexlab\endcsname\relax\def\natexlab#1{#1}\fi
\providecommand{\url}[1]{\href{#1}{#1}}
\providecommand{\dodoi}[1]{doi:~\href{http://doi.org/#1}{\nolinkurl{#1}}}
\providecommand{\doeprint}[1]{\href{http://ascl.net/#1}{\nolinkurl{http://ascl.net/#1}}}
\providecommand{\doarXiv}[1]{\href{https://arxiv.org/abs/#1}{\nolinkurl{https://arxiv.org/abs/#1}}}

\bibitem[{{Aab} {et~al.}(2020){Aab}, {Abreu}, {Aglietta}, {Albury},
  {Allekotte}, {Almela}, {Alvarez Castillo}, {Alvarez-Mu{\~n}iz}, {Alves
  Batista}, {Anastasi}, {Anchordoqui}, {Andrada}, {Andringa}, {Aramo},
  {Ara{\'u}jo Ferreira}, {Asorey}, {Assis}, {Avila}, {Badescu}, {Bakalova},
  {Balaceanu}, {Barbato}, {Barreira Luz}, {Becker}, {Bellido}, {Berat},
  {Bertaina}, {Bertou}, {Biermann}, {Bister}, {Biteau}, {Blanco}, {Blazek},
  {Bleve}, {Boh{\'a}{\v{c}}ov{\'a}}, {Boncioli}, {Bonifazi}, {Bonneau
  Arbeletche}, {Borodai}, {Botti}, {Brack}, {Bretz}, {Briechle}, {Buchholz},
  {Bueno}, {Buitink}, {Buscemi}, {Caballero-Mora}, {Caccianiga}, {Calcagni},
  {Cancio}, {Canfora}, {Caracas}, {Carceller}, {Caruso}, {Castellina},
  {Catalani}, {Cataldi}, {Cazon}, {Cerda}, {Chinellato}, {Choi}, {Chudoba},
  {Chytka}, {Clay}, {Cobos Cerutti}, {Colalillo}, {Coleman}, {Coluccia},
  {Concei{\c{c}}{\~a}o}, {Condorelli}, {Consolati}, {Contreras}, {Convenga},
  {Covault}, {Dasso}, {Daumiller}, {Dawson}, {Day}, {de Almeida}, {de
  Jes{\'u}s}, {de Jong}, {De Mauro}, {de Mello Neto}, {De Mitri}, {de
  Oliveira}, {de Oliveira Franco}, {de Souza}, {De Vito}, {Debatin}, {del
  R{\'\i}o}, {Deligny}, {Dembinski}, {Dhital}, {Di Giulio}, {Di Matteo},
  {D{\'\i}az Castro}, {Dobrigkeit}, {D'Olivo}, {Dorosti}, {dos Anjos}, {Dova},
  {Ebr}, {Engel}, {Epicoco}, {Erdmann}, {Escobar}, {Etchegoyen}, {Falcke},
  {Farmer}, {Farrar}, {Fauth}, {Fazzini}, {Feldbusch}, {Fenu}, {Fick},
  {Figueira}, {Filip{\v{c}}i{\v{c}}}, {Fodran}, {Freire}, {Fujii}, {Fuster},
  {Galea}, {Galelli}, {Garc{\'\i}a}, {Garcia Vegas}, {Gemmeke}, {Gesualdi},
  {Gherghel-Lascu}, {Ghia}, {Giaccari}, {Giammarchi}, {Giller}, {Glombitza},
  {Gobbi}, {Gollan}, {Golup}, {G{\'o}mez Berisso}, {G{\'o}mez Vitale},
  {Gongora}, {Gonz{\'a}lez}, {Goos}, {G{\'o}ra}, {Gorgi}, {Gottowik}, {Grubb},
  {Guarino}, {Guedes}, {Guido}, {Hahn}, {Halliday}, {Hampel}, {Hansen},
  {Harari}, {Harvey}, {Haungs}, {Hebbeker}, {Heck}, {Hill}, {Hojvat},
  {H{\"o}randel}, {Horvath}, {Hrabovsk{\'y}}, {Huege}, {Hulsman}, {Insolia},
  {Isar}, {Johnsen}, {Jurysek}, {K{\"a}{\"a}p{\"a}}, {Kampert}, {Keilhauer},
  {Kemp}, {Klages}, {Kleifges}, {Kleinfeller}, {K{\"o}pke}, {Kukec Mezek},
  {Lago}, {LaHurd}, {Lang}, {Leigui de Oliveira}, {Lenok}, {Letessier-Selvon},
  {Lhenry-Yvon}, {Lo Presti}, {Lopes}, {L{\'o}pez}, {Lorek}, {Luce}, {Lucero},
  {Machado Payeras}, {Malacari}, {Mancarella}, {Mandat}, {Manning},
  {Manshanden}, {Mantsch}, {Marafico}, {Mariazzi}, {Mari{\c{s}}}, {Marsella},
  {Martello}, {Martinez}, {Mart{\'\i}nez Bravo}, {Mastrodicasa}, {Mathes},
  {Matthews}, {Matthiae}, {Mayotte}, {Mazur}, {Medina-Tanco}, {Melo},
  {Menshikov}, {Merenda}, {Michal}, {Micheletti}, {Miramonti}, {Mockler},
  {Mollerach}, {Montanet}, {Morello}, {Mostaf{\'a}}, {M{\"u}ller}, {Muller},
  {Mulrey}, {Mussa}, {Muzio}, {Namasaka}, {Nellen}, {Nguyen},
  {Niculescu-Oglinzanu}, {Niechciol}, {Nitz}, {Nosek}, {Novotny},
  {No{\v{z}}ka}, {Nucita}, {N{\'u}{\~n}ez}, {Palatka}, {Pallotta}, {Panetta},
  {Papenbreer}, {Parente}, {Parra}, {Pech}, {Pedreira},
  {P{\`E}{\textcopyright}kala}, {Pelayo}, {Pe{\~n}a-Rodriguez}, {Perez Armand},
  {Perlin}, {Perrone}, {Peters}, {Petrera}, {Pierog}, {Pimenta}, {Pirronello},
  {Platino}, {Pont}, {Pothast}, {Privitera}, {Prouza}, {Puyleart},
  {Querchfeld}, {Rautenberg}, {Ravignani}, {Reininghaus}, {Ridky}, {Riehn},
  {Risse}, {Ristori}, {Rizi}, {Rodrigues de Carvalho}, {Rodriguez Fernandez},
  {Rodriguez Rojo}, {Roncoroni}, {Roth}, {Roulet}, {Rovero}, {Ruehl}, {Saffi},
  {Saftoiu}, {Salamida}, {Salazar}, {Salina}, {Sanabria Gomez}, {S{\'a}nchez},
  {Santos}, {Santos}, {Sarazin}, {Sarmento}, {Sarmiento-Cano}, {Sato},
  {Savina}, {Sch{\"a}fer}, {Scherini}, {Schieler}, {Schimassek}, {Schimp},
  {Schl{\"u}ter}, {Schmidt}, {Scholten}, {Schov{\'a}nek}, {Schr{\"o}der},
  {Schr{\"o}der}, {Schulz}, {Sciutto}, {Scornavacche}, {Shellard}, {Sigl},
  {Silli}, {Sima}, {{\v{S}}m{\'\i}da}, {Sommers}, {Soriano}, {Souchard},
  {Squartini}, {Stadelmaier}, {Stanca}, {Stani{\v{c}}}, {Stasielak}, {Stassi},
  {Streich}, {Su{\'a}rez-Dur{\'a}n}, {Sudholz}, {Suomij{\"a}rvi}, {Supanitsky},
  {{\v{S}}up{\'\i}k}, {Szadkowski}, {Taboada}, {Tapia}, {Timmermans},
  {Tkachenko}, {Tobiska}, {Todero Peixoto}, {Tom{\'e}}, {Torralba Elipe},
  {Travaini}, {Travnicek}, {Trimarelli}, {Trini}, {Tueros}, {Ulrich}, {Unger},
  {Urban}, {Vaclavek}, {Vacula}, {Vald{\'e}s Galicia}, {Vali{\~n}o}, {Valore},
  {van Vliet}, {Varela}, {Vargas C{\'a}rdenas}, {V{\'a}squez-Ram{\'\i}rez},
  {Veberi{\v{c}}}, {Ventura}, {Vergara Quispe}, {Verzi}, {Vicha},
  {Villase{\~n}or}, {Vink}, {Vorobiov}, {Wahlberg}, {Watson}, {Weber},
  {Weindl}, {Wiencke}, {Wilczy{\'n}ski}, {Winchen}, {Wirtz}, {Wittkowski},
  {Wundheiler}, {Yushkov}, {Zapparrata}, {Zas}, {Zavrtanik}, {Zavrtanik},
  {Zehrer}, {Zepeda}, {Ziolkowski}, {Zuccarello}, \& {Pierre Auger
  Collaboration}}]{Aab20}
{Aab}, A., {Abreu}, P., {Aglietta}, M., {et~al.} 2020, \prd, 102, 062005,
  \dodoi{10.1103/PhysRevD.102.062005}

\bibitem[{{Aartsen} {et~al.}(2014){Aartsen}, {Ackermann}, {Adams}, {Aguilar},
  {Ahlers}, {Ahrens}, {Altmann}, {Anderson}, {Arguelles}, {Arlen},
  {Auffenberg}, {Bai}, {Barwick}, {Baum}, {Beatty}, {Becker Tjus}, {Becker},
  {BenZvi}, {Berghaus}, {Berley}, {Bernardini}, {Bernhard}, {Besson}, {Binder},
  {Bindig}, {Bissok}, {Blaufuss}, {Blumenthal}, {Boersma}, {Bohm}, {Bose},
  {B{\"o}ser}, {Botner}, {Brayeur}, {Bretz}, {Brown}, {Casey}, {Casier},
  {Chirkin}, {Christov}, {Christy}, {Clark}, {Classen}, {Clevermann},
  {Coenders}, {Cowen}, {Cruz Silva}, {Danninger}, {Daughhetee}, {Davis}, {Day},
  {de Andr{\'e}}, {De Clercq}, {De Ridder}, {Desiati}, {de Vries}, {de With},
  {DeYoung}, {D{\'\i}az-V{\'e}lez}, {Dunkman}, {Eagan}, {Eberhardt},
  {Eichmann}, {Eisch}, {Euler}, {Evenson}, {Fadiran}, {Fazely}, {Fedynitch},
  {Feintzeig}, {Felde}, {Feusels}, {Filimonov}, {Finley}, {Fischer-Wasels},
  {Flis}, {Franckowiak}, {Frantzen}, {Fuchs}, {Gaisser}, {Gallagher},
  {Gerhardt}, {Gier}, {Gladstone}, {Gl{\"u}senkamp}, {Goldschmidt}, {Golup},
  {Gonzalez}, {Goodman}, {G{\'o}ra}, {Grandmont}, {Grant}, {Gretskov}, {Groh},
  {Gro{\ss}}, {Ha}, {Haack}, {Haj Ismail}, {Hallen}, {Hallgren}, {Halzen},
  {Hanson}, {Hebecker}, {Heereman}, {Heinen}, {Helbing}, {Hellauer}, {Hellwig},
  {Hickford}, {Hill}, {Hoffman}, {Hoffmann}, {Homeier}, {Hoshina}, {Huang},
  {Huelsnitz}, {Hulth}, {Hultqvist}, {Hussain}, {Ishihara}, {Jacobi},
  {Jacobsen}, {Jagielski}, {Japaridze}, {Jero}, {Jlelati}, {Jurkovic},
  {Kaminsky}, {Kappes}, {Karg}, {Karle}, {Kauer}, {Kelley}, {Kheirandish},
  {Kiryluk}, {Kl{\"a}s}, {Klein}, {K{\"o}hne}, {Kohnen}, {Kolanoski}, {Koob},
  {K{\"o}pke}, {Kopper}, {Kopper}, {Koskinen}, {Kowalski}, {Kriesten},
  {Krings}, {Kroll}, {Kunnen}, {Kurahashi}, {Kuwabara}, {Labare}, {Larsen},
  {Larson}, {Lesiak-Bzdak}, {Leuermann}, {Leute}, {L{\"u}nemann},
  {Mac{\'\i}as}, {Madsen}, {Maggi}, {Maruyama}, {Mase}, {Matis}, {McNally},
  {Meagher}, {Meli}, {Meures}, {Miarecki}, {Middell}, {Middlemas}, {Milke},
  {Miller}, {Mohrmann}, {Montaruli}, {Morse}, {Nahnhauer}, {Naumann},
  {Niederhausen}, {Nowicki}, {Nygren}, {Obertacke}, {Odrowski}, {Olivas},
  {Omairat}, {O'Murchadha}, {Palczewski}, {Paul}, {Penek}, {Pepper}, {P{\'e}rez
  de los Heros}, {Pfendner}, {Pieloth}, {Pinat}, {Posselt}, {Price},
  {Przybylski}, {P{\"u}tz}, {Quinnan}, {R{\"a}del}, {Rameez}, {Rawlins},
  {Redl}, {Rees}, {Reimann}, {Resconi}, {Rhode}, {Richman}, {Riedel},
  {Robertson}, {Rodrigues}, {Rongen}, {Rott}, {Ruhe}, {Ruzybayev}, {Ryckbosch},
  {Saba}, {Sander}, {Santander}, {Sarkar}, {Schatto}, {Scheriau}, {Schmidt},
  {Schmitz}, {Schoenen}, {Sch{\"o}neberg}, {Sch{\"o}nwald}, {Schukraft},
  {Schulte}, {Schulz}, {Seckel}, {Sestayo}, {Seunarine}, {Shanidze},
  {Sheremata}, {Smith}, {Soldin}, {Spiczak}, {Spiering}, {Stamatikos},
  {Stanev}, {Stanisha}, {Stasik}, {Stezelberger}, {Stokstad}, {St{\"o}{\ss}l},
  {Strahler}, {Str{\"o}m}, {Strotjohann}, {Sullivan}, {Taavola}, {Taboada},
  {Tamburro}, {Tepe}, {Ter-Antonyan}, {Terliuk}, {Te{\v{s}}i{\'c}}, {Tilav},
  {Toale}, {Tobin}, {Tosi}, {Tselengidou}, {Unger}, {Usner}, {Vallecorsa}, {van
  Eijndhoven}, {Vandenbroucke}, {van Santen}, {Vehring}, {Voge}, {Vraeghe},
  {Walck}, {Wallraff}, {Weaver}, {Wellons}, {Wendt}, {Westerhoff}, {Whelan},
  {Whitehorn}, {Wichary}, {Wiebe}, {Wiebusch}, {Williams}, {Wissing}, {Wolf},
  {Wood}, {Woschnagg}, {Xu}, {Xu}, {Yanez}, {Yodh}, {Yoshida}, {Zarzhitsky},
  {Ziemann}, {Zierke}, {Zoll}, \& {IceCube Collaboration}}]{IceCube2014PRL}
{Aartsen}, M.~G., {Ackermann}, M., {Adams}, J., {et~al.} 2014, \prl, 113,
  101101, \dodoi{10.1103/PhysRevLett.113.101101}

\bibitem[{{Abdo} {et~al.}(2010){Abdo}, {Ackermann}, {Agudo}, {Ajello}, {Aller},
  {Aller}, {Angelakis}, {Arkharov}, {Axelsson}, {Bach}, {Baldini}, {Ballet},
  {Barbiellini}, {Bastieri}, {Baughman}, {Bechtol}, {Bellazzini}, {Benitez},
  {Berdyugin}, {Berenji}, {Blandford}, {Bloom}, {Boettcher}, {Bonamente},
  {Borgland}, {Bregeon}, {Brez}, {Brigida}, {Bruel}, {Burnett}, {Burrows},
  {Buson}, {Caliandro}, {Calzoletti}, {Cameron}, {Capalbi}, {Caraveo},
  {Carosati}, {Casandjian}, {Cavazzuti}, {Cecchi}, {{\c{C}}elik}, {Charles},
  {Chaty}, {Chekhtman}, {Chen}, {Chiang}, {Chincarini}, {Ciprini}, {Claus},
  {Cohen-Tanugi}, {Colafrancesco}, {Cominsky}, {Conrad}, {Costamante},
  {Cutini}, {D'ammando}, {Deitrick}, {D'Elia}, {Dermer}, {de Angelis}, {de
  Palma}, {Digel}, {Donnarumma}, {Silva}, {Drell}, {Dubois}, {Dultzin},
  {Dumora}, {Falcone}, {Farnier}, {Favuzzi}, {Fegan}, {Focke}, {Forn{\'e}},
  {Fortin}, {Frailis}, {Fuhrmann}, {Fukazawa}, {Funk}, {Fusco}, {G{\'o}mez},
  {Gargano}, {Gasparrini}, {Gehrels}, {Germani}, {Giebels}, {Giglietto},
  {Giommi}, {Giordano}, {Giuliani}, {Glanzman}, {Godfrey}, {Grenier},
  {Gronwall}, {Grove}, {Guillemot}, {Guiriec}, {Gurwell}, {Hadasch},
  {Hanabata}, {Harding}, {Hayashida}, {Hays}, {Healey}, {Heidt}, {Hiriart},
  {Horan}, {Hoversten}, {Hughes}, {Itoh}, {Jackson}, {J{\'o}hannesson},
  {Johnson}, {Johnson}, {Jorstad}, {Kadler}, {Kamae}, {Katagiri}, {Kataoka},
  {Kawai}, {Kennea}, {Kerr}, {Kimeridze}, {Kn{\"o}dlseder}, {Kocian},
  {Kopatskaya}, {Koptelova}, {Konstantinova}, {Kovalev}, {Kovalev},
  {Kurtanidze}, {Kuss}, {Lande}, {Larionov}, {Latronico}, {Leto}, {Lindfors},
  {Longo}, {Loparco}, {Lott}, {Lovellette}, {Lubrano}, {Madejski}, {Makeev},
  {Marchegiani}, {Marscher}, {Marshall}, {Max-Moerbeck}, {Mazziotta},
  {McConville}, {McEnery}, {Meurer}, {Michelson}, {Mitthumsiri}, {Mizuno},
  {Moiseev}, {Monte}, {Monzani}, {Morselli}, {Moskalenko}, {Murgia},
  {Nestoras}, {Nilsson}, {Nizhelsky}, {Nolan}, {Norris}, {Nuss}, {Ohsugi},
  {Ojha}, {Omodei}, {Orlando}, {Ormes}, {Osborne}, {Ozaki}, {Pacciani},
  {Padovani}, {Pagani}, {Page}, {Paneque}, {Panetta}, {Parent}, {Pasanen},
  {Pavlidou}, {Pelassa}, {Pepe}, {Perri}, {Pesce-Rollins}, {Piranomonte},
  {Piron}, {Pittori}, {Porter}, {Puccetti}, {Rahoui}, {Rain{\`o}}, {Raiteri},
  {Rando}, {Razzano}, {Reimer}, {Reimer}, {Reposeur}, {Richards}, {Ritz},
  {Rochester}, {Rodriguez}, {Romani}, {Ros}, {Roth}, {Roustazadeh}, {Ryde},
  {Sadrozinski}, {Sadun}, {Sanchez}, {Sander}, {Saz Parkinson}, {Scargle},
  {Sellerholm}, {Sgr{\`o}}, {Shaw}, {Sigua}, {Siskind}, {Smith}, {Smith},
  {Spandre}, {Spinelli}, {Starck}, {Stevenson}, {Stratta}, {Strickman},
  {Suson}, {Tajima}, {Takahashi}, {Takahashi}, {Takalo}, {Tanaka}, {Thayer},
  {Thayer}, {Thompson}, {Tibaldo}, {Torres}, {Tosti}, {Tramacere}, {Uchiyama},
  {Usher}, {Vasileiou}, {Verrecchia}, {Vilchez}, {Villata}, {Vitale}, {Waite},
  {Wang}, {Winer}, {Wood}, {Ylinen}, {Zensus}, {Zhekanis}, \&
  {Ziegler}}]{Fermi10}
{Abdo}, A.~A., {Ackermann}, M., {Agudo}, I., {et~al.} 2010, \apj, 716, 30,
  \dodoi{10.1088/0004-637X/716/1/30}

\bibitem[{{Acciari} {et~al.}(2020){Acciari}, {Ansoldi}, {Antonelli}, {Arbet
  Engels}, {Baack}, {Babi{\'c}}, {Banerjee}, {Barres de Almeida}, {Barrio},
  {Becerra Gonz{\'a}lez}, {Bednarek}, {Bellizzi}, {Bernardini}, {Berti},
  {Besenrieder}, {Bhattacharyya}, {Bigongiari}, {Biland}, {Blanch}, {Bonnoli},
  {Bo{\v{s}}njak}, {Busetto}, {Carosi}, {Ceribella}, {Cerruti}, {Chai},
  {Chilingarian}, {Cikota}, {Colak}, {Colin}, {Colombo}, {Contreras},
  {Cortina}, {Covino}, {D'Elia}, {da Vela}, {Dazzi}, {de Angelis}, {de Lotto},
  {Del Puppo}, {Delfino}, {Delgado}, {Depaoli}, {di Pierro}, {di Venere}, {Do
  Souto Espi{\~n}eira}, {Prester}, {Donini}, {Dorner}, {Doro}, {Elsaesser},
  {Ramazani}, {Fattorini}, {Ferrara}, {Foffano}, {Fonseca}, {Font}, {Fruck},
  {Fukami}, {Garc{\'\i}a L{\'o}pez}, {Garczarczyk}, {Gasparyan}, {Gaug},
  {Giglietto}, {Giordano}, {Gliwny}, {Godinovi{\'c}}, {Green}, {Hadasch},
  {Hahn}, {Hassan}, {Herrera}, {Hoang}, {Hrupec}, {H{\"u}tten}, {Inada},
  {Inoue}, {Ishio}, {Iwamura}, {Jouvin}, {Kajiwara}, {Kerszberg}, {Kobayashi},
  {Kubo}, {Kushida}, {Lamastra}, {Lelas}, {Leone}, {Lindfors}, {Lombardi},
  {Longo}, {L{\'o}pez}, {L{\'o}pez-Coto}, {L{\'o}pez-Oramas}, {Loporchio},
  {Machado de Oliveira Fraga}, {Maggio}, {Majumdar}, {Makariev}, {Mallamaci},
  {Maneva}, {Manganaro}, {Mannheim}, {Maraschi}, {Mariotti}, {Mart{\'\i}nez},
  {Mazin}, {Mender}, {Mi{\'c}anovi{\'c}}, {Miceli}, {Miener}, {Minev},
  {Miranda}, {Mirzoyan}, {Molina}, {Moralejo}, {Morcuende}, {Moreno},
  {Moretti}, {Munar-Adrover}, {Neustroev}, {Nigro}, {Nilsson}, {Ninci},
  {Nishijima}, {Noda}, {Nogu{\'e}s}, {Nozaki}, {Ohtani}, {Oka}, {Otero-Santos},
  {Palatiello}, {Paneque}, {Paoletti}, {Paredes}, {Pavleti{\'c}}, {Pe{\~n}il},
  {Peresano}, {Persic}, {Moroni}, {Prandini}, {Puljak}, {Rhode}, {Rib{\'o}},
  {Rico}, {Righi}, {Rugliancich}, {Saha}, {Sahakyan}, {Saito}, {Sakurai},
  {Satalecka}, {Schleicher}, {Schmidt}, {Schweizer}, {Sitarek},
  {{\v{S}}nidari{\'c}}, {Sobczynska}, {Spolon}, {Stamerra}, {Strom}, {Strzys},
  {Suda}, {Suri{\'c}}, {Takahashi}, {Tavecchio}, {Temnikov}, {Terzi{\'c}},
  {Teshima}, {Torres-Alb{\`a}}, {Tosti}, {van Scherpenberg}, {Vanzo}, {Vazquez
  Acosta}, {Ventura}, {Verguilov}, {Vigorito}, {Vitale}, {Vovk}, {Will},
  {Zari{\'c}}, {MAGIC Collaboration}, {Finke}, {D'Ammando}, {Balokovi{\'c}},
  {Madejski}, {Mori}, {Puccetti}, {Leto}, {Perri}, {Verrecchia}, {Villata},
  {Raiteri}, {Agudo}, {Bachev}, {Berdyugin}, {Blinov}, {Chanishvili}, {Chen},
  {Chigladze}, {Damljanovic}, {Eswaraiah}, {Grishina}, {Ibryamov}, {Jordan},
  {Jorstad}, {Joshi}, {Kopatskaya}, {Kurtanidze}, {Kurtanidze}, {Larionova},
  {Larionova}, {Larionov}, {Latev}, {Lin}, {Marscher}, {Mokrushina},
  {Morozova}, {Nikolashvili}, {Semkov}, {Smith}, {Strigachev}, {Troitskaya},
  {Troitsky}, {Vince}, {Barnes}, {G{\"u}ver}, {Moody}, {Sadun}, {Hovatta},
  {Richards}, {Max-Moerbeck}, {Readhead}, {L{\"a}hteenm{\"a}ki}, {Tornikoski},
  {Tammi}, {Ramakrishnan}, \& {Reinthal}}]{2020ApJS..248...29A}
{Acciari}, V.~A., {Ansoldi}, S., {Antonelli}, L.~A., {et~al.} 2020, \apjs, 248,
  29, \dodoi{10.3847/1538-4365/ab89b5}

\bibitem[{{Acciari} {et~al.}(2022){Acciari}, {Aniello}, {Ansoldi}, {Antonelli},
  {Arbet Engels}, {Artero}, {Asano}, {Baack}, {Babi{\'c}}, {Baquero}, {Barres
  de Almeida}, {Barrio}, {Batkovi{\'c}}, {Becerra Gonz{\'a}lez}, {Bednarek},
  {Bernardini}, {Bernardos}, {Berti}, {Besenrieder}, {Bhattacharyya},
  {Bigongiari}, {Biland}, {Blanch}, {B{\"o}kenkamp}, {Bonnoli},
  {Bo{\v{s}}njak}, {Busetto}, {Carosi}, {Ceribella}, {Cerruti}, {Chai},
  {Chilingarian}, {Cikota}, {Colombo}, {Contreras}, {Cortina}, {Covino},
  {D'Amico}, {D'Elia}, {Vela}, {Dazzi}, {De Angelis}, {De Lotto}, {Del Popolo},
  {Delfino}, {Delgado}, {Mendez}, {Depaoli}, {Di Pierro}, {Di Venere}, {Do
  Souto Espi{\~n}eira}, {Dominis Prester}, {Donini}, {Dorner}, {Doro},
  {Elsaesser}, {Fallah Ramazani}, {Fari{\~n}a}, {Fattorini}, {Font}, {Fruck},
  {Fukami}, {Fukazawa}, {Garc{\'\i}a L{\'o}pez}, {Garczarczyk}, {Gasparyan},
  {Gaug}, {Giglietto}, {Giordano}, {Gliwny}, {Godinovi{\'c}}, {Green}, {Green},
  {Hadasch}, {Hahn}, {Hassan}, {Heckmann}, {Herrera}, {Hoang}, {Hrupec},
  {H{\"u}tten}, {Inada}, {Iotov}, {Ishio}, {Iwamura}, {Jim{\'e}nez
  Mart{\'\i}nez}, {Jormanainen}, {Jouvin}, {Kerszberg}, {Kobayashi}, {Kubo},
  {Kushida}, {Lamastra}, {Lelas}, {Leone}, {Lindfors}, {Linhoff}, {Lombardi},
  {Longo}, {L{\'o}pez-Coto}, {L{\'o}pez-Moya}, {L{\'o}pez-Oramas}, {Loporchio},
  {Machado de Oliveira Fraga}, {Maggio}, {Majumdar}, {Makariev}, {Mallamaci},
  {Maneva}, {Manganaro}, {Mannheim}, {Mariotti}, {Mart{\'\i}nez}, {Mas
  Aguilar}, {Mazin}, {Menchiari}, {Mender}, {Mi{\'c}anovi{\'c}}, {Miceli},
  {Miener}, {Miranda}, {Mirzoyan}, {Molina}, {Moralejo}, {Morcuende}, {Moreno},
  {Moretti}, {Nakamori}, {Nava}, {Neustroev}, {Nievas Rosillo}, {Nigro},
  {Nilsson}, {Nishijima}, {Noda}, {Nozaki}, {Ohtani}, {Oka}, {Otero-Santos},
  {Paiano}, {Palatiello}, {Paneque}, {Paoletti}, {Paredes}, {Pavleti{\'c}},
  {Pe{\~n}il}, {Persic}, {Pihet}, {Prada Moroni}, {Prandini}, {Priyadarshi},
  {Puljak}, {Rhode}, {Rib{\'o}}, {Rico}, {Righi}, {Rugliancich}, {Sahakyan},
  {Saito}, {Sakurai}, {Satalecka}, {Saturni}, {Schleicher}, {Schmidt},
  {Schmuckermaier}, {Schweizer}, {Sitarek}, {{\v{S}}nidari{\'c}}, {Sobczynska},
  {Spolon}, {Stamerra}, {Stri{\v{s}}kovi{\'c}}, {Strom}, {Strzys}, {Suda},
  {Suri{\'c}}, {Takahashi}, {Takeishi}, {Tavecchio}, {Temnikov}, {Terzi{\'c}},
  {Teshima}, {Tosti}, {Truzzi}, {Tutone}, {Ubach}, {van Scherpenberg}, {Vanzo},
  {Vazquez Acosta}, {Ventura}, {Verguilov}, {Viale}, {Vigorito}, {Vitale},
  {Vovk}, {Will}, {Wunderlich}, {Yamamoto}, {Zari{\'c}}, {Hodges}, {Hovatta},
  {Kiehlmann}, {Liodakis}, {Max-Moerbeck}, {Pearson}, {Readhead}, {Reeves},
  {L{\"a}hteenm{\"a}ki}, {Tornikoski}, {Tammi}, {D'Ammando}, \&
  {Marchini}}]{Acciari22}
{Acciari}, V.~A., {Aniello}, T., {Ansoldi}, S., {et~al.} 2022, \apj, 927, 197,
  \dodoi{10.3847/1538-4357/ac531d}

\bibitem[{{Acharyya} {et~al.}(2023){Acharyya}, {Adams}, {Archer}, {Bangale},
  {Bartkoske}, {Batista}, {Benbow}, {Brill}, {Buckley}, {Christiansen},
  {Chromey}, {Errando}, {Falcone}, {Feng}, {Foote}, {Fortson}, {Furniss},
  {Gallagher}, {Hanlon}, {Hanna}, {Hervet}, {Hinrichs}, {Hoang}, {Holder},
  {Humensky}, {Jin}, {Kaaret}, {Kertzman}, {Kherlakian}, {Kieda}, {Kleiner},
  {Korzoun}, {Kumar}, {Lang}, {Lundy}, {Maier}, {McGrath}, {Millard}, {Millis},
  {Mooney}, {Moriarty}, {Mukherjee}, {O'Brien}, {Ong}, {Pohl}, {Pueschel},
  {Quinn}, {Ragan}, {Reynolds}, {Ribeiro}, {Roache}, {Sadeh}, {Sadun}, {Saha},
  {Santander}, {Sembroski}, {Shang}, {Splettstoesser}, {Talluri}, {Tucci},
  {Vassiliev}, {Weinstein}, {Williams}, {Wong}, {Woo}, {Aharonian},
  {Aschersleben}, {Backes}, {Martins}, {Batzofin}, {Becherini}, {Berge},
  {Bernl{\"o}hr}, {Bi}, {B{\"o}ttcher}, {Boisson}, {Bolmont}, {de Bony de
  Lavergne}, {Borowska}, {Bouyahiaoui}, {Bradascio}, {Breuhaus}, {Brose},
  {Brun}, {Bruno}, {Bulik}, {Burger-Scheidlin}, {Caroff}, {Casanova}, {Cecil},
  {Celic}, {Cerruti}, {Chand}, {Chandra}, {Chen}, {Chibueze}, {Chibueze},
  {Cotter}, {Dai}, {Mbarubucyeye}, {Djannati-Ata{\"\i}}, {Dmytriiev},
  {Doroshenko}, {Einecke}, {Ernenwein}, {de Clairfontaine}, {Filipovic},
  {Fontaine}, {F{\"u}{\ss}ling}, {Funk}, {Gabici}, {Ghafourizadeh}, {Giavitto},
  {Glawion}, {Glicenstein}, {Goswami}, {Grolleron}, {Haerer}, {Hinton},
  {Holch}, {Holler}, {Horns}, {Jamrozy}, {Jankowsky}, {Joshi}, {Jung-Richardt},
  {Kasai}, {Katarzy{\'n}ski}, {Khatoon}, {Kh{\'e}lifi}, {Klepser},
  {Klu{\'z}niak}, {Kosack}, {Kostunin}, {Lang}, {Le Stum}, {Lemi{\`e}re},
  {Lenain}, {Leuschner}, {Lohse}, {Luashvili}, {Lypova}, {Mackey}, {Malyshev},
  {Marandon}, {Marchegiani}, {Marcowith}, {Mart{\'\i}-Devesa}, {Marx},
  {Mitchell}, {Moderski}, {Mohrmann}, {Montanari}, {Moulin}, {Murach},
  {Nakashima}, {Niemiec}, {Noel}, {O'Brien}, {Olivera-Nieto}, {de Ona
  Wilhelmi}, {Ostrowski}, {Panny}, {Panter}, {Peron}, {Prokhorov},
  {P{\"u}hlhofer}, {Punch}, {Quirrenbach}, {Reichherzer}, {Reimer}, {Reimer},
  {Ren}, {Renaud}, {Rieger}, {Rudak}, {Ruiz-Velasco}, {Sahakian}, {Santangelo},
  {Sasaki}, {Sch{\"a}fer}, {Sch{\"u}ssler}, {Schutte}, {Schwanke}, {Shapopi},
  {Specovius}, {Spencer}, {Stawarz}, {Steenkamp}, {Steinmassl}, {Sushch},
  {Suzuki}, {Takahashi}, {Tanaka}, {Terrier}, {van Eldik}, {Vecchi}, {Veh},
  {Venter}, {Vink}, {White}, {Wierzcholska}, {Wong}, {Zacharias}, {Zargaryan},
  {Zdziarski}, {Zech}, {Zouari}, {{\.Z}ywucka}, {Mori}, \& {H.~E.~S.~S.
  Collaboration}}]{Acharyya23}
{Acharyya}, A., {Adams}, C.~B., {Archer}, A., {et~al.} 2023, \apj, 954, 70,
  \dodoi{10.3847/1538-4357/ace327}

\bibitem[{{Aguilar} {et~al.}(2021){Aguilar}, {Ali Cavasonza}, {Ambrosi},
  {Arruda}, {Attig}, {Barao}, {Barrin}, {Bartoloni}, {Ba{\c{s}}e{\u{g}}mez-du
  Pree}, {Bates}, {Battiston}, {Behlmann}, {Beischer}, {Berdugo}, {Bertucci},
  {Bindi}, {de Boer}, {Bollweg}, {Borgia}, {Boschini}, {Bourquin}, {Bueno},
  {Burger}, {Burger}, {Burmeister}, {Cai}, {Capell}, {Casaus}, {Castellini},
  {Cervelli}, {Chang}, {Chen}, {Chen}, {Chen}, {Cheng}, {Chou}, {Chouridou},
  {Choutko}, {Chung}, {Clark}, {Coignet}, {Consolandi}, {Contin}, {Corti},
  {Cui}, {Dadzie}, {Dai}, {Delgado}, {Della Torre}, {Demirk{\"o}z}, {Derome},
  {Di Falco}, {Di Felice}, {D{\'\i}az}, {Dimiccoli}, {von Doetinchem}, {Dong},
  {Donnini}, {Duranti}, {Egorov}, {Eline}, {Feng}, {Fiandrini}, {Fisher},
  {Formato}, {Freeman}, {Galaktionov}, {G{\'a}mez}, {Garc{\'\i}a-L{\'o}pez},
  {Gargiulo}, {Gast}, {Gebauer}, {Gervasi}, {Giovacchini}, {G{\'o}mez-Coral},
  {Gong}, {Goy}, {Grabski}, {Grandi}, {Graziani}, {Guo}, {Haino}, {Han},
  {Hashmani}, {He}, {Heber}, {Hsieh}, {Hu}, {Huang}, {Hungerford}, {Incagli},
  {Jang}, {Jia}, {Jinchi}, {Kanishev}, {Khiali}, {Kim}, {Kirn}, {Konyushikhin},
  {Kounina}, {Kounine}, {Koutsenko}, {Kuhlman}, {Kulemzin}, {La Vacca},
  {Laudi}, {Laurenti}, {Lazzizzera}, {Lebedev}, {Lee}, {Lee}, {Leluc}, {Li},
  {Li}, {Li}, {Li}, {Li}, {Li}, {Light}, {Lin}, {Lippert}, {Liu}, {Lu}, {Lu},
  {Luebelsmeyer}, {Luo}, {Lyu}, {Machate}, {Ma{\~n}{\'a}}, {Mar{\'\i}n},
  {Marquardt}, {Martin}, {Mart{\'\i}nez}, {Masi}, {Maurin}, {Menchaca-Rocha},
  {Meng}, {Mo}, {Molero}, {Mott}, {Mussolin}, {Ni}, {Nikonov}, {Nozzoli},
  {Oliva}, {Orcinha}, {Palermo}, {Palmonari}, {Paniccia}, {Pashnin},
  {Pauluzzi}, {Pensotti}, {Phan}, {Plyaskin}, {Pohl}, {Porter}, {Qi}, {Qin},
  {Qu}, {Quadrani}, {Rancoita}, {Rapin}, {Reina Conde}, {Rosier-Lees},
  {Rozhkov}, {Rozza}, {Sagdeev}, {Schael}, {Schmidt}, {Schulz von Dratzig},
  {Schwering}, {Seo}, {Shan}, {Shi}, {Siedenburg}, {Solano}, {Song},
  {Sonnabend}, {Sun}, {Sun}, {Tacconi}, {Tang}, {Tang}, {Tian}, {Ting}, {Ting},
  {Tomassetti}, {Torsti}, {T{\"u}ys{\"u}z}, {Urban}, {Usoskin}, {Vagelli},
  {Vainio}, {Valente}, {Valtonen}, {V{\'a}zquez Acosta}, {Vecchi}, {Velasco},
  {Vialle}, {Wang}, {Wang}, {Wang}, {Wang}, {Wang}, {Wang}, {Wei}, {Weng},
  {Wu}, {Xiong}, {Xu}, {Yan}, {Yang}, {Yi}, {Yu}, {Yu}, {Zannoni}, {Zhang},
  {Zhang}, {Zhang}, {Zhang}, {Zhang}, {Zhao}, {Zheng}, {Zhuang}, {Zhukov},
  {Zichichi}, {Zimmermann}, {Zuccon}, \& {AMS Collaboration}}]{AMS21}
{Aguilar}, M., {Ali Cavasonza}, L., {Ambrosi}, G., {et~al.} 2021, \physrep,
  894, 1, \dodoi{10.1016/j.physrep.2020.09.003}

\bibitem[{{Aharonian} {et~al.}(2007){Aharonian}, {Akhperjanian}, {Bazer-Bachi},
  {Behera}, {Beilicke}, {Benbow}, {Berge}, {Bernl{\"o}hr}, {Boisson}, {Bolz},
  {Borrel}, {Boutelier}, {Braun}, {Brion}, {Brown}, {B{\"u}hler},
  {B{\"u}sching}, {Bulik}, {Carrigan}, {Chadwick}, {Clapson}, {Chounet},
  {Coignet}, {Cornils}, {Costamante}, {Degrange}, {Dickinson},
  {Djannati-Ata{\"\i}}, {Domainko}, {Drury}, {Dubus}, {Dyks}, {Egberts},
  {Emmanoulopoulos}, {Espigat}, {Farnier}, {Feinstein}, {Fiasson},
  {F{\"o}rster}, {Fontaine}, {Funk}, {Funk}, {F{\"u}{\ss}ling}, {Gallant},
  {Giebels}, {Glicenstein}, {Gl{\"u}ck}, {Goret}, {Hadjichristidis}, {Hauser},
  {Hauser}, {Heinzelmann}, {Henri}, {Hermann}, {Hinton}, {Hoffmann}, {Hofmann},
  {Holleran}, {Hoppe}, {Horns}, {Jacholkowska}, {de Jager}, {Kendziorra},
  {Kerschhaggl}, {Kh{\'e}lifi}, {Komin}, {Kosack}, {Lamanna}, {Latham}, {Le
  Gallou}, {Lemi{\`e}re}, {Lemoine-Goumard}, {Lenain}, {Lohse}, {Martin},
  {Martineau-Huynh}, {Marcowith}, {Masterson}, {Maurin}, {McComb}, {Moderski},
  {Moulin}, {de Naurois}, {Nedbal}, {Nolan}, {Olive}, {Orford}, {Osborne},
  {Ostrowski}, {Panter}, {Pedaletti}, {Pelletier}, {Petrucci}, {Pita},
  {P{\"u}hlhofer}, {Punch}, {Ranchon}, {Raubenheimer}, {Raue}, {Rayner},
  {Renaud}, {Ripken}, {Rob}, {Rolland}, {Rosier-Lees}, {Rowell}, {Rudak},
  {Ruppel}, {Sahakian}, {Santangelo}, {Saug{\'e}}, {Schlenker}, {Schlickeiser},
  {Schr{\"o}der}, {Schwanke}, {Schwarzburg}, {Schwemmer}, {Shalchi}, {Sol},
  {Spangler}, {Stawarz}, {Steenkamp}, {Stegmann}, {Superina}, {Tam},
  {Tavernet}, {Terrier}, {van Eldik}, {Vasileiadis}, {Venter}, {Vialle},
  {Vincent}, {Vivier}, {V{\"o}lk}, {Volpe}, {Wagner}, {Ward}, \&
  {Zdziarski}}]{HESS2155}
{Aharonian}, F., {Akhperjanian}, A.~G., {Bazer-Bachi}, A.~R., {et~al.} 2007,
  \apjl, 664, L71, \dodoi{10.1086/520635}

\bibitem[{{Aharonian}(2000)}]{Aharonian00}
{Aharonian}, F.~A. 2000, \na, 5, 377, \dodoi{10.1016/S1384-1076(00)00039-7}

\bibitem[{{Aharonian} {et~al.}(1983){Aharonian}, {Atoian}, \&
  {Nagapetian}}]{Aha83}
{Aharonian}, F.~A., {Atoian}, A.~M., \& {Nagapetian}, A.~M. 1983, Astrofizika,
  19, 323

\bibitem[{{Atoyan} \& {Dermer}(2003)}]{Atoyan03}
{Atoyan}, A.~M., \& {Dermer}, C.~D. 2003, \apj, 586, 79, \dodoi{10.1086/346261}

\bibitem[{{Balokovi{\'c}} {et~al.}(2016){Balokovi{\'c}}, {Paneque}, {Madejski},
  {Furniss}, {Chiang}, {Ajello}, {Alexander}, {Barret}, {Blandford}, {Boggs},
  {Christensen}, {Craig}, {Forster}, {Giommi}, {Grefenstette}, {Hailey},
  {Harrison}, {Hornstrup}, {Kitaguchi}, {Koglin}, {Madsen}, {Mao}, {Miyasaka},
  {Mori}, {Perri}, {Pivovaroff}, {Puccetti}, {Rana}, {Stern}, {Tagliaferri},
  {Urry}, {Westergaard}, {Zhang}, {Zoglauer}, {NuSTAR Team}, {Archambault},
  {Archer}, {Barnacka}, {Benbow}, {Bird}, {Buckley}, {Bugaev}, {Cerruti},
  {Chen}, {Ciupik}, {Connolly}, {Cui}, {Dickinson}, {Dumm}, {Eisch}, {Falcone},
  {Feng}, {Finley}, {Fleischhack}, {Fortson}, {Griffin}, {Griffiths}, {Grube},
  {Gyuk}, {Huetten}, {H{\r{a}}kansson}, {Holder}, {Humensky}, {Johnson},
  {Kaaret}, {Kertzman}, {Khassen}, {Kieda}, {Krause}, {Krennrich}, {Lang},
  {Maier}, {McArthur}, {Meagher}, {Moriarty}, {Nelson}, {Nieto}, {Ong}, {Park},
  {Pohl}, {Popkow}, {Pueschel}, {Reynolds}, {Richards}, {Roache}, {Santander},
  {Sembroski}, {Shahinyan}, {Smith}, {Staszak}, {Telezhinsky}, {Todd}, {Tucci},
  {Tyler}, {Vincent}, {Weinstein}, {Wilhelm}, {Williams}, {Zitzer}, {VERITAS
  Collaboration}, {Ahnen}, {Ansoldi}, {Antonelli}, {Antoranz}, {Babic},
  {Banerjee}, {Bangale}, {Barres de Almeida}, {Barrio}, {Becerra Gonz{\'a}lez},
  {Bednarek}, {Bernardini}, {Biasuzzi}, {Biland}, {Blanch}, {Bonnefoy},
  {Bonnoli}, {Borracci}, {Bretz}, {Carmona}, {Carosi}, {Chatterjee}, {Clavero},
  {Colin}, {Colombo}, {Contreras}, {Cortina}, {Covino}, {Da Vela}, {Dazzi}, {De
  Angelis}, {De Lotto}, {de O{\~n}a Wilhelmi}, {Delgado Mendez}, {Di Pierro},
  {Dominis Prester}, {Dorner}, {Doro}, {Einecke}, {Elsaesser},
  {Fern{\'a}ndez-Barral}, {Fidalgo}, {Fonseca}, {Font}, {Frantzen}, {Fruck},
  {Galindo}, {Garc{\'\i}a L{\'o}pez}, {Garczarczyk}, {Garrido Terrats}, {Gaug},
  {Giammaria}, {Glawion (Eisenacher}, {Godinovi{\'c}}, {Gonz{\'a}lez
  Mu{\~n}oz}, {Guberman}, {Hahn}, {Hanabata}, {Hayashida}, {Herrera}, {Hose},
  {Hrupec}, {Hughes}, {Idec}, {Kodani}, {Konno}, {Kubo}, {Kushida}, {La
  Barbera}, {Lelas}, {Lindfors}, {Lombardi}, {Longo}, {L{\'o}pez},
  {L{\'o}pez-Coto}, {L{\'o}pez-Oramas}, {Lorenz}, {Majumdar}, {Makariev},
  {Mallot}, {Maneva}, {Manganaro}, {Mannheim}, {Maraschi}, {Marcote},
  {Mariotti}, {Mart{\'\i}nez}, {Mazin}, {Menzel}, {Miranda}, {Mirzoyan},
  {Moralejo}, {Moretti}, {Nakajima}, {Neustroev}, {Niedzwiecki}, {Nievas
  Rosillo}, {Nilsson}, {Nishijima}, {Noda}, {Orito}, {Overkemping}, {Paiano},
  {Palacio}, {Palatiello}, {Paoletti}, {Paredes}, {Paredes-Fortuny}, {Persic},
  {Poutanen}, {Prada Moroni}, {Prandini}, {Puljak}, {Rhode}, {Rib{\'o}},
  {Rico}, {Rodriguez Garcia}, {Saito}, {Satalecka}, {Scapin}, {Schultz},
  {Schweizer}, {Shore}, {Sillanp{\"a}{\"a}}, {Sitarek}, {Snidaric},
  {Sobczynska}, {Stamerra}, {Steinbring}, {Strzys}, {Takalo}, {Takami},
  {Tavecchio}, {Temnikov}, {Terzi{\'c}}, {Tescaro}, {Teshima}, {Thaele},
  {Torres}, {Toyama}, {Treves}, {Verguilov}, {Vovk}, {Ward}, {Will}, {Wu},
  {Zanin}, {MAGIC Collaboration}, {Perkins}, {Verrecchia}, {Leto},
  {B{\"o}ttcher}, {Villata}, {Raiteri}, {Acosta-Pulido}, {Bachev}, {Berdyugin},
  {Blinov}, {Carnerero}, {Chen}, {Chinchilla}, {Damljanovic}, {Eswaraiah},
  {Grishina}, {Ibryamov}, {Jordan}, {Jorstad}, {Joshi}, {Kopatskaya},
  {Kurtanidze}, {Kurtanidze}, {Larionova}, {Larionova}, {Larionov}, {Latev},
  {Lin}, {Marscher}, {Mokrushina}, {Morozova}, {Nikolashvili}, {Semkov},
  {Smith}, {Strigachev}, {Troitskaya}, {Troitsky}, {Vince}, {Barnes},
  {G{\"u}ver}, {Moody}, {Sadun}, {Sun}, {Hovatta}, {Richards}, {Max-Moerbeck},
  {Readhead}, {L{\"a}hteenm{\"a}ki}, {Tornikoski}, {Tammi}, {Ramakrishnan},
  {Reinthal}, {Angelakis}, {Fuhrmann}, {Myserlis}, {Karamanavis}, {Sievers},
  {Ungerechts}, \& {Zensus}}]{Balokovic16}
{Balokovi{\'c}}, M., {Paneque}, D., {Madejski}, G., {et~al.} 2016, \apj, 819,
  156, \dodoi{10.3847/0004-637X/819/2/156}

\bibitem[{{Banerjee} {et~al.}(2024){Banerjee}, {Macera}, {Ludovico De Santis},
  {Mei}, {Tissino}, {Oganesyan}, {Frederiks}, {Lysenko}, {Svinkin},
  {Tsvetkova}, \& {Branchesi}}]{2024arXiv240515855B}
{Banerjee}, B., {Macera}, S., {Ludovico De Santis}, A., {et~al.} 2024, arXiv
  e-prints, arXiv:2405.15855, \dodoi{10.48550/arXiv.2405.15855}

\bibitem[{{Baring} {et~al.}(1999){Baring}, {Ellison}, {Reynolds}, {Grenier}, \&
  {Goret}}]{Baring99}
{Baring}, M.~G., {Ellison}, D.~C., {Reynolds}, S.~P., {Grenier}, I.~A., \&
  {Goret}, P. 1999, \apj, 513, 311, \dodoi{10.1086/306829}

\bibitem[{Biehl {et~al.}(2018)Biehl, Boncioli, Fedynitch, \&
  Winter}]{Biehl:2017zlw}
Biehl, D., Boncioli, D., Fedynitch, A., \& Winter, W. 2018, Astron. Astrophys.,
  611, A101, \dodoi{10.1051/0004-6361/201731337}

\bibitem[{{Blumenthal} \& {Gould}(1970)}]{BG70}
{Blumenthal}, G.~R., \& {Gould}, R.~J. 1970, Reviews of Modern Physics, 42,
  237, \dodoi{10.1103/RevModPhys.42.237}

\bibitem[{{Boettcher} {et~al.}(1997){Boettcher}, {Mause}, \&
  {Schlickeiser}}]{Boettcher_1997}
{Boettcher}, M., {Mause}, H., \& {Schlickeiser}, R. 1997, \aap, 324, 395.
\newblock \doarXiv{astro-ph/9604003}

\bibitem[{{B{\"o}ttcher}(2005)}]{Boettcher05}
{B{\"o}ttcher}, M. 2005, \apj, 621, 176, \dodoi{10.1086/427430}

\bibitem[{{B{\"o}ttcher} \& {Dermer}(1998)}]{Bottcher98}
{B{\"o}ttcher}, M., \& {Dermer}, C.~D. 1998, \apjl, 499, L131,
  \dodoi{10.1086/311366}

\bibitem[{{B{\"o}ttcher} {et~al.}(2013){B{\"o}ttcher}, {Reimer}, {Sweeney}, \&
  {Prakash}}]{Boettcher_2013}
{B{\"o}ttcher}, M., {Reimer}, A., {Sweeney}, K., \& {Prakash}, A. 2013, \apj,
  768, 54, \dodoi{10.1088/0004-637X/768/1/54}

\bibitem[{{Brainerd}(1987)}]{Brainerd_1987}
{Brainerd}, J.~J. 1987, \apj, 320, 714, \dodoi{10.1086/165589}

\bibitem[{{Cerruti}(2020)}]{Cerruti20}
{Cerruti}, M. 2020, Galaxies, 8, 72, \dodoi{10.3390/galaxies8040072}

\bibitem[{{Cerruti} {et~al.}(2017){Cerruti}, {Benbow}, {Chen}, {Dumm},
  {Fortson}, \& {Shahinyan}}]{Cerruti17}
{Cerruti}, M., {Benbow}, W., {Chen}, X., {et~al.} 2017, \aap, 606, A68,
  \dodoi{10.1051/0004-6361/201730799}

\bibitem[{{Cerruti} {et~al.}(2019){Cerruti}, {Zech}, {Boisson}, {Emery},
  {Inoue}, \& {Lenain}}]{Cerruti19}
{Cerruti}, M., {Zech}, A., {Boisson}, C., {et~al.} 2019, \mnras, 483, L12,
  \dodoi{10.1093/mnrasl/sly210}

\bibitem[{{Cerruti} {et~al.}(2015){Cerruti}, {Zech}, {Boisson}, \&
  {Inoue}}]{Cerruti15}
{Cerruti}, M., {Zech}, A., {Boisson}, C., \& {Inoue}, S. 2015, \mnras, 448,
  910, \dodoi{10.1093/mnras/stu2691}

\bibitem[{Chang \& Cooper(1970)}]{chang1970a}
Chang, T., \& Cooper, G. 1970, Journal of Computational Physics, 6, 1,
  \dodoi{10.1016/0021-9991(70)90009-7}

\bibitem[{{Chatzis} {et~al.}(2024){Chatzis}, {Stathopoulos}, {Petropoulou}, \&
  {Vasilopoulos}}]{2024Univ...10..392C}
{Chatzis}, M., {Stathopoulos}, S.~I., {Petropoulou}, M., \& {Vasilopoulos}, G.
  2024, Universe, 10, 392, \dodoi{10.3390/universe10100392}

\bibitem[{{Cherenkov Telescope Array Consortium} {et~al.}(2019){Cherenkov
  Telescope Array Consortium}, {Acharya}, {Agudo}, {Al Samarai}, {Alfaro},
  {Alfaro}, {Alispach}, {Alves Batista}, {Amans}, {Amato}, {Ambrosi},
  {Antolini}, {Antonelli}, {Aramo}, {Araya}, {Armstrong}, {Arqueros},
  {Arrabito}, {Asano}, {Ashley}, {Backes}, {Balazs}, {Balbo}, {Ballester},
  {Ballet}, {Bamba}, {Barkov}, {Barres de Almeida}, {Barrio}, {Bastieri},
  {Becherini}, {Belfiore}, {Benbow}, {Berge}, {Bernardini}, {Bernardini},
  {Bernardos}, {Bernl{\"o}hr}, {Bertucci}, {Biasuzzi}, {Bigongiari}, {Biland},
  {Bissaldi}, {Biteau}, {Blanch}, {Blazek}, {Boisson}, {Bolmont}, {Bonanno},
  {Bonardi}, {Bonavolont{\`a}}, {Bonnoli}, {Bosnjak}, {B{\"o}ttcher},
  {Braiding}, {Bregeon}, {Brill}, {Brown}, {Brun}, {Brunetti}, {Buanes},
  {Buckley}, {Bugaev}, {B{\"u}hler}, {Bulgarelli}, {Bulik}, {Burton},
  {Burtovoi}, {Busetto}, {Canestrari}, {Capalbi}, {Capitanio}, {Caproni},
  {Caraveo}, {C{\'a}rdenas}, {Carlile}, {Carosi}, {Carqu{\'\i}n}, {Carr},
  {Casanova}, {Cascone}, {Catalani}, {Catalano}, {Cauz}, {Cerruti}, {Chadwick},
  {Chaty}, {Chaves}, {Chen}, {Chen}, {Chernyakova}, {Chikawa}, {Christov},
  {Chudoba}, {Cie{\'s}lar}, {Coco}, {Colafrancesco}, {Colin}, {Conforti},
  {Connaughton}, {Conrad}, {Contreras}, {Cortina}, {Costa}, {Costantini},
  {Cotter}, {Covino}, {Crocker}, {Cuadra}, {Cuevas}, {Cumani}, {D'A{\`\i}},
  {D'Ammando}, {D'Avanzo}, {D'Urso}, {Daniel}, {Davids}, {Dawson}, {Dazzi}, {De
  Angelis}, {de C{\'a}ssia dos Anjos}, {De Cesare}, {De Franco}, {de Gouveia
  Dal Pino}, {de la Calle}, {de los Reyes Lopez}, {De Lotto}, {De Luca}, {De
  Lucia}, {de Naurois}, {de O{\~n}a Wilhelmi}, {De Palma}, {De Persio}, {de
  Souza}, {Deil}, {Del Santo}, {Delgado}, {della Volpe}, {Di Girolamo}, {Di
  Pierro}, {Di Venere}, {D{\'\i}az}, {Dib}, {Diebold}, {Djannati-Ata{\"\i}},
  {Dom{\'\i}nguez}, {Dominis Prester}, {Dorner}, {Doro}, {Drass}, {Dravins},
  {Dubus}, {Dwarkadas}, {Ebr}, {Eckner}, {Egberts}, {Einecke}, {Ekoume},
  {Els{\"a}sser}, {Ernenwein}, {Espinoza}, {Evoli}, {Fairbairn},
  {Falceta-Goncalves}, {Falcone}, {Farnier}, {Fasola}, {Fedorova}, {Fegan},
  {Fernandez-Alonso}, {Fern{\'a}ndez-Barral}, {Ferrand}, {Fesquet},
  {Filipovic}, {Fioretti}, {Fontaine}, {Fornasa}, {Fortson}, {Freixas
  Coromina}, {Fruck}, {Fujita}, {Fukazawa}, {Funk}, {F{\"u}{\ss}ling},
  {Gabici}, {Gadola}, {Gallant}, {Garcia}, {Garcia L{\'o}pez}, {Garczarczyk},
  {Gaskins}, {Gasparetto}, {Gaug}, {Gerard}, {Giavitto}, {Giglietto}, {Giommi},
  {Giordano}, {Giro}, {Giroletti}, {Giuliani}, {Glicenstein}, {Gnatyk},
  {Godinovic}, {Goldoni}, {G{\'o}mez-Vargas}, {Gonz{\'a}lez}, {Gonz{\'a}lez},
  {G{\"o}tz}, {Graham}, {Grandi}, {Granot}, {Green}, {Greenshaw}, {Griffiths},
  {Gunji}, {Hadasch}, {Hara}, {Hardcastle}, {Hassan}, {Hayashi}, {Hayashida},
  {Heller}, {Helo}, {Hermann}, {Hinton}, {Hnatyk}, {Hofmann}, {Holder},
  {Horan}, {H{\"o}randel}, {Horns}, {Horvath}, {Hovatta}, {Hrabovsky},
  {Hrupec}, {Humensky}, {H{\"u}tten}, {Iarlori}, {Inada}, {Inome}, {Inoue},
  {Inoue}, {Inoue}, {Iocco}, {Ioka}, {Iori}, {Ishio}, {Iwamura}, {Jamrozy},
  {Janecek}, {Jankowsky}, {Jean}, {Jung-Richardt}, {Jurysek}, {Kaaret},
  {Karkar}, {Katagiri}, {Katz}, {Kawanaka}, {Kazanas}, {Kh{\'e}lifi}, {Kieda},
  {Kimeswenger}, {Kimura}, {Kisaka}, {Knapp}, {Kn{\"o}dlseder}, {Koch},
  {Kohri}, {Komin}, {Kosack}, {Kraus}, {Krause}, {Krau{\ss}}, {Kubo}, {Kukec
  Mezek}, {Kuroda}, {Kushida}, {La Palombara}, {Lamanna}, {Lang}, {Lapington},
  {Le Blanc}, {Leach}, {Lees}, {Lefaucheur}, {Leigui de Oliveira}, {Lenain},
  {Lico}, {Limon}, {Lindfors}, {Lohse}, {Lombardi}, {Longo}, {L{\'o}pez},
  {L{\'o}pez-Coto}, {Lu}, {Lucarelli}, {Luque-Escamilla}, {Lyard}, {Maccarone},
  {Maier}, {Majumdar}, {Malaguti}, {Mandat}, {Maneva}, {Manganaro}, {Mangano},
  {Marcowith}, {Mar{\'\i}n}, {Markoff}, {Mart{\'\i}}, {Martin},
  {Mart{\'\i}nez}, {Mart{\'\i}nez}, {Masetti}, {Masuda}, {Maurin}, {Maxted},
  {Mazin}, {Medina}, {Melandri}, {Mereghetti}, {Meyer}, {Minaya}, {Mirabal},
  {Mirzoyan}, {Mitchell}, {Mizuno}, {Moderski}, {Mohammed}, {Mohrmann},
  {Montaruli}, {Moralejo}, {Morcuende-Parrilla}, {Mori}, {Morlino}, {Morris},
  {Morselli}, {Moulin}, {Mukherjee}, {Mundell}, {Murach}, {Muraishi}, {Murase},
  {Nagai}, {Nagataki}, {Nagayoshi}, {Naito}, {Nakamori}, {Nakamura}, {Niemiec},
  {Nieto}, {Niko{\l}ajuk}, {Nishijima}, {Noda}, {Nosek}, {Novosyadlyj},
  {Nozaki}, {O'Brien}, {Oakes}, {Ohira}, {Ohishi}, {Ohm}, {Okazaki}, {Okumura},
  {Ong}, {Orienti}, {Orito}, {Osborne}, {Ostrowski}, {Otte}, {Oya}, {Padovani},
  {Paizis}, {Palatiello}, {Palatka}, {Paoletti}, {Paredes}, {Pareschi},
  {Parsons}, {Pe'er}, {Pech}, {Pedaletti}, {Perri}, {Persic}, {Petrashyk},
  {Petrucci}, {Petruk}, {Peyaud}, {Pfeifer}, {Piano}, {Pisarski}, {Pita},
  {Pohl}, {Polo}, {Pozo}, {Prandini}, {Prast}, {Principe}, {Prokhorov},
  {Prokoph}, {Prouza}, {P{\"u}hlhofer}, {Punch}, {P{\"u}rckhauer}, {Queiroz},
  {Quirrenbach}, {Rain{\`o}}, {Razzaque}, {Reimer}, {Reimer}, {Reisenegger},
  {Renaud}, {Rezaeian}, {Rhode}, {Ribeiro}, {Rib{\'o}}, {Richtler}, {Rico},
  {Rieger}, {Riquelme}, {Rivoire}, {Rizi}, {Rodriguez}, {Rodriguez Fernandez},
  {Rodr{\'\i}guez V{\'a}zquez}, {Rojas}, {Romano}, {Romeo}, {Rosado}, {Rovero},
  {Rowell}, {Rudak}, {Rugliancich}, {Rulten}, {Sadeh}, {Safi-Harb}, {Saito},
  {Sakaki}, {Sakurai}, {Salina}, {S{\'a}nchez-Conde}, {Sandaker}, {Sandoval},
  {Sangiorgi}, {Sanguillon}, {Sano}, {Santander}, {Sarkar}, {Satalecka},
  {Saturni}, {Schioppa}, {Schlenstedt}, {Schneider}, {Schoorlemmer},
  {Schovanek}, {Schulz}, {Schussler}, {Schwanke}, {Sciacca}, {Scuderi},
  {Seitenzahl}, {Semikoz}, {Sergijenko}, {Servillat}, {Shalchi}, {Shellard},
  {Sidoli}, {Siejkowski}, {Sillanp{\"a}{\"a}}, {Sironi}, {Sitarek}, {Sliusar},
  {Slowikowska}, {Sol}, {Stamerra}, {Stani{\v{c}}}, {Starling}, {Stawarz},
  {Stefanik}, {Stephan}, {Stolarczyk}, {Stratta}, {Straumann}, {Suomijarvi},
  {Supanitsky}, {Tagliaferri}, {Tajima}, {Tavani}, {Tavecchio}, {Tavernet},
  {Tayabaly}, {Tejedor}, {Temnikov}, {Terada}, {Terrier}, {Terzic}, {Teshima},
  {Testa}, {Thoudam}, {Tian}, {Tibaldo}, {Tluczykont}, {Todero Peixoto},
  {Tokanai}, {Tomastik}, {Tonev}, {Tornikoski}, {Torres}, {Torresi}, {Tosti},
  {Tothill}, {Tovmassian}, {Travnicek}, {Trichard}, {Trifoglio}, {Troyano
  Pujadas}, {Tsujimoto}, {Umana}, {Vagelli}, {Vagnetti}, {Valentino},
  {Vallania}, {Valore}, {van Eldik}, {Vandenbroucke}, {Varner}, {Vasileiadis},
  {Vassiliev}, {V{\'a}zquez Acosta}, {Vecchi}, {Vega}, {Vercellone}, {Veres},
  {Vergani}, {Verzi}, {Vettolani}, {Viana}, {Vigorito}, {Villanueva}, {Voelk},
  {Vollhardt}, {Vorobiov}, {Vrastil}, {Vuillaume}, {Wagner}, {Wagner},
  {Walter}, {Ward}, {Warren}, {Watson}, {Werner}, {White}, {White},
  {Wierzcholska}, {Wilcox}, {Will}, {Williams}, {Wischnewski}, {Wood},
  {Yamamoto}, {Yamazaki}, {Yanagita}, {Yang}, {Yoshida}, {Yoshiike},
  {Yoshikoshi}, {Zacharias}, {Zaharijas}, {Zampieri}, {Zandanel}, {Zanin},
  {Zavrtanik}, {Zavrtanik}, {Zdziarski}, {Zech}, {Zechlin}, {Zhdanov},
  {Ziegler}, \& {Zorn}}]{scienceCTA}
{Cherenkov Telescope Array Consortium}, {Acharya}, B.~S., {Agudo}, I., {et~al.}
  2019, {Science with the Cherenkov Telescope Array}, \dodoi{10.1142/10986}

\bibitem[{{Dermer} \& {Menon}(2009)}]{2009herb.book.....D}
{Dermer}, C.~D., \& {Menon}, G. 2009, {High Energy Radiation from Black Holes:
  Gamma Rays, Cosmic Rays, and Neutrinos}

\bibitem[{Dermer \& Menon(2009)}]{dermer2009high}
Dermer, C.~D., \& Menon, G. 2009, High Energy Radiation from Black Holes
  (Princeton, NJ: Princeton University Press)

\bibitem[{{Di Gesu} {et~al.}(2022){Di Gesu}, {Donnarumma}, {Tavecchio},
  {Agudo}, {Barnounin}, {Cibrario}, {Di Lalla}, {Di Marco}, {Escudero},
  {Errando}, {Jorstad}, {Kim}, {Kouch}, {Liodakis}, {Lindfors}, {Madejski},
  {Marshall}, {Marscher}, {Middei}, {Muleri}, {Myserlis}, {Negro}, {Omodei},
  {Pacciani}, {Paggi}, {Perri}, {Puccetti}, {Antonelli}, {Bachetti}, {Baldini},
  {Baumgartner}, {Bellazzini}, {Bianchi}, {Bongiorno}, {Bonino}, {Brez},
  {Bucciantini}, {Capitanio}, {Castellano}, {Cavazzuti}, {Ciprini}, {Costa},
  {De Rosa}, {Del Monte}, {Doroshenko}, {Dov{\v{c}}iak}, {Ehlert}, {Enoto},
  {Evangelista}, {Fabiani}, {Ferrazzoli}, {Garcia}, {Gunji}, {Hayashida},
  {Heyl}, {Iwakiri}, {Karas}, {Kitaguchi}, {Kolodziejczak}, {Krawczynski}, {La
  Monaca}, {Latronico}, {Maldera}, {Manfreda}, {Marin}, {Marinucci}, {Massaro},
  {Matt}, {Mitsuishi}, {Mizuno}, {Ng}, {O'Dell}, {Oppedisano}, {Papitto},
  {Pavlov}, {Peirson}, {Pesce-Rollins}, {Petrucci}, {Pilia}, {Possenti},
  {Poutanen}, {Ramsey}, {Rankin}, {Ratheesh}, {Romani}, {Sgr{\`o}}, {Slane},
  {Soffitta}, {Spandre}, {Tamagawa}, {Taverna}, {Tawara}, {Tennant}, {Thomas},
  {Tombesi}, {Trois}, {Tsygankov}, {Turolla}, {Vink}, {Weisskopf}, {Wu}, {Xie},
  \& {Zane}}]{digesu22}
{Di Gesu}, L., {Donnarumma}, I., {Tavecchio}, F., {et~al.} 2022, \apjl, 938,
  L7, \dodoi{10.3847/2041-8213/ac913a}

\bibitem[{{Diltz} {et~al.}(2015){Diltz}, {B{\"o}ttcher}, \&
  {Fossati}}]{Diltz15}
{Diltz}, C., {B{\"o}ttcher}, M., \& {Fossati}, G. 2015, \apj, 802, 133,
  \dodoi{10.1088/0004-637X/802/2/133}

\bibitem[{{Dimitrakoudis} {et~al.}(2012){Dimitrakoudis}, {Mastichiadis},
  {Protheroe}, \& {Reimer}}]{DMPR12}
{Dimitrakoudis}, S., {Mastichiadis}, A., {Protheroe}, R.~J., \& {Reimer}, A.
  2012, \aap, 546, A120, \dodoi{10.1051/0004-6361/201219770}

\bibitem[{{Dimitrakoudis} {et~al.}(2014){Dimitrakoudis}, {Petropoulou}, \&
  {Mastichiadis}}]{DPM14}
{Dimitrakoudis}, S., {Petropoulou}, M., \& {Mastichiadis}, A. 2014,
  Astroparticle Physics, 54, 61, \dodoi{10.1016/j.astropartphys.2013.10.005}

\bibitem[{{Drury} {et~al.}(1994){Drury}, {Aharonian}, \& {Voelk}}]{Drury94}
{Drury}, L.~O., {Aharonian}, F.~A., \& {Voelk}, H.~J. 1994, \aap, 287, 959.
\newblock \doarXiv{astro-ph/9305037}

\bibitem[{{Fichet de Clairfontaine} {et~al.}(2023){Fichet de Clairfontaine},
  {Buson}, {Pfeiffer}, {Marchesi}, {Azzollini}, {Baghmanyan}, {Tramacere},
  {Barbano}, \& {Oswald}}]{2023ApJ...958L...2F}
{Fichet de Clairfontaine}, G., {Buson}, S., {Pfeiffer}, L., {et~al.} 2023,
  \apjl, 958, L2, \dodoi{10.3847/2041-8213/ad0644}

\bibitem[{{Florou} {et~al.}(2021){Florou}, {Petropoulou}, \&
  {Mastichiadis}}]{Florou21}
{Florou}, I., {Petropoulou}, M., \& {Mastichiadis}, A. 2021, \mnras, 505, 1367,
  \dodoi{10.1093/mnras/stab1285}

\bibitem[{Gao {et~al.}(2019)Gao, Fedynitch, Winter, \& Pohl}]{Gao:2018mnu}
Gao, S., Fedynitch, A., Winter, W., \& Pohl, M. 2019, Nature Astron., 3, 88,
  \dodoi{10.1038/s41550-018-0610-1}

\bibitem[{{Gao} {et~al.}(2017){Gao}, {Pohl}, \& {Winter}}]{Gao_2017}
{Gao}, S., {Pohl}, M., \& {Winter}, W. 2017, \apj, 843, 109,
  \dodoi{10.3847/1538-4357/aa7754}

\bibitem[{{Gasparyan} {et~al.}(2022){Gasparyan}, {B{\'e}gu{\'e}}, \&
  {Sahakyan}}]{soprano}
{Gasparyan}, S., {B{\'e}gu{\'e}}, D., \& {Sahakyan}, N. 2022, \mnras, 509,
  2102, \dodoi{10.1093/mnras/stab2688}

\bibitem[{{Gould}(1979)}]{Gould79}
{Gould}, R.~J. 1979, \aap, 76, 306

\bibitem[{Hummer {et~al.}(2010)Hummer, Ruger, Spanier, \&
  Winter}]{Hummer:2010vx}
Hummer, S., Ruger, M., Spanier, F., \& Winter, W. 2010, Astrophys. J., 721,
  630, \dodoi{10.1088/0004-637X/721/1/630}

\bibitem[{{IceCube Collaboration} {et~al.}(2018{\natexlab{a}}){IceCube
  Collaboration}, {Aartsen}, {Ackermann}, {Adams}, {Aguilar}, {Ahlers},
  {Ahrens}, {Al Samarai}, {Altmann}, {Andeen}, {Anderson}, {Ansseau}, {Anton},
  {Arg{\"u}elles}, {Auffenberg}, {Axani}, {Bagherpour}, {Bai}, {Barron},
  {Barwick}, {Baum}, {Bay}, {Beatty}, {Becker Tjus}, {Becker}, {BenZvi},
  {Berley}, {Bernardini}, {Besson}, {Binder}, {Bindig}, {Blaufuss}, {Blot},
  {Bohm}, {B{\"o}rner}, {Bos}, {B{\"o}ser}, {Botner}, {Bourbeau}, {Bourbeau},
  {Bradascio}, {Braun}, {Brenzke}, {Bretz}, {Bron}, {Brostean-Kaiser},
  {Burgman}, {Busse}, {Carver}, {Cheung}, {Chirkin}, {Christov}, {Clark},
  {Classen}, {Coenders}, {Collin}, {Conrad}, {Coppin}, {Correa}, {Cowen},
  {Cross}, {Dave}, {Day}, {de Andr{\'e}}, {De Clercq}, {DeLaunay}, {Dembinski},
  {De Ridder}, {Desiati}, {de Vries}, {de Wasseige}, {de With}, {DeYoung},
  {D{\'\i}az-V{\'e}lez}, {di Lorenzo}, {Dujmovic}, {Dumm}, {Dunkman}, {Dvorak},
  {Eberhardt}, {Ehrhardt}, {Eichmann}, {Eller}, {Evenson}, {Fahey}, {Fazely},
  {Felde}, {Filimonov}, {Finley}, {Flis}, {Franckowiak}, {Friedman}, {Fritz},
  {Gaisser}, {Gallagher}, {Gerhardt}, {Ghorbani}, {Glauch}, {Gl{\"u}senkamp},
  {Goldschmidt}, {Gonzalez}, {Grant}, {Griffith}, {Haack}, {Hallgren},
  {Halzen}, {Hanson}, {Hebecker}, {Heereman}, {Helbing}, {Hellauer},
  {Hickford}, {Hignight}, {Hill}, {Hoffman}, {Hoffmann}, {Hoinka},
  {Hokanson-Fasig}, {Hoshina}, {Huang}, {Huber}, {Hultqvist}, {H{\"u}nnefeld},
  {Hussain}, {In}, {Iovine}, {Ishihara}, {Jacobi}, {Japaridze}, {Jeong},
  {Jero}, {Jones}, {Kalaczynski}, {Kang}, {Kappes}, {Kappesser}, {Karg},
  {Karle}, {Katz}, {Kauer}, {Keivani}, {Kelley}, {Kheirandish}, {Kim}, {Kim},
  {Kintscher}, {Kiryluk}, {Kittler}, {Klein}, {Koirala}, {Kolanoski},
  {K{\"o}pke}, {Kopper}, {Kopper}, {Koschinsky}, {Koskinen}, {Kowalski},
  {Krings}, {Kroll}, {Kr{\"u}ckl}, {Kunwar}, {Kurahashi}, {Kuwabara},
  {Kyriacou}, {Labare}, {Lanfranchi}, {Larson}, {Lauber}, {Leonard},
  {Lesiak-Bzdak}, {Leuermann}, {Liu}, {Lozano Mariscal}, {Lu}, {L{\"u}nemann},
  {Luszczak}, {Madsen}, {Maggi}, {Mahn}, {Mancina}, {Maruyama}, {Mase},
  {Maunu}, {Meagher}, {Medici}, {Meier}, {Menne}, {Merino}, {Meures},
  {Miarecki}, {Micallef}, {Moment{\'e}}, {Montaruli}, {Moore}, {S}, {Morse},
  {Moulai}, {Nahnhauer}, {Nakarmi}, {Naumann}, {Neer}, {Niederhausen},
  {Nowicki}, {Nygren}, {Obertacke Pollmann}, {Olivas}, {O'Murchadha},
  {O'Sullivan}, {Palczewski}, {Pandya}, {Pankova}, {Peiffer}, {Pepper},
  {P{\'e}rez de los Heros}, {Pieloth}, {Pinat}, {Plum}, {Price}, {Przybylski},
  {Raab}, {R{\"a}del}, {Rameez}, {Rauch}, {Rawlins}, {Rea}, {Reimann},
  {Relethford}, {Relich}, {Resconi}, {Rhode}, {Richman}, {Robertson}, {Rongen},
  {Rott}, {Ruhe}, {Ryckbosch}, {Rysewyk}, {Safa}, {S{\"a}lzer}, {Sanchez
  Herrera}, {Sandrock}, {Sandroos}, {Santander}, {Sarkar}, {Sarkar},
  {Satalecka}, {Schlunder}, {Schmidt}, {Schneider}, {Schoenen},
  {Sch{\"o}neberg}, {Schumacher}, {Sclafani}, {Seckel}, {Seunarine},
  {Soedingrekso}, {Soldin}, {Song}, {Spiczak}, {Spiering}, {Stachurska},
  {Stamatikos}, {Stanev}, {Stasik}, {Stein}, {Stettner}, {Steuer},
  {Stezelberger}, {Stokstad}, {St{\"o}{\ss}l}, {Strotjohann}, {Stuttard},
  {Sullivan}, {Sutherland}, {Taboada}, {Tatar}, {Tenholt}, {Ter-Antonyan},
  {Terliuk}, {Tilav}, {Toale}, {Tobin}, {Toennis}, {Toscano}, {Tosi},
  {Tselengidou}, {Tung}, {Turcati}, {Turley}, {Ty}, {Unger}, {Usner},
  {Vandenbroucke}, {Van Driessche}, {van Eijk}, {van Eijndhoven}, {Vanheule},
  {van Santen}, {Vogel}, {Vraeghe}, {Walck}, {Wallace}, {Wallraff}, {Wandler},
  {Wandkowsky}, {Waza}, {Weaver}, {Weiss}, {Wendt}, {Werthebach}, {Westerhoff},
  {Whelan}, {Whitehorn}, {Wiebe}, {Wiebusch}, {Wille}, {Williams}, {Wills},
  {Wolf}, {Wood}, {Wood}, {Woschnagg}, {Xu}, {Xu}, {Xu}, {Yanez}, {Yodh},
  {Yoshida}, {Yuan}, {Fermi-LAT Collaboration}, {Abdollahi}, {Ajello},
  {Angioni}, {Baldini}, {Ballet}, {Barbiellini}, {Bastieri}, {Bechtol},
  {Bellazzini}, {Berenji}, {Bissaldi}, {Blandford}, {Bonino}, {Bottacini},
  {Bregeon}, {Bruel}, {Buehler}, {Burnett}, {Burns}, {Buson}, {Cameron},
  {Caputo}, {Caraveo}, {Cavazzuti}, {Charles}, {Chen}, {Cheung}, {Chiang},
  {Chiaro}, {Ciprini}, {Cohen-Tanugi}, {Conrad}, {Costantin}, {Cutini},
  {D'Ammando}, {de Palma}, {Digel}, {Di Lalla}, {Di Mauro}, {Di Venere},
  {Dom{\'\i}nguez}, {Favuzzi}, {Franckowiak}, {Fukazawa}, {Funk}, {Fusco},
  {Gargano}, {Gasparrini}, {Giglietto}, {Giomi}, {Giommi}, {Giordano},
  {Giroletti}, {Glanzman}, {Green}, {Grenier}, {Grondin}, {Guiriec}, {Harding},
  {Hayashida}, {Hays}, {Hewitt}, {Horan}, {J{\'o}hannesson}, {Kadler},
  {Kensei}, {Kocevski}, {Krauss}, {Kreter}, {Kuss}, {La Mura}, {Larsson},
  {Latronico}, {Lemoine-Goumard}, {Li}, {Longo}, {Loparco}, {Lovellette},
  {Lubrano}, {Magill}, {Maldera}, {Malyshev}, {Manfreda}, {Mazziotta},
  {McEnery}, {Meyer}, {Michelson}, {Mizuno}, {Monzani}, {Morselli},
  {Moskalenko}, {Negro}, {Nuss}, {Ojha}, {Omodei}, {Orienti}, {Orlando},
  {Palatiello}, {Paliya}, {Perkins}, {Persic}, {Pesce-Rollins}, {Piron},
  {Porter}, {Principe}, {Rain{\`o}}, {Rando}, {Rani}, {Razzano}, {Razzaque},
  {Reimer}, {Reimer}, {Renault-Tinacci}, {Ritz}, {Rochester}, {Saz Parkinson},
  {Sgr{\`o}}, {Siskind}, {Spandre}, {Spinelli}, {Suson}, {Tajima}, {Takahashi},
  {Tanaka}, {Thayer}, {Thompson}, {Tibaldo}, {Torres}, {Torresi}, {Tosti},
  {Troja}, {Valverde}, {Vianello}, {Vogel}, {Wood}, {Wood}, {Zaharijas}, {MAGIC
  Collaboration}, {Ahnen}, {Ansoldi}, {Antonelli}, {Arcaro}, {Baack},
  {Babi{\'c}}, {Banerjee}, {Bangale}, {Barres de Almeida}, {Barrio}, {Becerra
  Gonz{\'a}lez}, {Bednarek}, {Bernardini}, {Berti}, {Bhattacharyya}, {Biland},
  {Blanch}, {Bonnoli}, {Carosi}, {Carosi}, {Ceribella}, {Chatterjee}, {Colak},
  {Colin}, {Colombo}, {Contreras}, {Cortina}, {Covino}, {Cumani}, {Da Vela},
  {Dazzi}, {De Angelis}, {De Lotto}, {Delfino}, {Delgado}, {Di Pierro},
  {Dom{\'\i}nguez}, {Dominis Prester}, {Dorner}, {Doro}, {Einecke},
  {Elsaesser}, {Fallah Ramazani}, {Fern{\'a}ndez-Barral}, {Fidalgo}, {Foffano},
  {Pfrang}, {Fonseca}, {Font}, {Franceschini}, {Fruck}, {Galindo}, {Gallozzi},
  {Garc{\'\i}a L{\'o}pez}, {Garczarczyk}, {Gaug}, {Giammaria}, {Godinovi{\'c}},
  {Gora}, {Guberman}, {Hadasch}, {Hahn}, {Hassan}, {Hayashida}, {Herrera},
  {Hose}, {Hrupec}, {Inoue}, {Ishio}, {Konno}, {Kubo}, {Kushida}, {Lelas},
  {Lindfors}, {Lombardi}, {Longo}, {L{\'o}pez}, {Maggio}, {Majumdar},
  {Makariev}, {Maneva}, {Manganaro}, {Mannheim}, {Maraschi}, {Mariotti},
  {Mart{\'\i}nez}, {Masuda}, {Mazin}, {Minev}, {M}, {Mirzoyan}, {Moralejo},
  {Moreno}, {Moretti}, {Nagayoshi}, {Neustroev}, {Niedzwiecki}, {Nievas
  Rosillo}, {Nigro}, {Nilsson}, {Ninci}, {Nishijima}, {Noda}, {Nogu{\'e}s},
  {Paiano}, {Palacio}, {Paneque}, {Paoletti}, {Paredes}, {Pedaletti},
  {Peresano}, {Persic}, {Prada Moroni}, {Prandini}, {Puljak}, {Rodriguez
  Garcia}, {Reichardt}, {Rhode}, {Rib{\'o}}, {Rico}, {Righi}, {Rugliancich},
  {Saito}, {Satalecka}, {Schweizer}, {Sitarek}, {{\v{S}}nidaric ́},
  {Sobczynska}, {Stamerra}, {Strzys}, {Suri{\'c}}, {Takahashi}, {Tavecchio},
  {Temnikov}, {Terzi{\'c}}, {Teshima}, {Torres-Alb{\`a}}, {Treves},
  {Tsujimoto}, {Vanzo}, {Vazquez Acosta}, {Vovk}, {Ward}, {Will}, {S}, {Zaric
  ́}, {AGILE Team}, {Lucarelli}, {Tavani}, {Piano}, {Donnarumma}, {Pittori},
  {Verrecchia}, {Barbiellini}, {Bulgarelli}, {Caraveo}, {Cattaneo},
  {Colafrancesco}, {Costa}, {Di Cocco}, {Ferrari}, {Gianotti}, {Giuliani},
  {Lipari}, {Mereghetti}, {Morselli}, {Pacciani}, {Paoletti}, {Parmiggiani},
  {Pellizzoni}, {Picozza}, {Pilia}, {Rappoldi}, {Trois}, {Vercellone},
  {Vittorini}, {ASAS-SN Team}, {Stanek}, {Kochanek}, {Beacom}, {Thompson},
  {Holoien}, {Dong}, {Prieto}, {Shappee}, {Holmbo}, {HAWC Collaboration},
  {Abeysekara}, {Albert}, {Alfaro}, {Alvarez}, {Arceo},
  {Arteaga-Vel{\'a}zquez}, {Avila Rojas}, {Ayala Solares}, {Becerril},
  {Belmont-Moreno}, {Bernal}, {Caballero-Mora}, {Capistr{\'a}n},
  {Carrami{\~n}ana}, {Casanova}, {Castillo}, {Cotti}, {Cotzomi}, {Couti{\~n}o
  de Le{\'o}n}, {De Le{\'o}n}, {De la Fuente}, {Diaz Hernandez}, {Dichiara},
  {Dingus}, {DuVernois}, {D{\'\i}az-V{\'e}lez}, {Ellsworth}, {Engel},
  {Fiorino}, {Fleischhack}, {Fraija}, {Garc{\'\i}a-Gonz{\'a}lez}, {Garfias},
  {Gonz{\'a}lez Mu{\~n}oz}, {Gonz{\'a}lez}, {Goodman}, {Hampel-Arias},
  {Harding}, {Hernand ez}, {Hona}, {Hueyotl-Zahuantitla}, {Hui},
  {H{\"u}ntemeyer}, {Iriarte}, {Jardin-Blicq}, {Joshi}, {Kaufmann}, {Kunde},
  {Lara}, {Lauer}, {Lee}, {Lennarz}, {Le{\'o}n Vargas}, {Linnemann},
  {Longinotti}, {Luis-Raya}, {Luna-Garc{\'\i}a}, {Malone}, {Marinelli},
  {Martinez}, {Martinez-Castellanos}, {Mart{\'\i}nez-Castro},
  {Mart{\'\i}nez-Huerta}, {Matthews}, {Miranda-Romagnoli}, {Moreno},
  {Mostaf{\'a}}, {Nayerhoda}, {Nellen}, {Newbold}, {Nisa}, {Noriega-Papaqui},
  {Pelayo}, {Pretz}, {P{\'e}rez-P{\'e}rez}, {Ren}, {Rho}, {Rivi{\`e}re},
  {Rosa-Gonz{\'a}lez}, {Rosenberg}, {Ruiz-Velasco}, {Ruiz-Velasco}, {Salesa
  Greus}, {Sandoval}, {Schneider}, {Schoorlemmer}, {Sinnis}, {Smith},
  {Springer}, {Surajbali}, {Tibolla}, {Tollefson}, {Torres}, {Villase{\~n}or},
  {Weisgarber}, {Werner}, {Yapici}, {Gaurang}, {Zepeda}, {Zhou}, {{\'A}lvarez},
  {H.~E.~S.~S. Collaboration}, {Abdalla}, {Ang{\"u}ner}, {Armand}, {Backes},
  {Becherini}, {Berge}, {B{\"o}ttcher}, {Boisson}, {Bolmont}, {Bonnefoy},
  {Bordas}, {Brun}, {B{\"u}chele}, {Bulik}, {Caroff}, {Carosi}, {Casanova},
  {Cerruti}, {Chakraborty}, {Chandra}, {Chen}, {Colafrancesco}, {Davids},
  {Deil}, {Devin}, {Djannati-Ata{\"\i}}, {Egberts}, {Emery}, {Eschbach},
  {Fiasson}, {Fontaine}, {Funk}, {F{\"u}{\ss}ling}, {Gallant}, {Gat{\'e}},
  {Giavitto}, {Glawion}, {Glicenstein}, {Gottschall}, {Grondin}, {Haupt},
  {Henri}, {Hinton}, {Hoischen}, {Holch}, {Huber}, {Jamrozy}, {Jankowsky},
  {Jankowsky}, {Jouvin}, {Jung-Richardt}, {Kerszberg}, {Kh{\'e}lifi}, {King},
  {Klepser}, {Kluz ́niak}, {Komin}, {Kraus}, {Lefaucheur}, {Lemi{\`e}re},
  {Lemoine-Goumard}, {Lenain}, {Leser}, {Lohse}, {L{\'o}pez-Coto}, {Lorentz},
  {Lypova}, {Marandon}, {Guillem Mart{\'\i}-Devesa}, {Maurin}, {Mitchell},
  {Moderski}, {Mohamed}, {Mohrmann}, {Moulin}, {Murach}, {de Naurois},
  {Niederwanger}, {Niemiec}, {Oakes}, {O'Brien}, {Ohm}, {Ostrowski}, {Oya},
  {Panter}, {Parsons}, {Perennes}, {Piel}, {Pita}, {Poireau}, {Priyana Noel},
  {Prokoph}, {P{\"u}hlhofer}, {Quirrenbach}, {Raab}, {Rauth}, {Renaud},
  {Rieger}, {Rinchiuso}, {Romoli}, {Rowell}, {Rudak}, {Sasaki}, {Sanchez},
  {Schlickeiser}, {Sch{\"u}ssler}, {Schulz}, {Schwanke}, {Seglar-Arroyo},
  {Shafi}, {Simoni}, {Sol}, {Stegmann}, {Steppa}, {Tavernier}, {Taylor},
  {Tiziani}, {Trichard}, {Tsirou}, {van Eldik}, {van Rensburg}, {van Soelen},
  {Veh}, {Vincent}, {Voisin}, {Wagner}, {Wagner}, {Wierzcholska}, {Zanin},
  {Zdziarski}, {Zech}, {Ziegler}, {Zorn}, {{\.Z}ywucka}, {INTEGRAL Team},
  {Savchenko}, {Ferrigno}, {Bazzano}, {Diehl}, {Kuulkers}, {Laurent},
  {Mereghetti}, {Natalucci}, {Panessa}, {Rodi}, {Ubertini}, {Kanata}, Teams,
  {Morokuma}, {Ohta}, {Tanaka}, {Mori}, {Yamanaka}, {Kawabata}, {Utsumi},
  {Nakaoka}, {Kawabata}, {Nagashima}, {Yoshida}, {Matsuoka}, {Itoh}, {Kapteyn
  Team}, {Keel}, {Liverpool Telescope Team}, {Copperwheat}, {Steele},
  {Swift/NuSTAR Team}, {Cenko}, {Cowen}, {DeLaunay}, {Evans}, {Fox}, {Keivani},
  {Kennea}, {Marshall}, {Osborne}, {Santander}, {Tohuvavohu}, {Turley},
  {VERITAS Collaboration}, {Abeysekara}, {Archer}, {Benbow}, {Bird}, {Brill},
  {Brose}, {Buchovecky}, {Buckley}, {Bugaev}, {Christiansen}, {Connolly},
  {Cui}, {Daniel}, {Errando}, {Falcone}, {Feng}, {Finley}, {Fortson},
  {Furniss}, {Gueta}, {H{\"u}tten}, {Hervet}, {Hughes}, {Humensky}, {Johnson},
  {Kaaret}, {Kar}, {Kelley-Hoskins}, {Kertzman}, {Kieda}, {Krause},
  {Krennrich}, {Kumar}, {Lang}, {Lin}, {Maier}, {McArthur}, {Moriarty},
  {Mukherjee}, {Nieto}, {O'Brien}, {Ong}, {Otte}, {Park}, {Petrashyk}, {Pohl},
  {Popkow}, {Pueschel}, {Quinn}, {Ragan}, {Reynolds}, {Richards}, {Roache},
  {Rulten}, {Sadeh}, {Santander}, {Scott}, {Sembroski}, {Shahinyan}, {Sushch},
  {Tr{\'e}panier}, {Tyler}, {Vassiliev}, {Wakely}, {Weinstein}, {Wells},
  {Wilcox}, {Wilhelm}, {Williams}, {Zitzer}, {VLA/B Team}, {Tetarenko},
  {Kimball}, {Miller-Jones}, \& {Sivakoff}}]{0506science1}
{IceCube Collaboration}, {Aartsen}, M.~G., {Ackermann}, M., {et~al.}
  2018{\natexlab{a}}, Science, 361, eaat1378, \dodoi{10.1126/science.aat1378}

\bibitem[{{IceCube Collaboration} {et~al.}(2018{\natexlab{b}}){IceCube
  Collaboration}, {Aartsen}, {Ackermann}, {Adams}, {Aguilar}, {Ahlers},
  {Ahrens}, {Samarai}, {Altmann}, {Andeen}, {Anderson}, {Ansseau}, {Anton},
  {Arg{\"u}elles}, {Arsioli}, {Auffenberg}, {Axani}, {Bagherpour}, {Bai},
  {Barron}, {Barwick}, {Baum}, {Bay}, {Beatty}, {Becker Tjus}, {Becker},
  {BenZvi}, {Berley}, {Bernardini}, {Besson}, {Binder}, {Bindig}, {Blaufuss},
  {Blot}, {Bohm}, {B{\"o}rner}, {Bos}, {B{\"o}ser}, {Botner}, {Bourbeau},
  {Bourbeau}, {Bradascio}, {Braun}, {Brenzke}, {Bretz}, {Bron},
  {Brostean-Kaiser}, {Burgman}, {Busse}, {Carver}, {Cheung}, {Chirkin},
  {Christov}, {Clark}, {Classen}, {Coenders}, {Collin}, {Conrad}, {Coppin},
  {Correa}, {Cowen}, {Cross}, {Dave}, {Day}, {de Andr{\'e}}, {De Clercq},
  {DeLaunay}, {Dembinski}, {DeRidder}, {Desiati}, {de Vries}, {de Wasseige},
  {de With}, {DeYoung}, {D{\'\i}az-V{\'e}lez}, {di Lorenzo}, {Dujmovic},
  {Dumm}, {Dunkman}, {Dvorak}, {Eberhardt}, {Ehrhardt}, {Eichmann}, {Eller},
  {Evenson}, {Fahey}, {Fazely}, {Felde}, {Filimonov}, {Finley}, {Flis},
  {Franckowiak}, {Friedman}, {Fritz}, {Gaisser}, {Gallagher}, {Gerhardt},
  {Ghorbani}, {Giommi}, {Glauch}, {Gl{\"u}senkamp}, {Goldschmidt}, {Gonzalez},
  {Grant}, {Griffith}, {Haack}, {Hallgren}, {Halzen}, {Hanson}, {Hebecker},
  {Heereman}, {Helbing}, {Hellauer}, {Hickford}, {Hignight}, {Hill}, {Hoffman},
  {Hoffmann}, {Hoinka}, {Hokanson-Fasig}, {Hoshina}, {Huang}, {Huber},
  {Hultqvist}, {H{\"u}nnefeld}, {Hussain}, {In}, {Iovine}, {Ishihara},
  {Jacobi}, {Japaridze}, {Jeong}, {Jero}, {Jones}, {Kalaczynski}, {Kang},
  {Kappes}, {Kappesser}, {Karg}, {Karle}, {Katz}, {Kauer}, {Keivani}, {Kelley},
  {Kheirandish}, {Kim}, {Kim}, {Kintscher}, {Kiryluk}, {Kittler}, {Klein},
  {Koirala}, {Kolanoski}, {K{\"o}pke}, {Kopper}, {Kopper}, {Koschinsky},
  {Koskinen}, {Kowalski}, {Krammer}, {Krings}, {Kroll}, {Kr{\"u}ckl}, {Kunwar},
  {Kurahashi}, {Kuwabara}, {Kyriacou}, {Labare}, {Lanfranchi}, {Larson},
  {Lauber}, {Leonard}, {Lesiak-Bzdak}, {Leuermann}, {Liu}, {Lozano Mariscal},
  {Lu}, {L{\"u}nemann}, {Luszczak}, {Madsen}, {Maggi}, {Mahn}, {Mancina},
  {Maruyama}, {Mase}, {Maunu}, {Meagher}, {Medici}, {Meier}, {Menne}, {Merino},
  {Meures}, {Miarecki}, {Micallef}, {Moment{\'e}}, {Montaruli}, {Moore},
  {Morse}, {Moulai}, {Nahnhauer}, {Nakarmi}, {Naumann}, {Neer}, {Niederhausen},
  {Nowicki}, {Nygren}, {Obertacke Pollmann}, {Olivas}, {O'Murchadha},
  {O'Sullivan}, {Padovani}, {Palczewski}, {Pandya}, {Pankova}, {Peiffer},
  {Pepper}, {P{\'e}rez de los Heros}, {Pieloth}, {Pinat}, {Plum}, {Price},
  {Przybylski}, {Raab}, {R{\"a}del}, {Rameez}, {Rawlins}, {Rea}, {Reimann},
  {Relethford}, {Relich}, {Resconi}, {Rhode}, {Richman}, {Robertson}, {Rongen},
  {Rott}, {Ruhe}, {Ryckbosch}, {Rysewyk}, {Safa}, {Sahakyan}, {S{\"a}lzer},
  {Sanchez Herrera}, {Sandrock}, {Sandroos}, {Santander}, {Sarkar}, {Sarkar},
  {Satalecka}, {Schlunder}, {Schmidt}, {Schneider}, {Schoenen},
  {Sch{\"o}neberg}, {Schumacher}, {Sclafani}, {Seckel}, {Seunarine},
  {Soedingrekso}, {Soldin}, {Song}, {Spiczak}, {Spiering}, {Stachurska},
  {Stamatikos}, {Stanev}, {Stasik}, {Stettner}, {Steuer}, {Stezelberger},
  {Stokstad}, {St{\"o}{\ss}l}, {Strotjohann}, {Stuttard}, {Sullivan},
  {Sutherland}, {Taboada}, {Tatar}, {Tenholt}, {Ter-Antonyan}, {Terliuk},
  {Tilav}, {Toale}, {Tobin}, {Toennis}, {Toscano}, {Tosi}, {Tselengidou},
  {Tung}, {Turcati}, {Turley}, {Ty}, {Unger}, {Usner}, {Vandenbroucke}, {Van
  Driessche}, {van Eijk}, {van Eijndhoven}, {Vanheule}, {van Santen}, {Vogel},
  {Vraeghe}, {Walck}, {Wallace}, {Wallraff}, {Wandler}, {Wandkowsky}, {Waza},
  {Weaver}, {Weiss}, {Wendt}, {Werthebach}, {Westerhoff}, {Whelan},
  {Whitehorn}, {Wiebe}, {Wiebusch}, {Wille}, {Williams}, {Wills}, {Wolf},
  {Wood}, {Wood}, {Woschnagg}, {Xu}, {Xu}, {Xu}, {Yanez}, {Yodh}, {Yoshida}, \&
  {Yuan}}]{0506science2}
---. 2018{\natexlab{b}}, Science, 361, 147, \dodoi{10.1126/science.aat2890}

\bibitem[{{IceCube Collaboration} {et~al.}(2022){IceCube Collaboration},
  {Abbasi}, {Ackermann}, {Adams}, {Aguilar}, {Ahlers}, {Ahrens}, {Alameddine},
  {Alispach}, {Alves}, {Amin}, {Andeen}, {Anderson}, {Anton}, {Arg{\"u}elles},
  {Ashida}, {Axani}, {Bai}, {Balagopal}, {Barbano}, {Barwick}, {Bastian},
  {Basu}, {Baur}, {Bay}, {Beatty}, {Becker}, {Becker Tjus}, {Bellenghi},
  {Benzvi}, {Berley}, {Bernardini}, {Besson}, {Binder}, {Bindig}, {Blaufuss},
  {Blot}, {Boddenberg}, {Bontempo}, {Borowka}, {B{\"o}ser}, {Botner},
  {B{\"o}ttcher}, {Bourbeau}, {Bradascio}, {Braun}, {Brinson}, {Bron},
  {Brostean-Kaiser}, {Browne}, {Burgman}, {Burley}, {Busse}, {Campana},
  {Carnie-Bronca}, {Chen}, {Chen}, {Chirkin}, {Choi}, {Clark}, {Clark},
  {Classen}, {Coleman}, {Collin}, {Conrad}, {Coppin}, {Correa}, {Cowen},
  {Cross}, {Dappen}, {Dave}, {de Clercq}, {Delaunay}, {Delgado L{\'o}pez},
  {Dembinski}, {Deoskar}, {Desai}, {Desiati}, {de Vries}, {de Wasseige}, {de
  With}, {Deyoung}, {Diaz}, {D{\'\i}az-V{\'e}lez}, {Dittmer}, {Dujmovic},
  {Dunkman}, {Duvernois}, {Dvorak}, {Ehrhardt}, {Eller}, {Engel}, {Erpenbeck},
  {Evans}, {Evenson}, {Fan}, {Fazely}, {Fedynitch}, {Feigl}, {Fiedlschuster},
  {Fienberg}, {Filimonov}, {Finley}, {Fischer}, {Fox}, {Franckowiak},
  {Friedman}, {Fritz}, {F{\"u}rst}, {Gaisser}, {Gallagher}, {Ganster},
  {Garcia}, {Garrappa}, {Gerhardt}, {Ghadimi}, {Glaser}, {Glauch},
  {Gl{\"u}senkamp}, {Goldschmidt}, {Gonzalez}, {Goswami}, {Grant},
  {Gr{\'e}goire}, {Griswold}, {G{\"u}nther}, {Gutjahr}, {Haack}, {Hallgren},
  {Halliday}, {Halve}, {Halzen}, {Hanson}, {Hardin}, {Harnisch}, {Haungs},
  {Hebecker}, {Helbing}, {Henningsen}, {Hettinger}, {Hickford}, {Hignight},
  {Hill}, {Hill}, {Hoffman}, {Hoffmann}, {Hokanson-Fasig}, {Hoshina}, {Huang},
  {Huber}, {Huber}, {Hultqvist}, {H{\"u}nnefeld}, {Hussain}, {Hymon}, {in},
  {Iovine}, {Ishihara}, {Jansson}, {Japaridze}, {Jeong}, {Jin}, {Jones},
  {Kang}, {Kang}, {Kang}, {Kappes}, {Kappesser}, {Kardum}, {Karg}, {Karl},
  {Karle}, {Katz}, {Kauer}, {Kellermann}, {Kelley}, {Kheirandish}, {Kin},
  {Kintscher}, {Kiryluk}, {Klein}, {Koirala}, {Kolanoski}, {Kontrimas},
  {K{\"o}pke}, {Kopper}, {Kopper}, {Koskinen}, {Koundal}, {Kovacevich},
  {Kowalski}, {Kozynets}, {Kun}, {Kurahashi}, {Lad}, {Lagunas Gualda},
  {Lanfranchi}, {Larson}, {Lauber}, {Lazar}, {Lee}, {Leonard},
  {Leszczy{\'n}ska}, {Li}, {Lincetto}, {Liu}, {Liubarska}, {Lohfink}, {Lozano
  Mariscal}, {Lu}, {Lucarelli}, {Ludwig}, {Luszczak}, {Lyu}, {Ma}, {Madsen},
  {Mahn}, {Makino}, {Mancina}, {Mari{\c{s}}}, {Martinez-Soler}, {Maruyama},
  {Mase}, {McElroy}, {McNally}, {Mead}, {Meagher}, {Mechbal}, {Medina},
  {Meier}, {Meighen-Berger}, {Micallef}, {Mockler}, {Montaruli}, {Moore},
  {Morse}, {Moulai}, {Naab}, {Nagai}, {Nahnhauer}, {Naumann}, {Necker},
  {Nguyen}, {Niederhausen}, {Nisa}, {Nowicki}, {Nygren}, {Obertack},
  {Pollmann}, {Oehler}, {Oeyen}, {Olivas}, {O'Sullivan}, {Pandya}, {Pankova},
  {Park}, {Parker}, {Paudel}, {Paul}, {P{\'e}rez de Los Heros}, {Peters},
  {Peterson}, {Philippen}, {Pieper}, {Pittermann}, {Pizzuto}, {Plum},
  {Popovych}, {Porcelli}, {Prado Rodriguez}, {Price}, {Pries}, {Przybylski},
  {Rack-Helleis}, {Raissi}, {Rameez}, {Rawlins}, {Rea}, {Rehman},
  {Reichherzer}, {Reimann}, {Renzi}, {Resconi}, {Reusch}, {Rhode}, {Richman},
  {Riedel}, {Roberts}, {Robertson}, {Roellinghoff}, {Rongen}, {Rott}, {Ruhe},
  {Ryckbosch}, {Rysewyk Cantu}, {Safa}, {Saffer}, {Sanchez Herrera},
  {Sandrock}, {Sandroos}, {Santander}, {Sarkar}, {Sarkar}, {Satalecka},
  {Schaufel}, {Schieler}, {Schindler}, {Schmidt}, {Schneider}, {Schneider},
  {Schr{\"o}der}, {Schumacher}, {Schwefer}, {Sclafani}, {Seckel}, {Seunarine},
  {Sharma}, {Shefali}, {Silva}, {Skrzypek}, {Smithers}, {Snihur},
  {Soedingrekso}, {Soldin}, {Spannfellner}, {Spiczak}, {Spiering},
  {Stachurska}, {Stamatikos}, {Stanev}, {Stein}, {Stettner}, {Steuer},
  {Stezelberger}, {Stokstad}, {St{\"u}rwald}, {Stuttard}, {Sullivan},
  {Taboada}, {Ter-Antonyan}, {Tilav}, {Tischbein}, {Tollefson}, {T{\"o}nnis},
  {Toscano}, {Tosi}, {Trettin}, {Tselengidou}, {Tung}, {Turcati}, {Turcotte},
  {Turley}, {Twagirayezu}, {Ty}, {Unland Elorrieta}, {Valtonen-Mattila},
  {Vandenbroucke}, {van Eijndhoven}, {Vannerom}, {van Santen}, {Verpoest},
  {Walck}, {Watson}, {Weaver}, {Weigel}, {Weindl}, {Weiss}, {Weldert}, {Wendt},
  {Werthebach}, {Weyrauch}, {Whitehorn}, {Wiebusch}, {Williams}, {Wolf},
  {Woschnagg}, {Wrede}, {Wulff}, {Xu}, {Yanez}, {Yoshida}, {Yu}, {Yuan},
  {Zhangan}, \& {Zhelnin}}]{1068science}
{IceCube Collaboration}, {Abbasi}, R., {Ackermann}, M., {et~al.} 2022, Science,
  378, 538, \dodoi{10.1126/science.abg3395}

\bibitem[{{Inoue} {et~al.}(2022){Inoue}, {Cerruti}, {Murase}, \&
  {Liu}}]{Inoue1068}
{Inoue}, S., {Cerruti}, M., {Murase}, K., \& {Liu}, R.-Y. 2022, arXiv e-prints,
  arXiv:2207.02097, \dodoi{10.48550/arXiv.2207.02097}

\bibitem[{{Inoue} \& {Takahara}(1996)}]{Inoue_1996}
{Inoue}, S., \& {Takahara}, F. 1996, \apj, 463, 555, \dodoi{10.1086/177270}

\bibitem[{{Jim{\'e}nez Fern{\'a}ndez} \& {van Eerten}(2021)}]{katu}
{Jim{\'e}nez Fern{\'a}ndez}, B., \& {van Eerten}, H. 2021, arXiv e-prints,
  arXiv:2104.08207, \dodoi{10.48550/arXiv.2104.08207}

\bibitem[{{Jones}(1968)}]{Jones68}
{Jones}, F.~C. 1968, Physical Review, 167, 1159,
  \dodoi{10.1103/PhysRev.167.1159}

\bibitem[{{Karavola} \& {Petropoulou}(2024)}]{karavola}
{Karavola}, D., \& {Petropoulou}, M. 2024, \jcap, 2024, 006,
  \dodoi{10.1088/1475-7516/2024/07/006}

\bibitem[{{Karavola} {et~al.}(2025){Karavola}, {Petropoulou}, {Fiorillo},
  {Comisso}, \& {Sironi}}]{2025JCAP...04..075K}
{Karavola}, D., {Petropoulou}, M., {Fiorillo}, D.~F.~G., {Comisso}, L., \&
  {Sironi}, L. 2025, \jcap, 2025, 075, \dodoi{10.1088/1475-7516/2025/04/075}

\bibitem[{{Kataoka} {et~al.}(1999){Kataoka}, {Mattox}, {Quinn}, {Kubo},
  {Makino}, {Takahashi}, {Inoue}, {Hartman}, {Madejski}, {Sreekumar}, \&
  {Wagner}}]{Kataoka99}
{Kataoka}, J., {Mattox}, J.~R., {Quinn}, J., {et~al.} 1999, \apj, 514, 138,
  \dodoi{10.1086/306918}

\bibitem[{{Katarzy{\'n}ski} {et~al.}(2001){Katarzy{\'n}ski}, {Sol}, \&
  {Kus}}]{Kata01}
{Katarzy{\'n}ski}, K., {Sol}, H., \& {Kus}, A. 2001, \aap, 367, 809,
  \dodoi{10.1051/0004-6361:20000538}

\bibitem[{{Kelner} \& {Aharonian}(2008)}]{Kelner_2008}
{Kelner}, S.~R., \& {Aharonian}, F.~A. 2008, \prd, 78, 034013,
  \dodoi{10.1103/PhysRevD.78.034013}

\bibitem[{{Kelner} {et~al.}(2006){Kelner}, {Aharonian}, \&
  {Bugayov}}]{Kelner_2006}
{Kelner}, S.~R., {Aharonian}, F.~A., \& {Bugayov}, V.~V. 2006, \prd, 74,
  034018, \dodoi{10.1103/PhysRevD.74.034018}

\bibitem[{{Kirk} \& {Mastichiadis}(1992)}]{KM92}
{Kirk}, J.~G., \& {Mastichiadis}, A. 1992, \nat, 360, 135,
  \dodoi{10.1038/360135a0}

\bibitem[{{Klinger} {et~al.}(2024){Klinger}, {Yuan}, {Taylor}, \&
  {Winter}}]{2024arXiv240313902K}
{Klinger}, M., {Yuan}, C., {Taylor}, A.~M., \& {Winter}, W. 2024, arXiv
  e-prints, arXiv:2403.13902, \dodoi{10.48550/arXiv.2403.13902}

\bibitem[{Klinger {et~al.}(2024)Klinger, Rudolph, Rodrigues, Yuan,
  de~Clairfontaine, Fedynitch, Winter, Pohl, \& Gao}]{Klinger:2023zzv}
Klinger, M., Rudolph, A., Rodrigues, X., {et~al.} 2024, Astrophys. J. Suppl.,
  275, 4, \dodoi{10.3847/1538-4365/ad725c}

\bibitem[{{Krawczynski} {et~al.}(2004){Krawczynski}, {Hughes}, {Horan},
  {Aharonian}, {Aller}, {Aller}, {Boltwood}, {Buckley}, {Coppi}, {Fossati},
  {G{\"o}tting}, {Holder}, {Horns}, {Kurtanidze}, {Marscher}, {Nikolashvili},
  {Remillard}, {Sadun}, \& {Schr{\"o}der}}]{Krawczynski04}
{Krawczynski}, H., {Hughes}, S.~B., {Horan}, D., {et~al.} 2004, \apj, 601, 151,
  \dodoi{10.1086/380393}

\bibitem[{{Longair}(2011)}]{Longair_2011}
{Longair}, M.~S. 2011, {High Energy Astrophysics}

\bibitem[{{Mannheim}(1993)}]{Mannheim93}
{Mannheim}, K. 1993, \aap, 269, 67.
\newblock \doarXiv{astro-ph/9302006}

\bibitem[{{Mastichiadis}(1991)}]{Mastichiadis91}
{Mastichiadis}, A. 1991, \mnras, 253, 235, \dodoi{10.1093/mnras/253.2.235}

\bibitem[{{Mastichiadis} {et~al.}(2020){Mastichiadis}, {Florou}, {Kefala},
  {Boula}, \& {Petropoulou}}]{Mastichiadis20}
{Mastichiadis}, A., {Florou}, I., {Kefala}, E., {Boula}, S.~S., \&
  {Petropoulou}, M. 2020, arXiv e-prints, arXiv:2003.06956.
\newblock \doarXiv{2003.06956}

\bibitem[{{Mastichiadis} \& {Kirk}(1995)}]{MK95}
{Mastichiadis}, A., \& {Kirk}, J.~G. 1995, \aap, 295, 613

\bibitem[{{Mastichiadis} \& {Kirk}(1997)}]{MK97}
---. 1997, \aap, 320, 19.
\newblock \doarXiv{astro-ph/9610058}

\bibitem[{{Mastichiadis} {et~al.}(2013){Mastichiadis}, {Petropoulou}, \&
  {Dimitrakoudis}}]{MPD13}
{Mastichiadis}, A., {Petropoulou}, M., \& {Dimitrakoudis}, S. 2013, \mnras,
  434, 2684, \dodoi{10.1093/mnras/stt1210}

\bibitem[{{Mastichiadis} {et~al.}(2005){Mastichiadis}, {Protheroe}, \&
  {Kirk}}]{MPK05}
{Mastichiadis}, A., {Protheroe}, R.~J., \& {Kirk}, J.~G. 2005, \aap, 433, 765,
  \dodoi{10.1051/0004-6361:20042161}

\bibitem[{{McEnery} {et~al.}(2019){McEnery}, {van der Horst}, {Dominguez},
  {Moiseev}, {Marcowith}, {Harding}, {Lien}, {Giuliani}, {Inglis}, {Ansoldi},
  {Stamerra}, {Manousakis}, {Strong}, {Bambi}, {Patricelli}, {Baring},
  {Barrio}, {Bastieri}, {Fields}, {Beacom}, {Beckmann}, {Bednarek}, {Rani},
  {Boggs}, {Bolotnikov}, {Cenko}, {Buckley}, {Grefenstette}, {Hui}, {Pittori},
  {Prescod-Weinstein}, {Shrader}, {Gouiffes}, {Kierans}, {Wilson-Hodge},
  {D'Ammando}, {Castro}, {Kocveski}, {Gasparrini}, {Thompson}, {Williams}, {De
  Angelis}, {Bernard}, {Digel}, {Morcuende}, {Charles}, {Bissaldi}, {Hays},
  {Ferrara}, {Bozzo}, {Grove}, {Wulf}, {Bottacini}, {Caroli}, {Kislat},
  {Oikonomou}, {Giordano}, {Longo}, {Fryer}, {Fukazawa}, {Georganopoulos}, {De
  Nolfo}, {Vianello}, {Kanbach}, {Younes}, {Blumer}, {Hartmann}, {Hernanz},
  {Takahashi}, {Li}, {Agudo}, {Moskalenko}, {Stumke}, {Grenier}, {Smith},
  {Rodi}, {Perkins}, {Gelfand}, {Holder}, {Knodlseder}, {Kopp}, {Lenain},
  {{\'A}lvarez}, {Metcalfe}, {Krizmanic}, {Stephen}, {Hewitt}, {Mitchell},
  {Harding}, {Tomsick}, {Racusin}, {Finke}, {Kargaltsev}, {Klimenko},
  {Krawczynski}, {Smith}, {Kubo}, {Di Venere}, {Marcotulli}, {Lommler},
  {Parker}, {Baldini}, {Foffano}, {Zampieri}, {Tibaldo}, {Petropoulou},
  {Ajello}, {Meyer}, {L{\'o}pez}, {McConnell}, {Boettcher}, {Cardillo},
  {Martinez}, {Kerr}, {Mazziotta}, {McEnery}, {Di Mauro}, {Wood}, {Meyer},
  {Briggs}, {De Becker}, {Lovellette}, {Doro}, {Sanchez-Conde}, {Moss},
  {Mizuno}, {Rib{\'o}}, {Nakazawa}, {Neilson}, {Auricchio}, {Omodei},
  {Oberlack}, {Ohno}, {Orlando}, {Otte}, {Coppi}, {Bloser}, {Zhang}, {Laurent},
  {Pohl}, {Prandini}, {Shawhan}, {Caputo}, {Campana}, {Rando}, {Woolf},
  {Johnson}, {Mignani}, {Walter}, {Ojha}, {da Silva}, {Dietrich}, {Funk},
  {Zane}, {Anton}, {Buson}, {Cutini}, {Saz Parkinson}, {Schirato}, {Griffin},
  {Kaufmann}, {Stawarz}, {Ciprini}, {Del Sordo}, {Jones}, {Guiriec}, {Tajima},
  {Cheung}, {The}, {Venters}, {Porter}, {Linden}, {Barres}, {Paliya},
  {Bozhilov}, {Vestrand}, {Tatischeff}, {Chen}, {Wang}, {Tanaka}, {Uhm},
  {Zhang}, {Zimmer}, {Zoglauer}, \& {Wadiasingh}}]{AMEGO}
{McEnery}, J., {van der Horst}, A., {Dominguez}, A., {et~al.} 2019, in Bulletin
  of the American Astronomical Society, Vol.~51, 245.
\newblock \doarXiv{1907.07558}

\bibitem[{{Moderski} {et~al.}(2005){Moderski}, {Sikora}, {Coppi}, \&
  {Aharonian}}]{Moderski_2005}
{Moderski}, R., {Sikora}, M., {Coppi}, P.~S., \& {Aharonian}, F. 2005, \mnras,
  363, 954, \dodoi{10.1111/j.1365-2966.2005.09494.x}

\bibitem[{{M{\"u}cke} {et~al.}(2000){M{\"u}cke}, {Engel}, {Rachen},
  {Protheroe}, \& {Stanev}}]{Mucke_2000}
{M{\"u}cke}, A., {Engel}, R., {Rachen}, J.~P., {Protheroe}, R.~J., \& {Stanev},
  T. 2000, Computer Physics Communications, 124, 290,
  \dodoi{10.1016/S0010-4655(99)00446-4}

\bibitem[{{M{\"u}cke} \& {Protheroe}(2001)}]{Mucke01}
{M{\"u}cke}, A., \& {Protheroe}, R.~J. 2001, Astroparticle Physics, 15, 121,
  \dodoi{10.1016/S0927-6505(00)00141-9}

\bibitem[{{M{\"u}cke} {et~al.}(2003){M{\"u}cke}, {Protheroe}, {Engel},
  {Rachen}, \& {Stanev}}]{Mucke03}
{M{\"u}cke}, A., {Protheroe}, R.~J., {Engel}, R., {Rachen}, J.~P., \& {Stanev},
  T. 2003, Astroparticle Physics, 18, 593,
  \dodoi{10.1016/S0927-6505(02)00185-8}

\bibitem[{Murase \& Nagataki(2006)}]{Murase:2005hy}
Murase, K., \& Nagataki, S. 2006, Phys. Rev., D73, 063002,
  \dodoi{10.1103/PhysRevD.73.063002}

\bibitem[{{Padovani} \& {Giommi}(1995)}]{PadovaniGiommi}
{Padovani}, P., \& {Giommi}, P. 1995, \apj, 444, 567, \dodoi{10.1086/175631}

\bibitem[{{Padovani} {et~al.}(2017){Padovani}, {Alexander}, {Assef}, {De
  Marco}, {Giommi}, {Hickox}, {Richards}, {Smol{\v{c}}i{\'c}},
  {Hatziminaoglou}, {Mainieri}, \& {Salvato}}]{2017A&ARv..25....2P}
{Padovani}, P., {Alexander}, D.~M., {Assef}, R.~J., {et~al.} 2017, \aapr, 25,
  2, \dodoi{10.1007/s00159-017-0102-9}

\bibitem[{{Paglione} {et~al.}(1996){Paglione}, {Marscher}, {Jackson}, \&
  {Bertsch}}]{Paglione96}
{Paglione}, T. A.~D., {Marscher}, A.~P., {Jackson}, J.~M., \& {Bertsch}, D.~L.
  1996, \apj, 460, 295, \dodoi{10.1086/176969}

\bibitem[{{Petropoulou} {et~al.}(2016){Petropoulou}, {Coenders}, \&
  {Dimitrakoudis}}]{Petroetal16}
{Petropoulou}, M., {Coenders}, S., \& {Dimitrakoudis}, S. 2016, Astroparticle
  Physics, 80, 115, \dodoi{10.1016/j.astropartphys.2016.04.001}

\bibitem[{{Petropoulou} {et~al.}(2014{\natexlab{a}}){Petropoulou},
  {Dimitrakoudis}, {Mastichiadis}, \& {Giannios}}]{PDMG14}
{Petropoulou}, M., {Dimitrakoudis}, S., {Mastichiadis}, A., \& {Giannios}, D.
  2014{\natexlab{a}}, \mnras, 444, 2186, \dodoi{10.1093/mnras/stu1362}

\bibitem[{{Petropoulou} {et~al.}(2015{\natexlab{a}}){Petropoulou},
  {Dimitrakoudis}, {Padovani}, {Mastichiadis}, \& {Resconi}}]{Petroetal15}
{Petropoulou}, M., {Dimitrakoudis}, S., {Padovani}, P., {Mastichiadis}, A., \&
  {Resconi}, E. 2015{\natexlab{a}}, \mnras, 448, 2412,
  \dodoi{10.1093/mnras/stv179}

\bibitem[{{Petropoulou} {et~al.}(2014{\natexlab{b}}){Petropoulou}, {Giannios},
  \& {Dimitrakoudis}}]{PGD14}
{Petropoulou}, M., {Giannios}, D., \& {Dimitrakoudis}, S. 2014{\natexlab{b}},
  \mnras, 445, 570, \dodoi{10.1093/mnras/stu1757}

\bibitem[{{Petropoulou} {et~al.}(2014{\natexlab{c}}){Petropoulou}, {Lefa},
  {Dimitrakoudis}, \& {Mastichiadis}}]{Petroetal14}
{Petropoulou}, M., {Lefa}, E., {Dimitrakoudis}, S., \& {Mastichiadis}, A.
  2014{\natexlab{c}}, \aap, 562, A12, \dodoi{10.1051/0004-6361/201322833}

\bibitem[{{Petropoulou} \& {Mastichiadis}(2012)}]{PM12}
{Petropoulou}, M., \& {Mastichiadis}, A. 2012, \mnras, 421, 2325,
  \dodoi{10.1111/j.1365-2966.2012.20460.x}

\bibitem[{{Petropoulou} \& {Mastichiadis}(2018)}]{PM18}
---. 2018, \mnras, 477, 2917, \dodoi{10.1093/mnras/sty833}

\bibitem[{{Petropoulou} {et~al.}(2015{\natexlab{b}}){Petropoulou}, {Piran}, \&
  {Mastichiadis}}]{PPM15}
{Petropoulou}, M., {Piran}, T., \& {Mastichiadis}, A. 2015{\natexlab{b}},
  \mnras, 452, 3226, \dodoi{10.1093/mnras/stv1523}

\bibitem[{{Petropoulou} {et~al.}(2019){Petropoulou}, {Yuan}, {Chen}, \&
  {Mastichiadis}}]{Petroetal2019}
{Petropoulou}, M., {Yuan}, Y., {Chen}, A.~Y., \& {Mastichiadis}, A. 2019, \apj,
  883, 66, \dodoi{10.3847/1538-4357/ab3856}

\bibitem[{{Petropoulou} {et~al.}(2020){Petropoulou}, {Murase}, {Santander},
  {Buson}, {Tohuvavohu}, {Kawamuro}, {Vasilopoulos}, {Negoro}, {Ueda},
  {Siegel}, {Keivani}, {Kawai}, {Mastichiadis}, \&
  {Dimitrakoudis}}]{Petroetal20}
{Petropoulou}, M., {Murase}, K., {Santander}, M., {et~al.} 2020, \apj, 891,
  115, \dodoi{10.3847/1538-4357/ab76d0}

\bibitem[{{Protheroe} \& {Johnson}(1996)}]{PJ96}
{Protheroe}, R.~J., \& {Johnson}, P.~A. 1996, Astroparticle Physics, 4, 253,
  \dodoi{10.1016/0927-6505(95)00039-9}

\bibitem[{{Rodrigues} {et~al.}(2021){Rodrigues}, {Garrappa}, {Gao}, {Paliya},
  {Franckowiak}, \& {Winter}}]{Rodrigues:2020fbu}
{Rodrigues}, X., {Garrappa}, S., {Gao}, S., {et~al.} 2021, \apj, 912, 54,
  \dodoi{10.3847/1538-4357/abe87b}

\bibitem[{Rodrigues {et~al.}(2024{\natexlab{a}})Rodrigues, Karl, Padovani,
  Giommi, Paiano, Falomo, Petropoulou, \& Oikonomou}]{Rodrigues:2024fhu}
Rodrigues, X., Karl, M., Padovani, P., {et~al.} 2024{\natexlab{a}}, Astron.
  Astrophys., 689, A147, \dodoi{10.1051/0004-6361/202450592}

\bibitem[{Rodrigues {et~al.}(2024{\natexlab{b}})Rodrigues, Paliya, Garrappa,
  Omeliukh, Franckowiak, \& Winter}]{Rodrigues:2023vbv}
Rodrigues, X., Paliya, V.~S., Garrappa, S., {et~al.} 2024{\natexlab{b}},
  Astron. Astrophys., 681, A119, \dodoi{10.1051/0004-6361/202347540}

\bibitem[{{Rudolph} {et~al.}(2022){Rudolph}, {Bo{\v{s}}njak}, {Palladino},
  {Sadeh}, \& {Winter}}]{Rudolph:2021cvn}
{Rudolph}, A., {Bo{\v{s}}njak}, {\v{Z}}., {Palladino}, A., {Sadeh}, I., \&
  {Winter}, W. 2022, \mnras, 511, 5823, \dodoi{10.1093/mnras/stac433}

\bibitem[{{Rudolph} {et~al.}(2023{\natexlab{a}}){Rudolph}, {Petropoulou},
  {Bo{\v{s}}njak}, \& {Winter}}]{2023ApJ...950...28R}
{Rudolph}, A., {Petropoulou}, M., {Bo{\v{s}}njak}, {\v{Z}}., \& {Winter}, W.
  2023{\natexlab{a}}, \apj, 950, 28, \dodoi{10.3847/1538-4357/acc861}

\bibitem[{{Rudolph} {et~al.}(2023{\natexlab{b}}){Rudolph}, {Petropoulou},
  {Winter}, \& {Bo{\v{s}}njak}}]{2023ApJ...944L..34R}
{Rudolph}, A., {Petropoulou}, M., {Winter}, W., \& {Bo{\v{s}}njak}, {\v{Z}}.
  2023{\natexlab{b}}, \apjl, 944, L34, \dodoi{10.3847/2041-8213/acb6d7}

\bibitem[{{Stathopoulos} \& {Petropoulou}(2025)}]{2025arXiv250708680S}
{Stathopoulos}, S.~I., \& {Petropoulou}, M. 2025, arXiv e-prints,
  arXiv:2507.08680, \dodoi{10.48550/arXiv.2507.08680}

\bibitem[{{Stathopoulos} {et~al.}(2024{\natexlab{a}}){Stathopoulos},
  {Petropoulou}, {Sironi}, \& {Giannios}}]{2024arXiv240601211S}
{Stathopoulos}, S.~I., {Petropoulou}, M., {Sironi}, L., \& {Giannios}, D.
  2024{\natexlab{a}}, arXiv e-prints, arXiv:2406.01211,
  \dodoi{10.48550/arXiv.2406.01211}

\bibitem[{{Stathopoulos} {et~al.}(2024{\natexlab{b}}){Stathopoulos},
  {Petropoulou}, {Vasilopoulos}, \& {Mastichiadis}}]{lehamoc}
{Stathopoulos}, S.~I., {Petropoulou}, M., {Vasilopoulos}, G., \&
  {Mastichiadis}, A. 2024{\natexlab{b}}, \aap, 683, A225,
  \dodoi{10.1051/0004-6361/202347277}

\bibitem[{{Stecker}(1979)}]{1979ApJ...228..919S}
{Stecker}, F.~W. 1979, ApJ, 228, 919, \dodoi{10.1086/156919}

\bibitem[{Thomas(1949)}]{thomas1949}
Thomas, L.~H. 1949, Watson Scientific Computing Laboratory Report, DSIR O--L

\bibitem[{{Urry} \& {Padovani}(1995)}]{Urry95}
{Urry}, C.~M., \& {Padovani}, P. 1995, \pasp, 107, 803, \dodoi{10.1086/133630}

\bibitem[{{V{\"o}lk} {et~al.}(1996){V{\"o}lk}, {Aharonian}, \&
  {Breitschwerdt}}]{Volk96}
{V{\"o}lk}, H.~J., {Aharonian}, F.~A., \& {Breitschwerdt}, D. 1996, \ssr, 75,
  279, \dodoi{10.1007/BF00195040}

\bibitem[{Vurm \& Poutanen(2009)}]{Vurm:2008ue}
Vurm, I., \& Poutanen, J. 2009, Astrophys. J., 698, 293,
  \dodoi{10.1088/0004-637X/698/1/293}

\bibitem[{{Wang} {et~al.}(2022){Wang}, {Liu}, {Petropoulou}, {Oikonomou},
  {Xue}, \& {Wang}}]{2022PhRvD.105b3005W}
{Wang}, Z.-R., {Liu}, R.-Y., {Petropoulou}, M., {et~al.} 2022, \prd, 105,
  023005, \dodoi{10.1103/PhysRevD.105.023005}

\bibitem[{{Yuan} \& {Winter}(2023)}]{2023ApJ...956...30Y}
{Yuan}, C., \& {Winter}, W. 2023, \apj, 956, 30,
  \dodoi{10.3847/1538-4357/acf615}

\bibitem[{{Yuan} {et~al.}(2024{\natexlab{a}}){Yuan}, {Winter}, \&
  {Lunardini}}]{2024ApJ...969..136Y}
{Yuan}, C., {Winter}, W., \& {Lunardini}, C. 2024{\natexlab{a}}, \apj, 969,
  136, \dodoi{10.3847/1538-4357/ad50a9}

\bibitem[{{Yuan} {et~al.}(2024{\natexlab{b}}){Yuan}, {Zhang}, {Winter}, \&
  {Murase}}]{2024ApJ...974..162Y}
{Yuan}, C., {Zhang}, B.~T., {Winter}, W., \& {Murase}, K. 2024{\natexlab{b}},
  \apj, 974, 162, \dodoi{10.3847/1538-4357/ad6c50}

\bibitem[{{Zacharias} {et~al.}(2022){Zacharias}, {Reimer}, {Boisson}, \&
  {Zech}}]{zacharias}
{Zacharias}, M., {Reimer}, A., {Boisson}, C., \& {Zech}, A. 2022, \mnras, 512,
  3948, \dodoi{10.1093/mnras/stac754}

\bibitem[{{Zech} {et~al.}(2017){Zech}, {Cerruti}, \& {Mazin}}]{Zech17}
{Zech}, A., {Cerruti}, M., \& {Mazin}, D. 2017, \aap, 602, A25,
  \dodoi{10.1051/0004-6361/201629997}

\bibitem[{{Zhang} \& {M{\'e}sz{\'a}ros}(2001)}]{Zhang01}
{Zhang}, B., \& {M{\'e}sz{\'a}ros}, P. 2001, \apj, 559, 110,
  \dodoi{10.1086/322400}

\end{thebibliography}
\bibliographystyle{aasjournal}

\appendix
\section{Semi-analytical calculation of neutrino spectra}
\label{app:analytical}

Throughout the paper, we compare the neutrino fluxes obtained with the five numerical codes to the neutrino fluxes obtained 
with a simple semi-analytical calculation in order to 
test the range of applicability of this method, which is frequently employed in the literature. 

\begin{figure}
    \centering
    \includegraphics[width=0.7\linewidth]{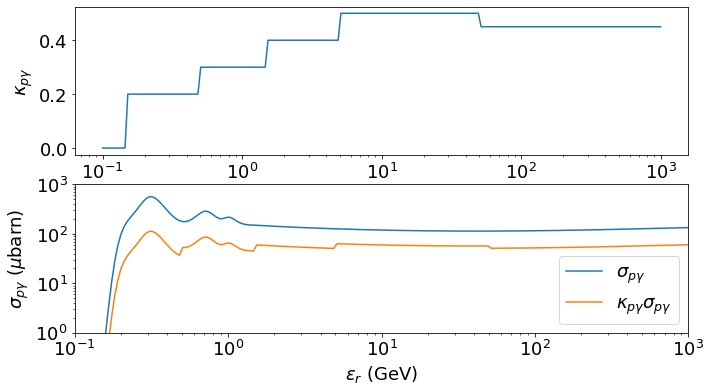}
    \caption{Parametrization of the inelasticity (top panel) and the cross section (bottom panel) for photomeson production obtained from GEANT~4 as a function of the interaction energy \citep{Murase:2005hy}.}
    \label{fig:sigma-pg}
\end{figure}

The photomeson production timescale for protons with Lorentz factor $\gamma_p$ interacting with an isotropic photon distribution with differential number density $n_{\rm ph}(\epsilon)$ is defined as~\citep{1979ApJ...228..919S},
\begin{equation} 
t_{\rm p\gamma}^{-1} (\gamma_p) = \frac{c}{2 \gamma_p^2}
\int^{\infty}_{\epsilon_{\rm th}} {\rm d}
\epsilon_{\gamma} \sigma_{\rm p \gamma}(\epsilon_{\gamma}) \kappa_{\rm p \gamma}(\epsilon_{\gamma})
\epsilon_{\gamma}
\int^{\infty}_{\epsilon_{\gamma}/(2\gamma_p)} {\rm d}
\epsilon \epsilon^{-2}
  n_{\rm ph}(\epsilon),
\label{eq:pgammaRate}
\end{equation}
where, $\epsilon_{\gamma}$ is the photon energy in the proton rest frame, 
$\epsilon_{\rm th} \sim 145$~MeV is the threshold energy for pion production, and
$\sigma_{\rm p \gamma}$ and $\kappa_{\rm p \gamma}$ are the cross section and inelasticity of photomeson interactions, respectively. 
For $\sigma_{\rm p \gamma}$ and $\kappa_{p\gamma}$ we use a fit to the energy dependent cross section and a multi-step parametrization of the inelasticity (shown in Fig.~\ref{fig:sigma-pg}) obtained by running GEANT~4  (see also \citealp{Murase:2005hy}).

In general, the fraction of energy converted to pions is estimated as,
$f_{\rm p \gamma} \equiv t_{\rm cool} / t_{\rm p\gamma}$, where $t_{\rm cool}$ is the 
proton energy loss cooling time, defined as
\begin{equation} 
t_{\rm cool}^{-1} \equiv t_{\rm cross}^{-1} + t_{p,{\rm syn}}^{-1} + t_{\rm p \gamma}^{-1} + t_{\rm BH}^{-1},
\end{equation}
where the synchrotron cooling time for protons with energy $\varepsilon_p$ 
in a magnetic field with strength $B$
is given by, $t_{p,{\rm syn}} = 6 \pi m_p^4c^3/(m_e^2 \sigma_T B^2
\varepsilon_p)$. 

The crossing time, $t_{\rm cross} = r_{b}/c$, approximates the adiabatic energy loss rate. The Bethe-Heitler energy loss timescale, $t_{\rm BH}$, is calculated as described in Appendix~A of \cite{lehamoc} (see Eq.~A.24).  

The per-flavor neutrino luminosity per logarithmic energy is estimated as,
\begin{equation}
\varepsilon_{\nu}L_{\varepsilon_{\nu}} \approx \frac{1}{8} f_{p\gamma}(\varepsilon_{p}) \varepsilon_{p}L_{\varepsilon_{p}}
\label{eq:Lnu-analytic}
\end{equation}
where $\varepsilon_{p}L_{\varepsilon_{p}}$ is the injected proton luminosity per logarithmic energy
and the neutrinos are assumed to be produced with energy $\varepsilon_{\nu} \sim 0.05 \varepsilon_p$. 

\begin{figure}
    \centering
    \includegraphics[width=0.47\linewidth]{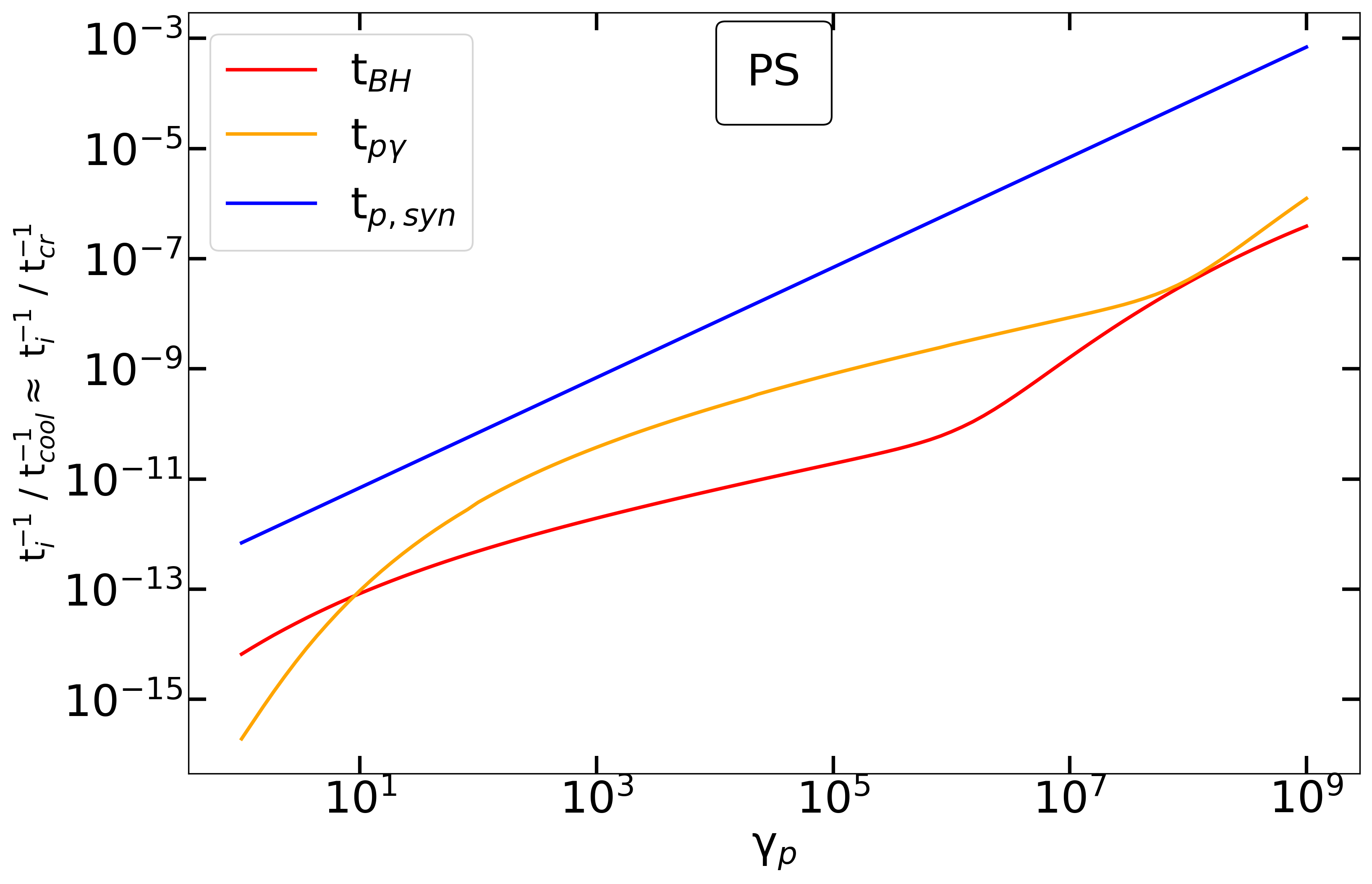}
    \hfill
    \includegraphics[width=0.47\linewidth]{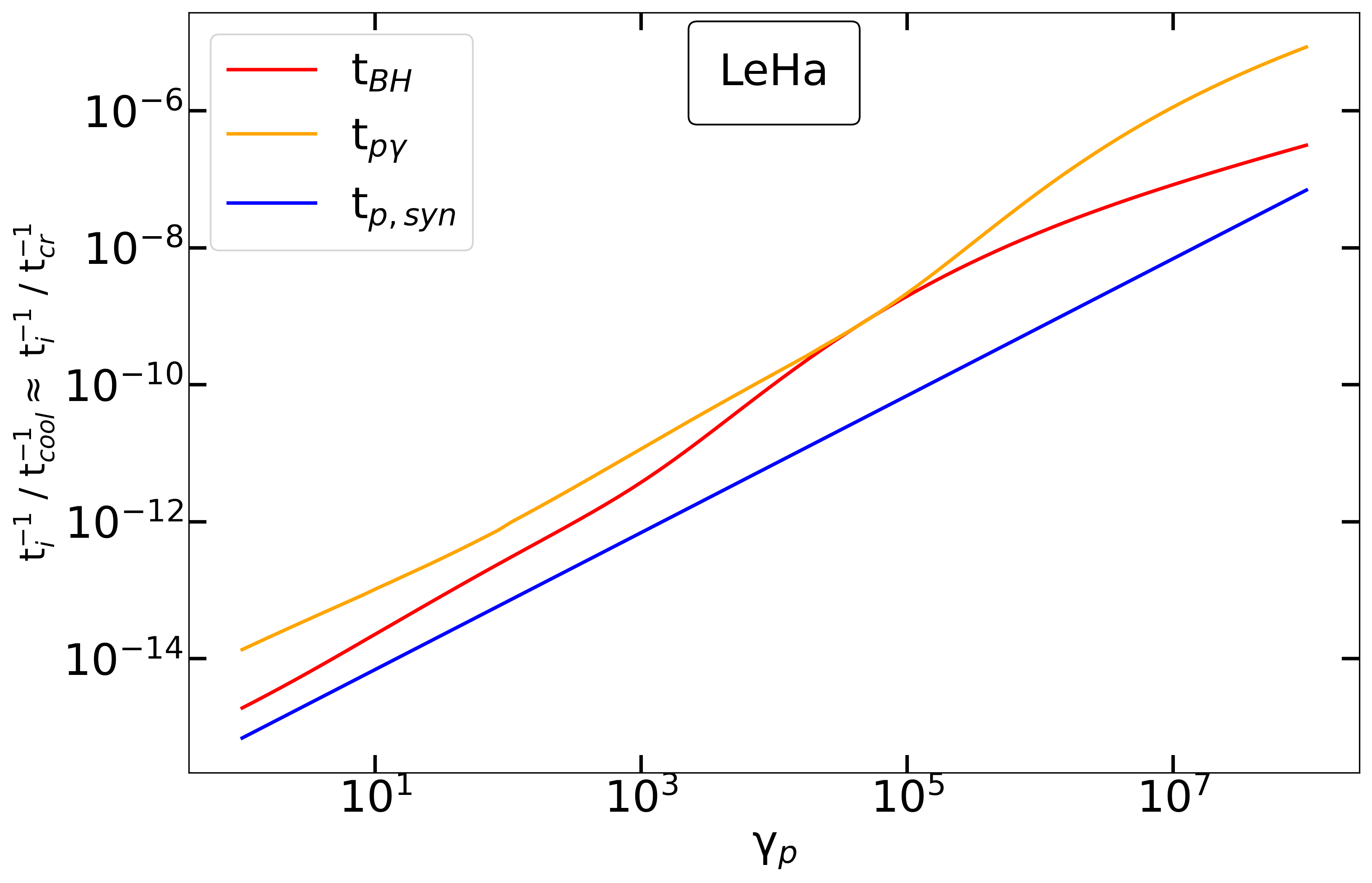}
    \caption{Normalized proton loss rates for the PS and LeHa scenarios. In both cases, the Bethe-Heitler, photomeson and proton synchrotron loss rates are much smaller than the proton escape rate, hence $t_{\rm cool} \approx t_{\rm cross}$.}
    \label{fig:proton-rates}
\end{figure}

For the semi-analytical calculation of the neutrino spectra in the PS and LeHa cases (displayed in Figs.~\ref{fig:psyn} and \ref{fig:lepto-hadronic}) we evaluated $t_{\rm cool}$ as described above using as target photon fields the one computed with the code \lehamoc. The proton loss rates for these two scenarios are presented in Fig.~\ref{fig:proton-rates}. The ordering of the loss rates $t^{-1}_{\rm p\gamma} \approx t^{-1}_{\rm BH} \ll t^{-1}_{\rm p,syn}$ for $\gamma_{p, \max}=10^8$ in the PS scenario is also reflected at the photon luminosity of the respective SED components (see Fig.~\ref{fig:psyn}). 
For the p$\gamma$-MONOGB and p$\gamma$-PLPL cases, where only photomeson interactions and proton escape were taken into account, the proton cooling timescale was computed as $t^{-1}_{\rm cool} = t_{\rm cross}^{-1} + t_{\rm p\gamma}^{-1}$. In all cases $\varepsilon_p L_{\varepsilon_p}$ was described by the proton distribution used as input in \lehamoc.

We checked that adopting a simpler, two-step function approximation for the single-pion resonance channel and the multipion channel respectively following the approach of~\citet{Atoyan03}
has a negligible effect on the semi-analytic results, in the sense that it is smaller than the difference between the semi-analytic results and the full numerical results. 

\section{Additional test cases}
\label{app:other}
\subsection{Electron cooling in the Klein-Nishina regime}\label{appendix:KN}

In this section we compare the codes in their treatment of radiative losses due to inverse Compton scatterings in the Klein-Nishina limit. In this regime, electrons lose a significant fraction of their energy per scattering, which can be thought of as a catastrophic energy-loss process. The treatment of energy losses in the deep Klein-Nishina regime differs among the codes under comparison. Currently, \lehamoc accounts only for cooling in the Thomson regime following Eqs.~(42)-(43) in \cite{MK95}. \paris uses a continuous energy loss scheme, which was introduced by \cite{Moderski_2005}, that approximates the reduction of electron cooling in the Klein-Nishina regime. 

\boet and \ath use Eq.~(2.56) from \cite{BG70}. In \ath there is also the option to treat Klein-Nishina cooling as a catastrophic energy loss term, following Eq.~(2.57) from \cite{BG70} -- see also Eq.~(45) in \cite{MK95}. 
As a test case we consider electrons injected into a spherical source of radius $10^{15}$~cm, with a power-law distribution of slope $p=1.9$,  starting from  $\gamma_{e,\min}=1$ and having an exponential cutoff at $\gamma_{e,\max}=10^6$. The electron injection compactness is $10^{-4}$.
 
Electrons inverse Compton scatter photons from a grey-body photon field of temperature $T=10^6$~K and compactness $\ell_{\gamma}=100$ (or $u_{\gamma}=3.7\times 10^5$~erg cm$^{-3}$). The typical target photon energy is $\epsilon_{GB}\sim 2.7 k_B T \simeq 233$~eV. For the adopted parameters, electrons with $\gamma_e \ll m_e c^2 / \epsilon_{GB} \sim 2193$ will experience Thomson cooling, while electrons from the high-energy tail of the distribution will undergo reduced cooling due scatterings in the Klein-Nishina regime. For the purposes of this comparison, we also switch off synchrotron radiation and $\gamma \gamma$ pair production.

\begin{figure}
    \centering
    \includegraphics[width=0.45\textwidth]{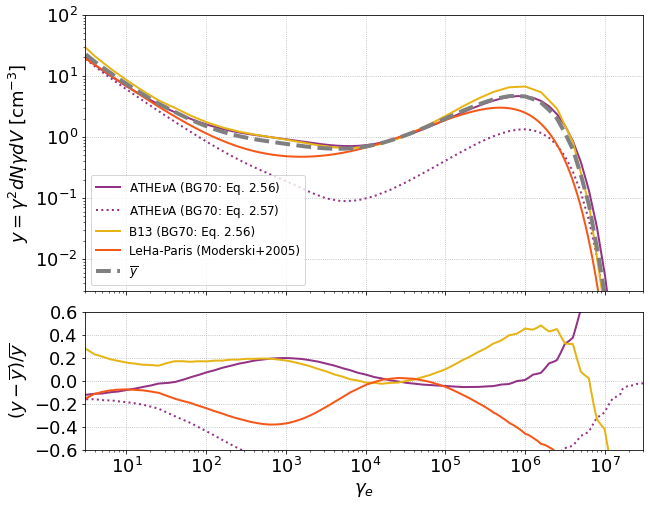}
    \hfill
    \includegraphics[width=0.45\textwidth]{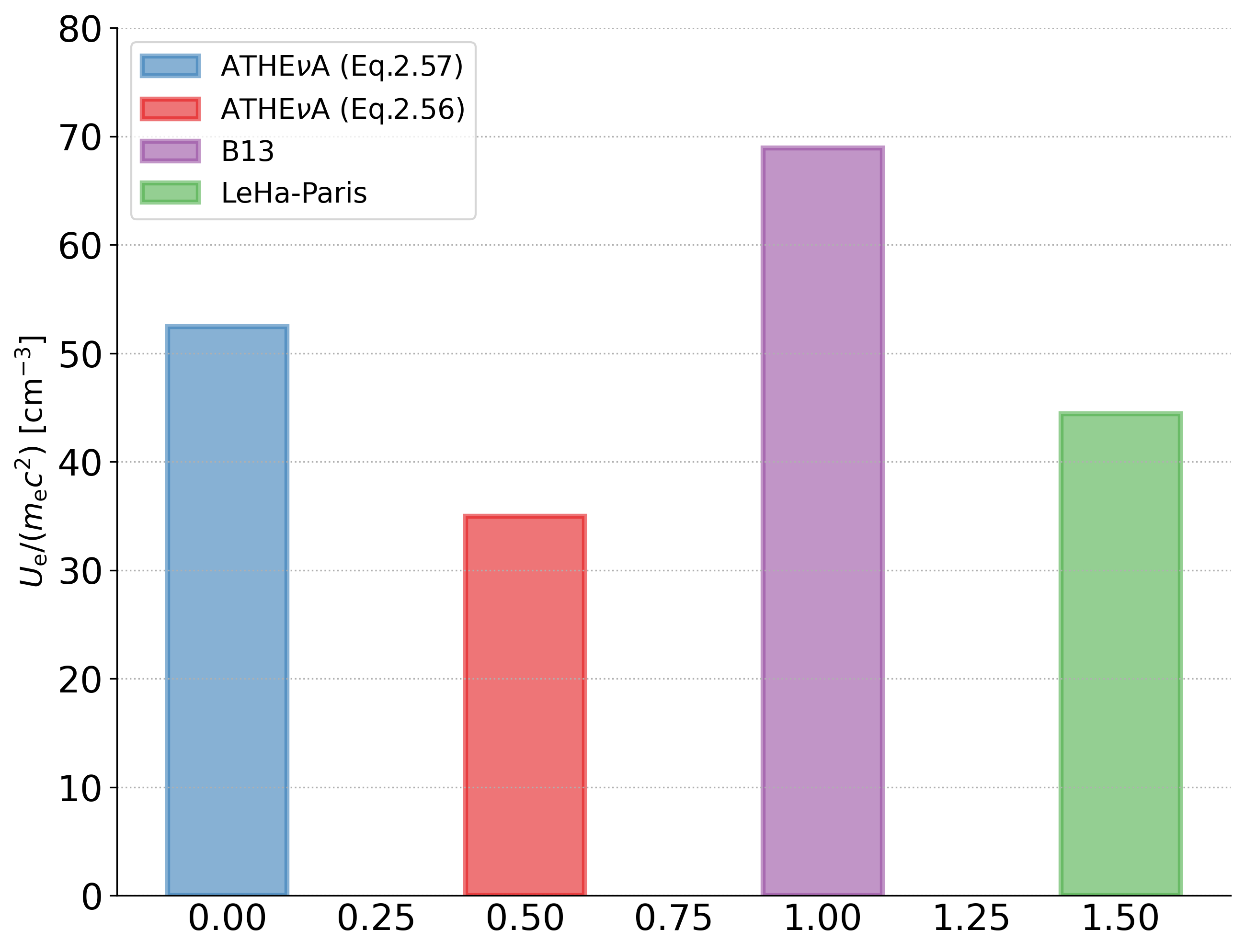}
    \caption{Effects of Klein-Nishina cooling. \textit{Left panel:} Steady state electron distributions. The calculation of the average does not include the dotted curve. \textit{Right panel:} Integrated energy density of relativistic electrons at steady state.
    }
    \label{fig:KN}
\end{figure}

We present the results for this test in Fig.~\ref{fig:KN}. The results of \boet and \ath, which are obtained using the same prescription for Klein-Nishina cooling, are similar for a wide range of Lorentz factors. In particular, their agreement is excellent (relative difference less than $\pm 5\%$) for $3 \cdot 10^3 \lesssim \gamma_e \lesssim 10^5$. The maximum difference between the two codes is $\sim 20\%$ and is found at the cutoffs. The catastrophic-loss approximation for the Klein-Nishina cooling (dotted line) overestimates electron cooling for $\gamma_e \gtrsim 10^2$. As a result, the energy density of the steady-state electron distribution is underestimated by a factor of $\sim10$ at $\gamma_e > 10^3$. The effect on the integrated energy density of the distribution is much smaller (see blue and red bars in the right panel), as this is driven by the lowest energy particles, where both prescriptions agree well. Finally, the approximation by \cite{Moderski_2005}, which is employed in \paris, yields similar results with the other two codes. However, it overestimates cooling at the highest energies (hence, the shift of the exponential cutoff of the distribution to lower energies), and close to the transition region between Thomson and Klein-Nishina cooling at $\gamma_e \sim 10^3$. 

\subsection{Mono-energetic protons on grey-body photons}\label{appendix:monoenergetic7}

In this section we provide the result for the mono-energetic proton test (p$\gamma$-MONOGB) with a proton energy higher by a factor of 10 compared to the one used in the main text ($\gamma_{p, \min} = 10^7$, $\gamma_{p, \max} = 10^{7.2}$). More energetic protons interact with photons from the peak of the grey-body distribution further away from the energy threshold of the interaction.

\begin{figure}[h!]
\centering
\includegraphics[width = .45 \textwidth]{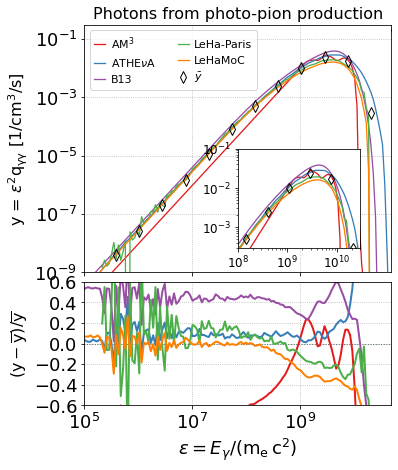}
\includegraphics[width = .45 \textwidth]{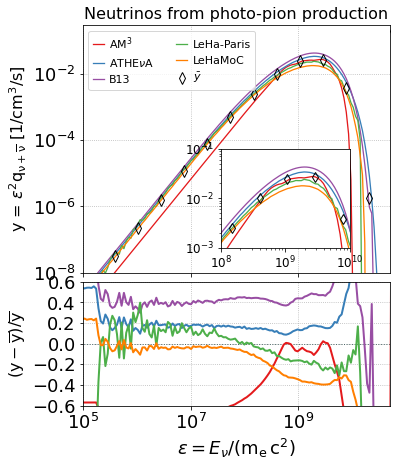}
\includegraphics[width = .45 \textwidth]{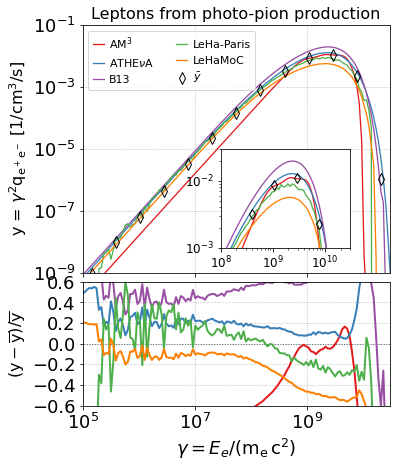}
\includegraphics[width = .47 \textwidth]{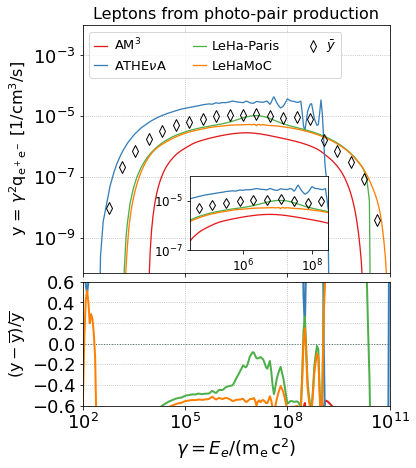}
\caption{Same as in Fig.~\ref{fig:hadronic_rates_mono_6} but for more energetic protons with Lorentz factors between $10^7$ and $10^{7.2}$.}
\label{fig:hadronic_rates_mono_7}
\end{figure}

As the interaction energy becomes higher we find larger differences among the five codes. While the secondary photopion production spectra of \ath and \paris are in good agreement (see also Fig.~\ref{fig:hadronic_rates_mono_6}), the \boet spectra are systematically higher than the mean spectrum by $\sim40\%$ for a wide range of energies. Moreover, the spectrum of leptons from \boet appears to have different spectral shape close the peak. We argue that these differences arise from the different implementation methods used to compute the photopion production spectra. \ath uses tabulated results of the Monte Carlo code {\sc sophia} and performs appropriate interpolation to match the proton and photon energy grids.  \paris actually executes {\sc sophia} and uses the direct output of the Monte Carlo simulation. \boet employs the formalism of \cite{Kelner_2008}, which provides analytical parametrizations to the {\sc sophia} results. Inspection of Figs. 1 to 3  in \cite{Kelner_2008} shows that these parametrizations deviate more as the energy of interaction increases. While the parametrized spectra describe very well the peak of the binned {\sc sophia} energy spectra, they slightly under- or over-estimate the {\sc sophia} results away from the peak. For \am, the discrepancy with the other models increases at higher interaction energies for the monochromatic case, because the original approach in \citet{Hummer:2010vx} not capture the interaction energy-dependence of the secondary re-distribution functions very well and multi-pion processes play a larger role at these energies, which are more difficult to describe in terms of multiplicities and re-distribution functions; see also discussion in main text.
The Bethe-Heitler injection also shows a larger dispersion compared to the test with $\gamma_{p, \min} = 10^6$, $\gamma_{p, \max} = 10^{6.2}$ shown in the main text. In this case as well, the overestimation by \ath is related to the very narrow proton distribution, as discussed in more details in Appendix \ref{app:BH}.

\subsection{Power-law protons on grey-body photons}
\begin{figure}
\centering
\includegraphics[width = .32 \textwidth]{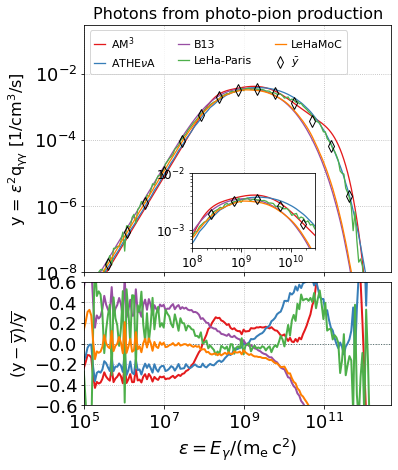}
\includegraphics[width = .32 \textwidth]{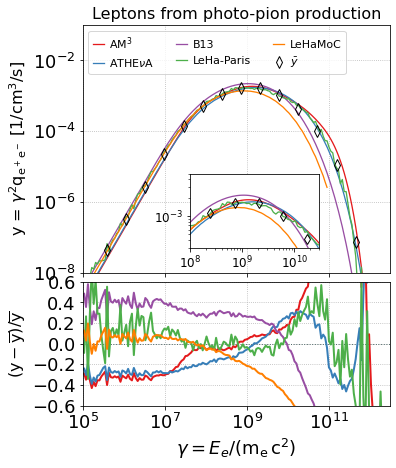}
\includegraphics[width = .32 \textwidth]{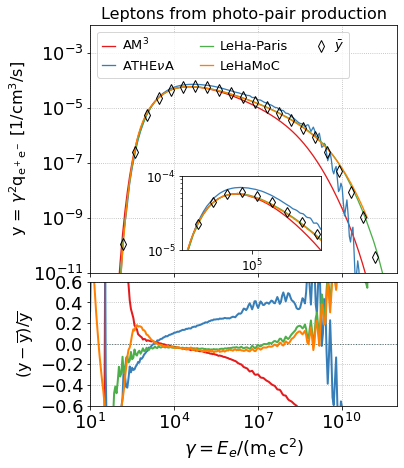}
\caption{Production rates per unit volume of secondaries from photopion production (left and middle panels) and photopair production (right panel) for the case of a power-law proton distribution interacting with a grey-body radiation field. Inset panels show a zoom into the energy range around the peak of the curves. The residuals of each code with respect to the mean value of the results are plotted in the bottom panels. The proton distribution and grey-body parameters are the same as in those used in models p$\gamma-$PLPL and p$\gamma-$MONOGB respectively (see Table~\ref{tab:parameters}).}
\label{fig:hadronic_rates_gb}
\end{figure}

In this section we present results for interactions between power-law protons with grey-body photons. The parameters used for the proton distribution are the same as in model p$\gamma-$PLPL. The grey-body radiation field has the same temperature and energy density as in the p$\gamma-$MONOGB model (see Table~\ref{tab:parameters}). The injection rates of secondaries from photo-pion and from photopair production are shown in Fig. \ref{fig:hadronic_rates_gb}. 

\newpage
\section{Bethe-Heitler numerical implementation}\label{app:BH}

\begin{figure}
    \centering
    \includegraphics[width=0.99\textwidth]{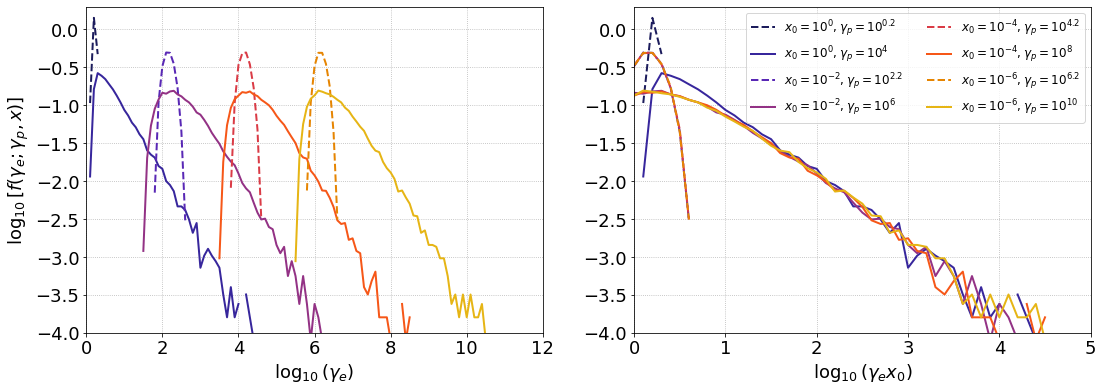}
    \caption{\textit{Left panel}: Electron/positron distributions, $\gamma_e \, dN/d\gamma_e$, from Bethe-Heitler pair production for various proton and target photon energies. \textit{Right panel}: Same as in the left panel, but after multiplying the electron/positron energies by $x_0$.}
    \label{fig:ray2}
\end{figure}

The numerical implementation of Bethe-Heitler in \ath is described in \cite{MPK05}, but some details, which are missing there, will be elucidated on here. A Monte Carlo code was used to calculate the electron-positron spectra, without differentiating between electrons and positrons. Four photon energies were used as targets: $x_0 = 10^{-6}, 10^{-4}, 10^{-2},$ and $10^0$, in units of $m_e c^2$. Those were allowed to interact with protons of energies ranging from slightly above $\gamma_p = x_0^{-1}$ (to satisfy the threshold requirement) up to $\gamma_p = 10^4 \ x_0$. 25,000 interactions were performed for each energy pair, resulting in a list of interaction rates and electron-positron distributions. However, the whole list is not used in the \ath code for the following reason. As we can see in Figure~\ref{fig:ray2}, for target photon energies of $x_0 = 10^{-6}, 10^{-4}$ and $10^{-2}$ the electron-positron distributions are practically identical for commensurate proton energies (so that the interaction energy, $\gamma_p x_0$, is conserved). There is only a shift in electron-positron Lorentz factors by a factor of $x_0$, but once that is corrected for (right panel), the distributions are identical. The only exception to this self-similar behavior is the pair distributions produced by proton interactions with very energetic photons ($x_0=1$). Therefore, to save on memory, \ath only uses the Monte Carlo results for $x_0=10^{-6}$ and, when considering the interaction of $\gamma_p$ protons with $x$ photons, retrieves the electron/positron distribution from $\gamma_{p0} = \gamma_p x/x_0$ protons interacting with  $x_0=10^{-6}$ photons. That distribution is then energy-shifted, so that any electron energy $\gamma_{e0}$ will be converted to $\gamma_e = \gamma_{e0} x_0/x$. Since the Monte Carlo calculations were performed with a logarithmic step of 0.1, an interpolation is performed to obtain more accurate values of interaction rates, since photons in \ath generally do not have energies in neat logarithmic multiples of 0.1. There is an obvious limitation to this method, as seen in Figure~\ref{fig:ray2}. The distributions are very different for the highest target photon energy ($x_0=1$), which serves as a target for very low proton energies. Therefore, \ath simply does not consider Bethe-Heitler pair production for $\gamma_p < 10^3$, to avoid this high photon energy regime. 

In \lehamoc the prescription of \cite{Kelner_2008} for the Bethe Heitler pair spectrum (see Eq. 62) has been implemented. This is valid in the ultrarelativistic regime for protons $\gamma_{p}\gg 1$ and for target energy photons $\epsilon$ satisfying the following inequalities, $m_pc^2\gg \gamma_p \epsilon \gg m_e c^2$. The computation of the innermost integral in Eq. 62 of \cite{Kelner_2008}, which involves the differential cross section, can significantly increase the simulation's execution time. To address this issue in \lehamoc,  the integral has been pre-calculated for various combinations of the electron and proton Lorentz factor $\gamma_e$, $\gamma_p$, and the energy of the photon in the proton's rest frame $\omega$, drawn from a large sample, based on the parameters of each simulation. The pre-calculated values are then stored in an array as described in Appendix A of \cite{lehamoc}.

In Fig.~\ref{fig:BH_mono_to_PL} we compare the pair production spectra obtained with \ath and \paris for proton distributions that become progressively wider in energy. We start from the effectively monoenergetic case discussed in the main text for $\gamma_p \in (10^6, 10^{6.2})$ (see $p\gamma$-MONOGB in Table~\ref{tab:parameters}) and decrease the lower energy of the distribution by 0.1 or 0.3 in logarithm. In all cases, the proton injection compactness (or energy density) is fixed, thus leading to a reduction of the normalization in the power-law energy spectra displayed on the right panel of the figure. We find that \ath overpredicts the energy injection spectra by a factor of 1.5 with respect to \paris when the proton energy distribution is very narrow, but the code results converge to a ratio of $\sim 1.1$ as the distribution becomes progressively wider (left hand-side panels). We argue that these differences arise from differences in the proton energy grid resolution. The latter is fixed in \ath (10 grid points per energy decade), while in \paris the resolution is much higher (see markers on the right panel of Fig.~\ref{fig:BH_mono_to_PL}). To test our hypothesis, we compute the injection spectra of secondaries using the Bethe-Heitler emissivity of \lehamoc, and the proton distributions of \ath and \paris as is. The results are presented in Fig.~\ref{fig:BH_mono_to_PL_lehamoc}. Despite the different implementations on the Bethe-Heitler emissivity among the three codes, the results are very similar when the same resolution is used for the proton distributions. While the fixed proton energy grid resolution used in \ath can overestimate the injection energy for very narrow proton distributions, the code conserved energy: the energy injection rate into pairs equals the energy loss rate at every time step, thus ensuring the internal self consistency of the numerical calculations.

\begin{figure}
    \centering
    \includegraphics[width=0.99\textwidth]{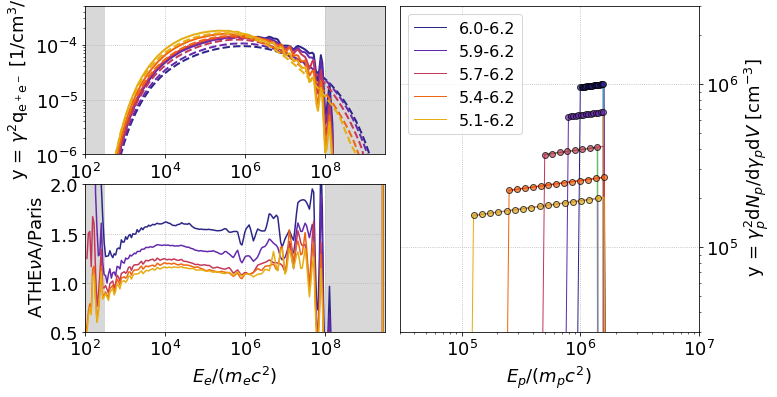}
    \caption{Photo-pair production spectra (top left) computed for interactions of protons with a progressively broader distribution (right panel) with photons from a grey-body distribution of temperature $10^6$~K and compactness $\ell_{\gamma} = 8.1\times10^{-6}$. The ratio of the \ath and \paris results for different pair energies is shown on the left bottom panel. }
    \label{fig:BH_mono_to_PL}
\end{figure}

\begin{figure}
    \centering
    \includegraphics[width=0.99\textwidth]{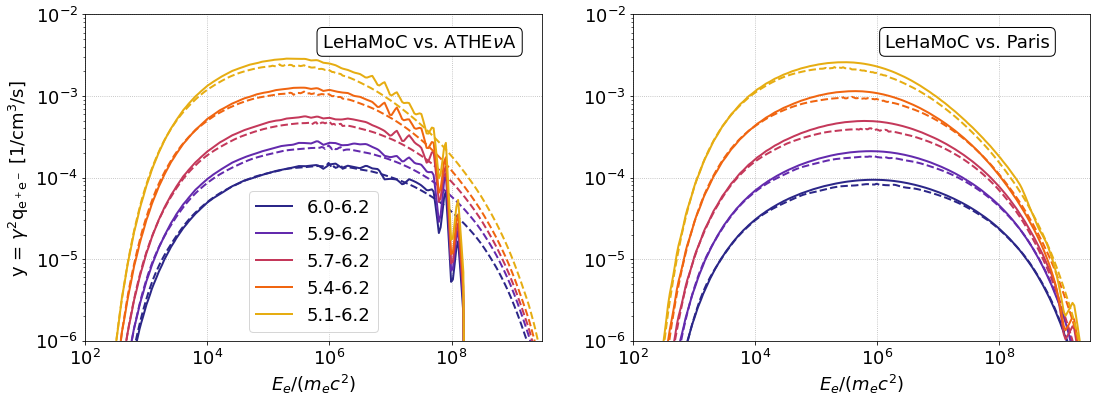}
    \caption{Photo-pair production spectra computed using the Bethe-Heitler emissivity from \lehamoc and the proton distributions from \ath (left) and \paris (right) as is (dashed lines). Solid lines show the output of \ath and \paris codes. Starting from the bottom, each curve has been shifted vertically by a factor of $2^n$ for clarity purposes. All parameters are the same as in Fig.~\ref{fig:BH_mono_to_PL}.}
    \label{fig:BH_mono_to_PL_lehamoc}
\end{figure}
\end{document}